\documentclass[11pt]{article}
\pdfoutput=1
\usepackage[utf8]{inputenc}
\usepackage{cite}
\usepackage{amsmath,amssymb,amsbsy,amstext,amsthm,simplewick,amsfonts}
\usepackage{graphicx}
\usepackage{wrapfig}
\usepackage{upgreek}
\usepackage{bm} 
\usepackage{framed}
\usepackage{bbm}
\usepackage{textcomp}
\usepackage{tikz}
\usepackage{pifont}
\usepackage{mathtools}
\usetikzlibrary{matrix,shapes,fit,tikzmark,calc}
\usepackage{adjustbox}
\usepackage{makecell}
\usepackage{tcolorbox}
\usepackage{physics}
\usepackage{empheq}
\usepackage[normalem]{ulem}
\usepackage{enumitem}
\usepackage{braket} 
\usepackage{array}
\usepackage{dsfont}
\usepackage{ulem}
\usepackage{bbold}
\usepackage{tocloft}

\numberwithin{equation}{section}

\def\be{\begin{equation}}
\def\ee{\end{equation}}

\def\e{\epsilon}

\def\ba{\begin{eqnarray}}
\def\ea{\end{eqnarray}}

\def\bfx{\textbf{x}}

\def\bfk{{\textbf{k}}}

\def\del{\partial}

\usepackage{caption}
\usepackage{subcaption}

\usepackage[framemethod=default]{mdframed}
\newmdenv[skipabove=7pt,
skipbelow=7pt,
rightline=false,
leftline=false,
topline=false,
bottomline=false,
backgroundcolor=gray!10,
linecolor=gray,
innerleftmargin=5pt,
innerrightmargin=5pt,
innertopmargin=5pt,
innerbottommargin=5pt,
leftmargin=0cm,
rightmargin=0cm,
linewidth=4pt]{eBox}
\newmdenv[skipabove=7pt,
skipbelow=7pt,
rightline=false,
leftline=false,
topline=false,
bottomline=false,
backgroundcolor=gray!10,
linecolor=gray,
innerleftmargin=5pt,
innerrightmargin=5pt,
innertopmargin=-5pt,
innerbottommargin=5pt,
leftmargin=0cm,
rightmargin=0cm,
linewidth=4pt]{eBox2}

\definecolor{blue3}{RGB}{31,119,180}
\definecolor{red3}{RGB}{214,39,40}
\definecolor{orange3}{RGB}{255,127,14}
\definecolor{green3}{RGB}{44,160,44}

\usepackage{colortbl}
\definecolor{lightgreen}{cmyk}{0.2, 0, 0.2, 0.2}
\definecolor{lightgray}{cmyk}{0.1,0.2,0,0.1}
\definecolor{lightgray2}{cmyk}{0.1,0.1,0,0.1}

\usepackage[linktocpage, colorlinks]{hyperref}
\hypersetup{linkcolor=teal,citecolor=orange3,urlcolor=green3,pdfencoding=auto}

\makeatletter
\newlength{\apb@width}
\newcommand{\autoparbox}[2][c]{\settowidth{\apb@width}{#2}\parbox[#1]{\apb@width}{#2}}

\makeatother


\def\rd{{\rm d}}

\renewcommand{\v}[1]{\ensuremath{\mathbf{#1}}} 

\def\nn{\nonumber}

\def\beq{\begin{equation}}
\def\eeq{\end{equation}}

\newcommand{\ccdot}{\! \cdot \!}

\allowdisplaybreaks[1]
\begin{document}


\begin{titlepage}
\setcounter{page}{1} \baselineskip=15.5pt 
\thispagestyle{empty}

\begin{center}
{\fontsize{22}{22}\centering \bf The graviton four-point function\\[6pt]
in de Sitter space}\\
\end{center}

\vskip 12pt
\begin{center}
\noindent
{\fontsize{12}{18}\selectfont James Bonifacio,\footnote{\tt bonifacio@phy.olemiss.edu}$^{,\ast}$ Harry Goodhew,\footnote{\tt hfg23@cam.ac.uk}$^{,\dagger}$ Austin Joyce,\footnote{\tt austinjoyce@uchicago.edu}$^{,\star}$\\[4pt]
Enrico Pajer,\footnote{\tt ep551@cam.ac.uk}$^{,\dagger}$ and David Stefanyszyn\footnote{\tt david.stefanyszyn@nottingham.ac.uk}$ ^{,\ddag}$ }
\end{center}

\begin{center}
\vskip 10pt
${}^\ast$\textit{Department of Physics and Astronomy, University of Mississippi, University, MS 38677, USA}\\[10pt]
${}^\dagger$\textit{Department of Applied Mathematics and Theoretical Physics, University of Cambridge, Wilberforce Road, Cambridge, CB3 0WA, UK}\\[10pt] 
${}^\star$\textit{Kavli Institute for Cosmological Physics, Department of Astronomy and Astrophysics,\\
University of Chicago, Chicago, IL 60637, USA } \\[10pt] 
${}^\ddag$\textit{School of Mathematical Sciences \& School of Physics and Astronomy,
University of Nottingham, University Park, Nottingham, NG7 2RD, UK} 
\end{center}


\vspace{1.4cm}

\noindent We compute the tree-level late-time graviton four-point correlation function, and the related quartic wavefunction coefficient, for Einstein gravity in de Sitter spacetime. We derive this result in several ways: by direct calculation, using the in-in formalism and the wavefunction of the universe; by a heuristic derivation leveraging the flat space wavefunction coefficient; and by using the boostless cosmological bootstrap, in particular the combination of the cosmological optical theorem, the amplitude limit, and the manifestly local test. We find agreement among the different methods.


\end{titlepage} 


\newpage
\setcounter{tocdepth}{2}
\setcounter{page}{2}
{
\tableofcontents
}



\setcounter{footnote}{0}

\newpage
\section{Introduction}

More than a century after its proposal, General Relativity continues to score successes by accurately describing an ever-increasing range of phenomena, all the way from cosmology to the spectra of gravitational waves from binary mergers and the shadows of black holes. Aside from its truly impressive predictive power, Einstein gravity is a remarkably rigid theoretical structure. In some sense, this is an inevitable consequence of the combined constraints of Lorentz invariance and quantum mechanics. Indeed, General Relativity is the unique theory of a massless spin-2 particle that mediates a $1/r^2$ force at long distances between matter sources~\cite{Weinberg:1965nx}.

\vskip4pt
Despite its myriad empirical confirmations, there is a somewhat disquieting feature of Einstein gravity. Since space and time themselves fluctuate, rigorously defining observable quantities is notoriously slippery. {\it Where} do we measure something, and {\it when}? One recourse is to turn to asymptotic observables, where we pin the spacetime at infinity and define quantities relative to this asymptotic metric. This approach has been very successful in both asymptotically flat spaces (where we study the $S$-matrix) and in asymptotically anti-de Sitter spaces (where we can study boundary correlators). These asymptotic observables provide another viewpoint on the theoretical inevitability of Einstein gravity and reveal highly non-obvious connections to gauge theories~\cite{Bern:2019prr}.
Essentially, these remarkable features are a consequence of the fact that
amplitudes involving massless particles are extremely constrained~\cite{Benincasa:2007xk,Schuster:2008nh,McGady:2013sga}.
For example, there is a finite number of possible three-particle amplitudes involving massless spin-2 fields that are consistent with Lorentz invariance. General Relativity is built from the most relevant one at long distances, and the constraints of 
consistent factorization of the four-graviton $S$-matrix then lead uniquely to General Relativity.\footnote{Similar results can also be obtained with the assumption of invariance under Lorentz boosts relaxed~\cite{Pajer:2020wnj,Hertzberg:2020yzl}. In the cosmological context, extensions of the graviton sector were studied for example in~\cite{Creminelli:2014wna,Bordin:2017hal,Bartolo:2017szm,Bordin:2020eui,Bartolo:2020gsh,Cabass:2021fnw}.}

\vskip4pt
In this paper, we study the dynamics of pure gravity in de Sitter space, which is the other natural choice of asymptotics where we could contemplate defining observables. Much as in flat space (or anti-de Sitter space), the symmetries of de Sitter space are very constraining, completely fixing the form of the graviton three-point function up to three overall constants \cite{Maldacena:2011nz}. The four-point function is therefore the first true probe of the dynamics of Einstein gravity, and our goal is to compute this object in exact de Sitter space.\footnote{An earlier computation appears in~\cite{Fu:2015vja}, but their result appears to be incomplete. See also \cite{Li:2018wkt} for a discussion of the exchange part of the correlator using spinor-helicity variables.} The anti-de Sitter space version of this result appears in \cite{Raju:2012zs} (see also \cite{Albayrak:2019yve} for a 5-point graviton exchange correlator in anti-de Sitter space).
In the context of flat space scattering amplitudes, the analogous object---the four graviton $S$-matrix---has served as a font of wisdom, illuminating hidden structures in amplitudes, and elucidating the constraints of theoretical consistency. Our hope is that similar insights will flow from the graviton four-point function in de Sitter space. One motivation is to better understand the
theoretical rigidity of General Relativity in cosmological spacetimes. Another is to gain a glimpse of some hidden simplicity of cosmological correlators from this highly constrained object.\footnote{Indeed, there has been much recent progress in the study of cosmological correlators, e.g.,~\cite{Raju:2012zr,Raju:2012zs,Raju:2011mp,Mata:2012bx,Bzowski:2011ab,Bzowski:2012ih,Bzowski:2013sza,Bzowski:2019kwd,Kundu:2014gxa,Kundu:2015xta,Arkani-Hamed:2015bza,Shukla:2016bnu,Arkani-Hamed:2017fdk,Arkani-Hamed:2018kmz,Baumann:2019oyu,Albayrak:2018tam,Albayrak:2019yve,Sleight:2019mgd,Sleight:2019hfp,Baumgart:2019clc,Gorbenko:2019rza,Mirbabayi:2020vyt,Cohen:2020php,Green:2020whw,Green:2020ebl,BBBB,Baumann:2020dch,Sleight:2020obc,Sleight:2021iix,Baumann:2021fxj,Albayrak:2020fyp,Albayrak:2020isk,Armstrong:2020woi,Meltzer:2021zin,Bonifacio:2021azc,Hogervorst:2021uvp,DiPietro:2021sjt,Sleight:2021plv,Goodhew:2021oqg,Premkumar:2021mlz,Bzowski:2022rlz,Pethybridge:2021rwf, Green:2022slj,Pimentel:2022fsc,Jazayeri:2022kjy,Armstrong:2022csc,Cabass:2022rhr,Armstrong:2022jsa,Armstrong:2022vgl,Mirbabayi:2022gnl}, and the four-graviton correlator provides additional theoretical data in the search for deeper underlying structures.} 

\vskip4pt
Given the complexity of this four-point function, we compute it in several complementary ways. 
The first, presented in Section~\ref{ss:lagrangian}, is a brute force calculation in the in-in formalism, working in the same gauge used in~\cite{Maldacena:2002vr} to compute the three-point function. After solving the constraint equations of General Relativity, we compute both the graviton trispectrum and the associated quartic graviton wavefunction coefficient. In Section~\ref{ss:dSgggglift} we show how this result could have been derived from the study of exchange correlators involving spinning particles with the help of a few heuristic derivations and explicit calculations. The idea here is to take the flat-space wavefunction coefficient and lift it to de Sitter space by acting with various differential operators, along the lines of~\cite{Benincasa:2019vqr,Baumann:2021fxj,Hillman:2021bnk}. Finally, in Section~\ref{s:dS} we present a third derivation using the boostless cosmological bootstrap~\cite{BBBB,MLT} (see \cite{Baumann:2022jpr,Benincasa:2022gtd} for reviews). In this three-step approach we employ three general results. In the first step, we fix all terms that are singular at vanishing partial energies in terms of three-point functions, as done in~\cite{MLT}, using the cosmological optical theorem~\cite{COT}. In the second step, we employ the flat-space amplitude to fix the residue of the leading total-energy pole~\cite{Maldacena:2011nz,Raju:2012zr}. Finally, all the remaining subleading total-energy poles are fixed by the manifestly local test~\cite{MLT}. All three methods produce the same four-point function, providing a useful cross-check.

\vskip4pt
To explain our approach with as few technical complications as possible, in Section~\ref{s:warmup} we present a warm-up computation of the four-point function of a massless scalar induced by the minimal coupling to gravity, using the same three methods outlined above. For both the scalar and graviton trispectrum, our different methods yield results that are identical up to terms that can be generated by field redefinitions. In Section~\ref{sec:conclusions} we conclude and speculate on future directions and lessons.


\paragraph{Notation and conventions:}\label{Conventions}
We work with the mostly positive metric signature $(-+++)$ and our Fourier convention is 
\be
f(\bfx)=\int \dfrac{\rd^3\bfk}{(2\pi)^3}{f}(\bfk)e^{i\bfk\cdot\bfx}\equiv\int_{\bfk}{f}(\bfk)e^{i\bfk\cdot\bfx} \label{FT} \,,
\ee
where bolded letters indicate three-dimensional spatial vectors.
To account for factors of $(2\pi)^3$, it is convenient to define the normalized Dirac delta function 
\be
    \hat \delta(\bfk)\equiv (2\pi)^3\delta^{(3)}(\bfk)\,.
\ee
We will typically denote derivatives with respect to conformal time by a prime, e.g., $\phi'=\del_\eta \phi$.
The wavefunction of the universe $\Psi$ at conformal time $\eta_0$ is parameterized as
\be
\label{eq:WFdef}
\Psi[\eta_{0},\gamma({\bf k})] = \text{exp}\left[-\sum_{n=2}^{\infty} \frac{1}{n!} \sum_{h_i=\pm} \int_{{\bfk}_{1}, \ldots, {{\bf{k}}}_{n}} \psi_{n}({\bf k_{1}}, \ldots, {\bf k_{n}})\, \hat \delta \left(\raisebox{.75pt}{$\sum$}\, \bfk_a \right) \gamma^{h_1}({\bf k}_{1}) \ldots \gamma^{h_n}({\bf k}_{n}) \right],
\ee
where the functions  $\psi_{n}$  are called wavefunction coefficients, and
$\gamma^h(\v{k})$ is the spin-$h$ Fourier mode of the graviton with $h=\pm 2$ (the dependence of $\psi_n$ on the polarizations $h_i$ is left implicit). 
In Fourier space, the graviton is parameterized in terms of $\gamma^\pm$ with helicity $\pm 2$ according to
\be\label{eq:gamma}
\gamma_{ij}(\eta,{\bf x})=\int_{\v{k}} e^{i\v{k}\cdot \v{x}} \sum_{h=\pm} \e_{ij}^{h}(\v{k}) \gamma_{\textbf{k}}^h(\eta) \,,
\ee
where the polarization tensors $\epsilon_{ij}^h$ satisfy
\begin{subequations}
\label{eq:pol-conditions}
\begin{align}\label{pol1}
\e_{ii}^{h}(\v{k})&=k^{i}\e^{h}_{ij}(\v{k})=0 & \text{(transverse and traceless)}\,,\\
\e_{ij}^{h}(\v{k})&=\e_{ji}^{h}(\v{k})& \text{(symmetric)} \,, \\
\e_{ij}^{h}(\v{k})\e_{jk}^{h}(\v{k})&=0 & \text{(lightlike)} \,, \\
\e^{h}_{ij}(\v{k})\e^{h'}_{ij}(\v{k})^{\ast}&=2\delta_{hh'} & \text{(normalization)}\,,  \\
\e_{ij}^h(\v{k})^{\ast}&=\e_{ij}^h(-\v{k}) &\text{($  \gamma_{ij}(x) $ is real)}  \,.\label{poln}
\end{align}
\end{subequations}
Throughout we denote three-dimensional polarisation vectors, used for writing wavefunction coefficients and correlators, by $\epsilon_{i}$, and four-dimensional polarisation vectors, used for writing scattering amplitudes, by $\varepsilon_{\mu}$. 

\vskip4pt
We also define primed correlators with the momentum-conserving delta function removed
\be
\langle {\cal{O}} {(\bfk_1)\dots {\cal{O}}(\bfk_n)} \rangle \equiv \langle {\cal{O}} {(\bfk_1)\dots {\cal{O} }({\bfk_n}) }\rangle' \hat \delta \left(\raisebox{.75pt}{$\sum$}\, \bfk_a \right)\,.
\ee
The three-momenta ${\bf k}_a$ have components $k_{a}^i$ and norms $k_a =|{\bf k}_a| $. The corresponding massless four-momenta have components $k_a^{\mu} = (k_a, k_a^i)$. We define $k_{ab} = k_a+k_b$. The elementary symmetric polynomials in three variables are written as
 \be \label{ESP}
    e_1=k_1+k_2+k_3, \qquad e_2=k_1 k_2 +k_1 k_3+k_2k_3, \qquad e_3=k_1 k_2 k_3,
\ee
while those in four variables are written as
 \begin{align} 
    \mathbb{e}_1&=k_1+k_2+k_3+k_{4}\,,  & \mathbb{e}_2&=k_1 k_2 +k_1 k_3+k_1k_4 + k_{2}k_{3}+k_{2}k_{4}+k_{3}k_{4}\,, \nn \\  \mathbb{e}_3&=k_1 k_2 k_3+k_1 k_2 k_4+k_1 k_3 k_4+k_2 k_3 k_4\,, & \mathbb{e}_4&=k_1 k_2 k_3 k_{4}.
\end{align}
We will often use $k_{T}$ in place of $e_{1}$ and $\mathbb{e}_{1}$ where no confusion arises. We also define
\be
 \quad s = | {\bf k}_1+{ \bf k}_2|, \qquad\quad t = | {\bf k}_1+{ \bf k}_3|, \qquad\quad u = | {\bf k}_1+{\bf k}_4|.
\ee
Using three-momentum conservation ($\sum_{a=1}^4 {\bf k}_{a}=0$), these satisfy 
 \be \label{NL_energy_relation}
 s^2+t^2+u^2=k_1^2+k_2^2+k_3^2+k_4^2.
 \ee
The Mandelstam variables for $k_{a}^{\mu}$ are defined such that
 \begin{align}
S &=-(k_1^{\mu} +k_2^{\mu})(k_{1, \mu} +k_{2, \mu})=2 k_1 k_2-2 {\bf k}_1 \cdot {\bf k}_2 =k_{12}^2 - s^2, \label{Stos} \\
T &=-(k_1^{\mu} +k_3^{\mu})(k_{1 \mu} +k_{3 \mu})=2 k_1 k_3-2 {\bf k}_1 \cdot {\bf k}_3 = k_{13}^2 - t^2, \label{Ttot} \\
U &=-(k_1^{\mu} +k_4^{\mu})(k_{1 \mu} +k_{4 \mu})=2 k_1 k_4-2 {\bf k}_1 \cdot {\bf k}_4 = k_{14}^2 - u^2. \label{Utou} 
 \end{align}
Four-momentum conservation ($\sum_{a=1}^4k_{a}^{\mu}=0$) implies that these satisfy $S+T+U=0$. 

\newpage
 
\section{Scalar trispectrum with graviton exchange}\label{s:warmup}

In this section, we discuss the tree-level four-point function for a massless spectator scalar induced by minimal coupling to pure gravity in exact de Sitter space. This calculation already contains most of the features that we will encounter in the graviton four-point function case while being algebraically simpler.
First, in Section~\ref{ss:ssgss}, we review the explicit bulk perturbation theory calculation, which was first done in Ref.~\cite{Seery:2006vu, Seery:2008ax}. Then, in Section~\ref{ss:dSs4}, we show how the same result can be derived from a more ``on-shell" perspective by lifting the corresponding flat-space wavefunction coefficient to de Sitter space. Finally, in Section~\ref{ss:blesss4}, we derive this result in a third way, using the boostless cosmological bootstrap. Namely, from the cosmological optical theorem and energy shifts, the amplitude limit, and the manifestly local test.

 
\subsection{Lagrangian calculation}\label{ss:ssgss}

We start with the action of a scalar field minimally coupled to gravity, 
\be \label{eq:minimal-coupling-action}
S = \int \rd^4 x \sqrt{-g} \left[ \frac{M_{\rm Pl}^2}{2}  \left(R^{(4)} -6 H^2\right)- \frac{1}{2} g^{\mu \nu} \partial_{\mu} \phi \partial_{\nu} \phi \right].
\ee
Our goal is to compute the four-point correlation function of the massless scalar $\phi$ that arises from the exchange of a graviton.
We first perform a $3+1$ decomposition in the ADM formalism, where the line element is written as
\be
\rd s^2 = -N^2 \rd t^2+h_{ij}(\rd x^i +N^i \rd t)(\rd x^j +N^j \rd t).
\ee
The extrinsic curvature is $K_{ij} = N^{-1} E_{ij}$, where
\be
E_{ij} = \frac{1}{2} ( \dot{h}_{ij} - \nabla_i N_j - \nabla_j N_i).
\ee
We also define $E = h^{ij} E_{ij}$. 
In terms of these variables, the action is given by
\be
\begin{aligned}
S =~&\frac{M_{\rm Pl}^2}{2}\int \rd t \rd^3 x\sqrt{h}\left(N R^{(3)}-6NH^2+N^{-1}\left(E_{ij}E^{ij}-E^2\right)\right)  \\
&- \frac{1}{2}\int \rd t \rd^3 x \frac{\sqrt{h}}{N}\left( - \dot{\phi}^2+2 N^i \dot{\phi} \partial_i \phi +(h^{ij} N^2 -N^i N^j) \partial_i \phi \partial_j \phi \right). \label{eq:3+1action}
\end{aligned}
\ee
In this formalism, the lapse, $N$, and shift, $N_i$, are Lagrange multipliers that enforce constraints. Conceptually, we solve these constraints order-by-order in perturbation theory and substitute their solution back into the action~\cite{Maldacena:2002vr}. The resulting action for the dynamical variables serves as the starting point for the field theory calculation of correlation functions.

\subsubsection{Constraint equations}

Varying the action~\eqref{eq:3+1action} with respect to the lapse and shift gives the constraint equations,
\begin{align}
&R^{(3)}-6H^2-\frac{1}{N^2}\left(E_{ij}E^{ij}-E^2\right) +\frac{1}{M_{\rm Pl}^{2} N^{2}}\left(- \dot{\phi}^2+2 N^i \dot{\phi} \partial_i \phi -(h^{ij}N^2+N^i N^j) \partial_i \phi \partial_j \phi \right)=0,\nonumber\\
&\nabla_j\left[N^{-1}\left(E^j_i-E\delta^j_i\right)\right]-M_{\rm Pl}^{-2} N^{-1} \left(\dot{\phi} \partial_i \phi-N^j \partial_i \phi \partial_j \phi \right)=0.
\label{eq:constraints}
\end{align}
We want to solve these constraints perturbatively for the fluctuations around the de Sitter background solution,
\be
\rd s^2= - \rd t^2 + a^2 \delta_{ij} \rd x^i \rd x^j, \qquad\qquad \phi = 0, \label{vanish}
\ee
where the scale factor is $a(t)=e^{ Ht}$. Since the scalar has a vanishing background, we will sometimes refer to it as a ``spectator" field to distinguish it from an inflaton, which would have a non-trivial time-dependent background. 

\vskip4pt
We fix the gauge such that the spatial metric takes the form 
\be \label{eq:gamma-def}
h_{ij} = a(t)^2\, (e^\gamma)_{ij}\,,
\ee
where $\gamma_{ij}$ is both transverse and traceless: $ \gamma_{ii}=\partial_i\gamma_{ij}=0$.
To facilitate finding the perturbative solution of the constraints, we expand the lapse and shift order by order in powers of the perturbations,
\begin{align}
N & = 1 + \alpha^{(1)} +\alpha^{(2)} + \dots , \\
N_i & = N_i^{(1)} + N_i^{(2)} + \dots ,
\end{align}
where the superscript labels the order in perturbations of the fields.
We also further decompose $N_i^{(n)}$ into a longitudinal piece and a transverse piece,\footnote{The notation $\psi^{(n)}$ for the longitudinal component of the shift is conventional in the literature. It should not be confused for a wavefunction coefficient---we hope that the distinction is clear from the context.}
\be
N_i^{(n)} =  \beta_{i}^{(n)} + \partial_i \psi^{(n)},
\ee
where $\beta_{i}^{(n)}$ is transverse: $\partial^i \beta_{i}^{(n)}=0$. (Note that indices on $\gamma_{ij},\ \partial_i$ and $\beta_i$ are raised and lowered using the flat metric $\delta_{ij}$.) We now substitute the decomposition of the lapse and shift into the equations~\eqref{eq:constraints} and solve them order-by-order.

\vskip4pt
To find the interactions to quartic order, it is sufficient to solve the constraint equations to second order (see, e.g.,~\cite{Chen:2006nt} or Appendix A of~\cite{Pajer:2016ieg}). At first order, the constraint equations are 
\begin{align}
12H^2 \alpha^{(1)} +4H e^{-2Ht} \partial^2 \psi^{(1)} & = 0, \\
4 H \partial_i \alpha^{(1)} - e^{-2 H t} \partial^2 \beta_{i}^{(1)} & =0.
\end{align}
These can be solved to find $\alpha^{(1)} =\psi^{(1)}=\beta_i^{(1)}=0$ \cite{Maldacena:2002vr}. The solutions are particularly simple because we are considering a spectator field with a vanishing background~\eqref{vanish}.
The constraint equations at second order read
\begin{align}
12H^2 \alpha^{(2)} +4H e^{-2Ht} \partial^2 \psi^{(2)} + \frac{1}{M_{\rm Pl}^2} \left( \dot{\phi}^2 + e^{-2 Ht} \partial_i \phi \partial_i \phi \right)+ \dots & = 0\,, \\
4 H  \partial_i \alpha^{(2)} -  e^{-2 H t} \partial^2 \beta_{i}^{(2)} - \frac{2}{M_{\rm Pl}^2} \dot{\phi} \partial_i \phi + \dots& =0\,,
\end{align}
where we have only shown the terms needed to compute the scalar trispectrum (the terms involving $\gamma_{ij}$ are written in Section~\ref{ss:constr}). We can solve these equations to find
\begin{align}
\alpha^{(2)} & = \frac{1}{2M_{\rm Pl}^2 H} \frac{1}{\partial^2} \partial_i( \dot{\phi} \partial_i \phi) + \dots, \\
\beta_i^{(2)} & = \frac{2}{M_{\rm Pl}^2} \frac{e^{2Ht}}{\partial^2} \left( \frac{\partial_i \partial_j}{\partial^2} -\delta_{ij} \right) ( \dot{\phi}  \partial_j \phi) + \dots, \\
\psi^{(2)} & =-\frac{e^{2 Ht}}{4 M_{\rm Pl}^2 H \partial^2} \left[ \frac{6 H}{\partial^2} \partial_i( \dot{\phi} \partial_i \phi)  +\dot{\phi}^2 +e^{-2 Ht} \partial_i \phi \partial_i \phi \right] + \dots,
\end{align}
where again we have only shown the terms needed for the scalar trispectrum.

\subsubsection{Lagrangian interactions to quartic order}

We now want to substitute the solution to the constraints back into the action to obtain the action governing fluctuations around the background up to quartic order.
Expanding~\eqref{eq:3+1action} gives the following interactions relevant for the computation of the scalar four-point function: 
\be
\begin{aligned}
S \supset & \int \rd t \rd^3 x \frac{e^{3 Ht} }{2} \left[ \dot{\phi}^2 - e^{-2 Ht} \partial_i \phi \partial_i\phi+ e^{-2 Ht} \gamma_{ij} \partial_i \phi \partial_j \phi \right]  \\
 &+  \int \rd t \rd^3 x \left[ -\frac{e^{ Ht} }{2} \alpha^{(2)} \partial_i \phi \partial_i \phi - e^{ Ht} \beta^{(2)}_i \dot{\phi} \partial_i\phi - e^{ Ht} \partial_i \psi^{(2)} \dot{\phi} \partial_i \phi-\frac{e^{ 3Ht} }{2} \alpha^{(2)} \dot{\phi}^2 \right]  \\ 
 &+   \int \rd t \rd^3 x  M_{\rm Pl}^2\left[  -3 e^{3 Ht} H^2 (\alpha^{(2)})^2-2 H e^{H t} \alpha^{(2)} \partial^2 \psi^{(2)} +\frac{1}{4} e^{-H t} (\partial_i \beta^{(2)}_j)^2 \right],
\end{aligned}
\label{eq:quartactionscalar}
\ee
where the last line contains the interactions of the constrained fields coming from the Einstein--Hilbert term. 

\vskip4pt
The scalar-graviton interaction that contributes to the scalar trispectrum at tree level is the cubic term $\gamma_{ij} \partial_i \phi \partial_j \phi$, which generates a graviton-exchange diagram, as computed in~\cite{Seery:2008ax}. The quartic terms in $\phi$ contribute to the contact term, as computed in~\cite{Seery:2006vu}. The second-order solutions to the constraints make their first appearance in these quartic terms. Substituting the solutions for the constrained fields, we obtain the following quartic scalar interactions:
\be \label{ScalarInteractions}
S \supset \int \rd t \rd^3 x  \frac{e^{3 Ht} }{4M_{\rm Pl}^2} \left[   \frac{\partial_i}{\partial^2}(  \dot{\phi} \partial_i \phi) \frac{\partial_j}{\partial^2}( \dot{\phi} \partial_j \phi) - \frac{1}{H}  \frac{\partial_j}{\partial^2}( \dot{\phi} \partial_j  \phi) \left( \dot{\phi}^2+e^{-2 Ht} \partial_i \phi \partial_i \phi \right) + 4\dot{\phi}\partial_i \phi \frac{1}{\partial^2}( \dot{\phi} \partial_i  \phi)\right].
\ee
Notice that at quartic order in perturbations, inverse Laplacians in interactions make their first appearance. Such interactions appear because we have integrated out auxiliary fields. The appearance at fourth order is a consequence of the fact that we are working with a spectator field. If the field had a nontrivial background, inverse Laplacians would have appeared already at cubic order~\cite{Maldacena:2002vr}. The presence of inverse Laplacians violates the assumptions underlying a useful constraint on wavefunction coefficients called the manifestly local test (MLT) \cite{MLT}. For a  quartic wavefunction coefficient, the MLT says that
 \be \label{MLT-0}
    \frac{\partial }{\partial k_{a}} \psi_{4}\Big|_{k_{a}=0}=0\, , \quad a=1, \dots, 4 .
\ee
One might expect that the wavefunction coefficients considered in this paper do not satisfy the MLT because of the presence of inverse Laplacians in the interactions. However, as we argue in Section \ref{ss:MLT}, the MLT does still hold for these wavefunction coefficients.


\subsubsection{Calculating wavefunction coefficients and correlators}

Once we have the action expanded to the appropriate order, we can calculate wavefunction coefficients and correlators following the standard approaches. As a cross-check of our computations, we calculate the relevant correlation function using both the in-in formalism and by computing the relevant wavefunction coefficients.

\paragraph{The wavefunction:}
In order to compute tree-level wavefunction coefficients, we note that given a collection of fields, $\phi$, the path integral
\be
\Psi[\varphi] \ =  \hspace{-0.4cm} \int\limits_{\substack{\phi(t) \,=\,\varphi\\ \hspace{-0.45cm}\phi(-\infty)\,=\,0}} 
\hspace{-0.5cm} \raisebox{-.05cm}{ ${\cal D} \phi\, e^{iS[\phi]}\,,$ }
\ee
formally solves the Schr\"odinger equation (where $\varphi$ is the field profile at time $t$ and the boundary condition at the initial time should be interpreted with the suitable $i\epsilon$ prescription). At tree-level, this path integral is approximated by the on-shell action evaluated on the classical solution with vacuum initial conditions and Dirichlet late-time boundary conditions. The on-shell action is then a functional of this boundary data (see, e.g.,~\cite{Maldacena:2002vr,Anninos:2014lwa,Goon:2018fyu,Baumann:2020dch,COT} for reviews of the formalism).

\vskip4pt
As a matter of practice, we can efficiently compute the wavefunction coefficients that appear in~\eqref{eq:WFdef} by employing Feynman rules analogous to those that compute flat space scattering amplitudes. The main difference is that we have to work in the time domain and integrate over all possible interaction times. Aside from this, there are two different propagators: external lines are associated to the so-called bulk-to-boundary propagator, $K_k(\eta)$, and internal lines are associated to the bulk-to-bulk propagator, $G_k(\eta,\eta')$. In this particular case, the bulk-to-boundary propagator for a massless scalar/graviton is given by
\be
    K_k(\eta)=(1- ik\eta)e^{ik\eta}\,,
    \label{eq:bbdyprop}
\ee
while the bulk-to-bulk propagator for a massless graviton is
\be
 G_k^{hh'}(\eta,\eta')=i\frac{H^2}{2k^3}\delta^{hh'}\theta(\eta-\eta')\left(K_{k}(\eta')K^{\ast}_{k}(\eta)-K_{k}(\eta')K_{k}(\eta)\right)+(\eta\leftrightarrow \eta')\,,
 \label{eq:bbulkprop}
\ee
where $h,h'$ are the helicities of the incoming/outgoing gravitons. In addition to these propagators, we can derive the relevant interaction vertices from the action~\eqref{eq:quartactionscalar} or~\eqref{ScalarInteractions} in the usual way.

\vskip4pt
Once one has the wavefunction, it can be used to compute equal-time correlation functions as in quantum mechanics:
\be
\langle\phi_1 \cdots \phi_N\rangle = \int{\cal D}\phi\, \phi_1\cdots\phi_N\left\lvert\Psi[\phi]\right\rvert^2\,.
\ee
This formula can be expanded perturbatively to give a relation between correlation functions and their corresponding wavefunction coefficients, as we will describe.

\paragraph{In-in correlators:}
We can also directly calculate in-in correlation functions in canonical quantization. In order to proceed with this formalism, 
we promote the fields to operators by expanding in terms of creation and annihilation operators,
\begin{align}
\phi_{{\bf k}}( \tau) & = f_k(\tau) a_{{\bf k}} + f^*_k(\tau) a_{-{\bf k}}^{ \dagger}, \\
\gamma^h_{{\bf k}}( \tau) & = f_k(\tau) a_{{\bf k}}^h + f^*_k(\tau) a_{-{\bf k}}^{h\, \dagger},
\end{align}
which have the commutation relations
\begin{align}
\big[ a_{{\bf k}}, a_{{\bf k}'}^{ \dagger} \big] & = \hat{\delta}({\bf k}-{\bf k}'), \\
\big[ a_{{\bf k}}^h, a_{{\bf k}'}^{h' \dagger} \big] & = \frac{2}{ M_{\rm Pl}^2}\hat{\delta}({\bf k}-{\bf k}') \delta_{h h'}.
\end{align}
The massless mode functions for both the scalar and graviton with Bunch--Davies vacuum conditions are
\be
 f_k(\tau) = \frac{H}{\sqrt{2 k^3}} \left( 1+ i k \tau \right) e^{-ik \tau}.
\ee
In-in correlators are then given by the analogue of Dyson's formula, written in terms of the interaction Hamiltonian $ \mathcal{H}_{\rm int}$,
\be
\langle \phi_1\cdots \phi_N\rangle = \langle 0\rvert \bar T\,e^{i\int_{-\infty}^\eta\rd\eta'{\cal H}_{\rm int}}\,\phi_1\cdots \phi_N\,T\,e^{i\int_{-\infty}^\eta\rd\eta'{\cal H}_{\rm int}}\lvert 0\rangle\,,
\ee
where $\lvert 0\rangle$ is the ``in" Fock vacuum. It is often convenient to expand out the exponentials and rearrange them into the commutator formula~\cite{Weinberg:2005vy}
\be
\begin{aligned}
\langle \phi_{{\bf k}_1}(\eta) \dots  \phi_{{\bf k}_n}(\eta) \rangle&= \sum_{N=0}^{\infty} i^N \int^{\eta}_{- \infty} \rd\eta_1 \int^{\eta_1}_{- \infty} \rd\eta_{2} \dots  \int^{\eta_{N-1}}_{- \infty}  \rd \eta_N   \\
& \hspace{.75cm}\times \left\langle \left[ \mathcal{H}_{\rm int} (\eta_N), \left[ \mathcal{H}_{\rm int} (\eta_{N-1}), \dots [ \mathcal{H}_{\rm int} (\eta_{1}),  \phi_{{\bf k}_1}( \eta) \dots \phi_{{\bf k}_n}( \eta) ]  \dots \right] \right] \right\rangle.
\end{aligned}
\ee
The late-time correlator is obtained by taking the $\eta \rightarrow 0$ limit,
\be
\langle \phi_{{\bf k}_1} \dots  \phi_{{\bf k}_n} \rangle \equiv \lim_{\eta \rightarrow 0} \langle \phi_{{\bf k}_1}(\eta) \dots  \phi_{{\bf k}_n}(\eta) \rangle,
\ee
and all trispectra of interest in this paper are finite in this limit.

\subsubsection{The power spectrum and bispectrum}

We start by computing the two and three-point objects that are needed to relate the four-point wavefunction coefficient to the analogous correlator.
The two-point correlator, or power spectrum, of a massless scalar is
\be
P(k) = \langle \phi_{{\bf k}} \phi_{-{\bf k}}\rangle'= \frac{H^2}{2 k^3},
\ee
while the power spectrum of the graviton with our normalizations is given by\footnote{In terms of $\gamma_{ij}$, the two-point function is
\be 
\lim_{\eta \rightarrow 0} \left\langle \gamma_{ij} ({\bf k}_1, \eta) \gamma_{kl} ({\bf k}_2, \eta)  \right\rangle = \frac{2 H^2}{M_{\rm Pl}^2 k_1^3} (2 \pi)^3 \delta^{(3)}\left( {\bf k}_1 + {\bf k}_2\right) \Pi_{ijkl}^{{\bf k}_1},
\ee
where $\Pi_{ijkl}^{{\bf k}_1}$ is the transverse-traceless spin-2 projector defined in~\eqref{eq:pol-sum}.}
\be \label{GravitonPS}
\langle \gamma^{h_1}_{{\bf k}} \gamma^{h_2}_{-{\bf k}}\rangle'=P_{\gamma}(k) \delta_{h_1 h_2}, \qquad\quad P_{\gamma}(k)= \frac{H^2}{M_{\rm Pl}^2 k^3}.
\ee
The real part of the cubic wavefunction coefficient for two scalars and one graviton is
\be \label{eq:psi3-scalar}
{\rm Re} \, \psi_{\phi \phi \gamma}({\bf k}_1, {\bf k}_2, {\bf k}_3) = \frac{1}{H^2 k_T^2} \left(e_3+e_2 k_T-k_T^3 \right) ( \epsilon_3 \ccdot {\bf k}_1)^2,
\ee
where the elementary symmetric polynomials $e_2$ and $e_3$ are defined in~\eqref{ESP} and we have written the symmetric polarization tensor as $\epsilon_{3, ij} = \epsilon_{3, i} \epsilon_{3, j}$.\footnote{This is just a convenient way to track the index contractions of a symmetric tensor.} From the boundary point of view, this wavefunction coefficient can be fixed by using the symmetries of de Sitter space, which act like conformal transformations on the boundary~\cite{Mata:2012bx}, or as the lowest order solution to the MLT~\cite{MLT}.
The cubic correlator, or bispectrum, is related to this by
\be
\langle \phi_{{\bf k}_1} \phi_{{\bf k}_2} \gamma_{{\bf k}_3} \rangle'= -2 P(k_1) P(k_2) P_{\gamma}(k_3)
{\rm Re} \, \psi_{\phi \phi \gamma}({\bf k}_1, {\bf k}_2, {\bf k}_3).
\ee
We can also obtain the bispectrum from an in-in calculation using the cubic interaction Hamiltonian and the formula
\be
\langle \phi_{{\bf k}_1} \phi_{{\bf k}_2} \gamma_{{\bf k}_3} \rangle= \lim_{\eta \rightarrow 0} i \int^{\eta}_{- \infty} \rd \eta_1 \left\langle \left[ \mathcal{H}^{(3)}_{\rm int} (\eta_1),  \phi_{{\bf k}_1} ( \eta) \phi_{{\bf k}_2}( \eta)  \gamma_{{\bf k}_3} ( \eta) \right]  \right\rangle.
\ee
We now turn to the computation of the four-point function.

\subsubsection{The quartic wavefunction coefficient}
There are two conceptually different contributions to the quartic wavefunction coefficient. The first arises from the exchange of a massless graviton between two pairs of scalars. The other contribution comes from the quartic vertices~\eqref{ScalarInteractions}. We obtain the full wavefunction coefficient by adding these together. For example, the contribution coming from graviton exchange in the $s$-channel along with the contact diagram with the same permutation symmetries is
\be
\psi_{\phi^4}^{(s)}(\{\bfk\})  =\frac{1}{6 H^2 M_{\rm Pl}^2} \left[ f^{(s)}_{(2,2)} s^4 \Pi^{(s)}_{2,2} + f^{(s)}_{(2,1)}s^2 \Pi^{(s)}_{2,1}+ f^{(s)}_{(2,0)} (E_L E_R - s k_T )\Pi^{(s)}_{2,0} + f_{c} \right].
\label{eq:schannelWFmasslessscalar}
\ee
The full permutation-invariant answer is then given by summing over channels:
\be
\psi_{\phi^4}(\{\bfk\}) =  \psi_{\phi^4}^{(s)}(\{\bfk\})+\psi_{\phi^4}^{(t)}(\{\bfk\})+\psi_{\phi^4}^{(u)}(\{\bfk\})\,.
\label{eq:fullmassless4pt}
\ee
In~\eqref{eq:schannelWFmasslessscalar},
$\{\bfk\}$ is shorthand for $\{\bfk_1,\bfk_2,\bfk_3,\bfk_4\}$ and $\Pi^{(s)}_{m,n}$ are contractions of projectors with external momenta whose explicit form is given in Appendix~\ref{app:PolSums}. The $f$s are functions defined by 
\begin{align}\label{f22}
f^{(s)}_{(2,2)} = &\,\frac{1}{k_{T}}-\frac{s^2}{k_{T}E_{L}E_{R}}+\frac{s k_{1}k_{2}}{k_{T}E_{L}^2E_{R}}+\frac{s k_{3}k_{4}}{k_{T}E_{R}^2E_{L}}+\frac{2s k_{1}k_{2}k_{3}k_{4}}{k_{T}^2 E_{L}^2 E_{R}^2} -\frac{s(k_{1}k_{2}+k_{3}k_{4})}{k_{T}^2 E_{L}E_{R}}
\nonumber \\ & + \frac{k_{1}k_{2}}{k_{T}^2E_{L}} +\frac{k_{3}k_{4}}{k_{T}^2E_{R}} + \frac{2k_{1}k_{2}k_{3}k_{4}}{k_{T}^3E_{L}E_{R}}, \\
\label{eq:f21}
f_{(2,1)}^{(s)} =& -\frac{2 k_{1}k_{2}k_{3}k_{4}}{k_{T}^3} - \frac{k_{12}k_{3}k_{4}+k_{34}k_{1}k_{2}}{k_{T}^2}-\frac{k_{12}k_{34}}{k_{T}}, 
\end{align}
\begin{align}
\label{eq:f20}
f^{(s)}_{(2,0)} = & -f^{(s)}_{(2,1)}, \\
f_c =~& -\frac{2k_1k_2k_3k_4(k_{12}k_{34}+s^2)}{k_T^3}-\frac{1}{k_T^2}(k_1k_2k_3+k_1k_2k_4+k_1k_3k_4+k_2k_3k_4)(k_{12}k_{34}+s^2)\nonumber\\
&-\frac{1}{4k_T}\bigg[
24k_1k_2k_3k_4+8(k_1k_3+k_2k_3+k_1k_4+k_2k_4)^2 \nonumber\\
&\hspace{1.3cm}+(k_1k_2+k_3k_4)\left[9(k_1k_3+k_2k_3+k_1k_4+k_2k_4)-12(k_1k_2+k_3k_4)\right] \nonumber\\
&\hspace{1.3cm}+\frac{1}{2}(k_{12}^2+k_{34}^2+2s^2)\Big(3(k_1k_2+k_3k_4)+4(k_1k_3+k_2k_3+k_1k_4+k_2k_4)\Big)
\bigg] \nonumber\\
&-\frac{3(k_1-k_2)^2(k_3-k_4)^2(k_1k_2+k_3k_4)}{4s^2k_T}-3(k_1k_2k_3+k_1k_2k_4+k_3k_4k_1+k_3k_4k_2)\nonumber\\
& +\frac{k_T}{8}\bigg[
21(k_1k_2+k_3k_4)+34(k_1k_3+k_2k_3+k_1k_4+k_2k_4) \nonumber\\
&\hspace{1.3cm}+3(k_{12}^2+k_{34}^2+2s^2)
-\frac{6}{s^2}(k_1-k_2)^2(k_3-k_4)^2\bigg]-\frac{9}{8}k_T^3\,,
\label{eq:fcdSmassless}
\end{align}
where we recall our notation
\begin{align}
    k_{ab} &= k_a+k_b\,, & s&=|{\bf k}_1+{\bf k}_2| \,, \\
    k_T&=k_1+k_2+k_3+k_4\,, & E_L &=k_1+k_2+s\,, E_R=k_3+k_4+s\,.
\end{align}
The expressions for $\psi_{\phi^4}^{(t)}(\{\bfk\})$ and $\psi_{\phi^4}^{(u)}(\{\bfk\})$ are obtained from $\psi_{\phi^4}^{(s)}(\{\bfk\})$ by permuting energies and momenta. 
This trispectrum was first computed in~\cite{Seery:2006vu,Seery:2008ax}, and the result~\eqref{eq:fullmassless4pt} matches their computation.

\vskip4pt
Note that the leading total-energy pole of~\eqref{eq:schannelWFmasslessscalar} is of degree $3$, which is what we expect if the interaction vertices have at most two derivatives~\cite{BBBB}. There are, however, three-derivative quartic self-interactions in the middle terms in~\eqref{ScalarInteractions}, so some cancellations have occurred. In fact, it is possible to integrate by parts and use the scalar's equation of motion to write these three derivative terms as two derivatives ones (up to boundary terms that do not contribute to the wavefunction or correlator), which makes it clear that there should be no total-energy poles of a higher degree. 

\vskip4pt
The trispectrum is related to the wavefunction coefficients by 
\be
\frac{\langle \prod_{a=1}^4\phi_{{\bf k}_a} \rangle'}{4\prod_{a=1}^4 P(k_a)} =-\frac{1}{2} {\rm Re \,} \psi_{\phi^4}(\{\bfk\})+\sum_{3 \,\, {\rm perms}} \sum_h   P_{\gamma}(s)
{\rm Re} \, \psi_{\phi \phi \gamma^h}({\bf k}_1, {\bf k}_2, -{\bf s}){\rm Re} \, \psi_{\phi \phi \gamma^h}({\bf k}_3, {\bf k}_4, {\bf s})\,,
\ee
where $\bf{s}=\bf{k}_1+\bf{k}_2$.
This expression can be reproduced directly from an in-in calculation, where
the contact contribution is given by 
\be
\left\langle \prod_{a=1}^4\phi_{{\bf k}_a} \right\rangle_{\rm c} = \lim_{\eta \rightarrow 0} i \int^{\eta}_{- \infty} \rd \eta_1 \left\langle \left[ \mathcal{H}^{(4)}_{\rm int} (\eta_1), \prod_{a=1}^4\phi_{{\bf k}_a} (\eta) \right]  \right\rangle,
\ee
and the exchange contribution is given by
\be
\left\langle \prod_{a=1}^4\phi_{{\bf k}_a} \right\rangle_{\rm exc.}= \lim_{\eta \rightarrow 0} i^2 \int^{\eta}_{- \infty} \rd \eta_1 \int^{\eta_1}_{- \infty}  \rd \eta_2 \left\langle \left[ \mathcal{H}^{(3)}_{\rm int} (\eta_2), \left[ \mathcal{H}^{(3)}_{\rm int} (\eta_1), \prod_{a=1}^4\phi_{{\bf k}_a}(\eta) \right]  \right]\right\rangle.
\ee
In these particular cases, the 
interaction Hamiltonians are just minus the corresponding Fourier-transformed interaction Lagrangians. 
 
\subsection{Lifting from flat space to de Sitter space}\label{ss:dSs4}

The direct calculation of the wavefunction coefficient~\eqref{eq:schannelWFmasslessscalar} has the benefit of being completely systematic, but the trade-off is that we were forced to introduce a redundant and somewhat complicated description of the bulk physics. In the end, few of these complications propagate all the way to the final answer, which is somewhat simpler than the procedure that gave rise to it. We are therefore motivated to find a more direct construction of~\eqref{eq:schannelWFmasslessscalar} by working at late times. In Section~\ref{ss:blesss4}, we will give a totally systematic (boostless) bootstrap construction of this wavefunction.\footnote{The wavefunction~\eqref{eq:schannelWFmasslessscalar} was bootstrapped in~\cite{Baumann:2020dch} utilizing the de Sitter symmetries of the answer to construct it via weight-shifting~\cite{Arkani-Hamed:2018kmz,Baumann:2019oyu,Karateev:2017jgd}. The difference in Section~\ref{ss:blesss4} is that we do not directly require invariance under de Sitter boosts (it is instead an output). This turns out to be algebraically simpler, which makes the generalization to the graviton four-point function easier.} Here instead we want to provide a somewhat more artisanal approach. What is lost in rigor is made up for by its simplicity. Essentially, the strategy is to take the corresponding flat space wavefunction as an input and transform it to a de Sitter wavefunction, in the spirit of~\cite{Benincasa:2019vqr,Baumann:2021fxj,Hillman:2021bnk}.

\vskip4pt
The starting point is the four-point wavefunction coefficient in flat space arising from four massless scalars exchanging a graviton in the $s$ channel:
\be
\label{eq:flatspacespin2exc}
\psi_{\phi^4}^{(s)\, \rm flat}(\{ {\bf k}\})	=\frac{1}{6}\left(\frac{1}{k_TE_LE_R}s^4\Pi^{(s)}_{2,2}-\frac{1}{k_T}s^2\Pi^{(s)}_{2,1}+\frac{1}{k_T}(E_LE_R-sk_T)\Pi^{(s)}_{2,0} +f_c^{\rm flat}\right)\,,
\ee
where we have defined the function
\be
f_c^{\rm flat}=-\frac{k_{12}k_{34}+s^2}{k_T} -\frac{3}{2}\left(\frac{k_{12}(k_3-k_4)^2}{s^2}+\frac{k_{34} (k_1-k_2)^2}{s^2}\right)+ \frac{3}{2}k_T\,.
\ee
For this discussion, we will just take~\eqref{eq:flatspacespin2exc} as given, but it can be systematically constructed by requiring that it have only singularities at physical locations, that these singularities have the correct residues, and that
it vanishes when any of the external momenta are taken to be soft~\cite{Baumann:2021fxj}. (Or, alternatively, it can just be computed directly.) Notice that the wavefunction~\eqref{eq:flatspacespin2exc} is given by energy-dependent form factors multiplying polarization sums, which in turn come from the exchanges of the different helicity modes of the graviton. We can understand the presence of these particular polarization sums from the fact that they reduce to Legendre polynomials in the limit $k_T\to 0$~\cite{Baumann:2020dch} (see Appendix \ref{app:PolSums}). It is conceptually useful to separate the helicity-0 contribution into the term multiplying the $\Pi_{2,0}$ polarization sum and the $f_c^{\rm flat}$ piece (which ensures that the soft limit vanishes), because they will behave somewhat differently when we lift them to de Sitter space.

\vskip4pt
We can write the $s$-channel de Sitter wavefunction in the same schematic form as~\eqref{eq:flatspacespin2exc}:
\be
\label{eq:dsspin2exc}
  \psi_{\phi^4}^{(s)}(\{ {\bf k}\}) =\frac{1}{6}\left(  f^{(s)}_{(2,2)} s^4 \Pi^{(s)}_{2,2} + f^{(s)}_{(2,1)}s^2 \Pi^{(s)}_{2,1}+ f^{(s)}_{(2,0)} (E_L E_R - s k_T )\Pi^{(s)}_{2,0} + f_{c}\right)\,.
\ee
Our goal is now to
obtain the various form factors from their flat space counterparts. The strategy is to compare the structure of the cut obtained from the cosmological optical theorem (COT) \cite{COT} between the flat space wavefunction and the corresponding de Sitter wavefunction (which only requires knowledge of lower-point functions). We then want to find a differential operator that transforms the flat space cut into that of the de Sitter wavefunction. Using this operator, we can transmute the highest helicity form factor to de Sitter space. We can then utilize the simple relation between the highest helicity form factor and those corresponding to lower helicity exchanges. 

\vskip4pt
Late-time wavefunction coefficients that arise from unitary time evolution in the bulk spacetime must satisfy the COT. This implies that the contribution to the quartic wavefunction that comes from the $s$-channel exchange of gravitons satisfies~\cite{COT,Goodhew:2021oqg} 
    \begin{align}
    \label{COT4Point}
    \psi_{\phi^4}^{(s)}(\{k\}, s)+\psi^{\ast\,(s)}_{\phi^4}(\{-k\}, s) &= \sum_h P_{\gamma}(s)\Big[ \psi_{\phi\phi\gamma^h}(k_{1},k_{2},s)+\psi^{\ast}_{\phi\phi\gamma^h}(-k_{1},-k_{2},s) \Big] \nonumber \\ &\hspace{1cm}\times \Big[ \psi_{\phi\phi\gamma^h}(k_{3},k_{4},s)+\psi^{\ast}_{\phi\phi\gamma^h}(-k_{3},-k_{4},s) \Big] \,,
    \end{align}
    where $P_{\gamma}(s)$ is the power spectrum of the exchanged field.\footnote{This expression assumes that the mode functions of external and internal fields obey Bunch--Davies vacuum conditions, but generalisations have also been derived~\cite{Cespedes:2020xqq}. Loop level cuts have also been derived in~\cite{Melville:2021lst}, and similar relations hold for higher-point diagrams and those with additional internal lines~\cite{Goodhew:2021oqg}. If the exchanged field has spin, only the highest-helicity components of the exchanged field contribute to this cutting rule \cite{Baumann:2021fxj}. This will play an important role for us since we are interested in wavefunction coefficients due to graviton exchange.} The COT shows that the cut of a four-point function is completely fixed by the three-point functions appearing in the corresponding Feynman diagram.

\vskip4pt
In order to construct the four-point function of massless scalars, we begin by taking the cut of~\eqref{eq:flatspacespin2exc}:
\be
\psi^{(s)\,\rm flat}_{\phi^4}(\{ k\}, \{ \bfk \}, s)+\psi^{\ast\,(s)\,{\rm flat}}_{\phi^4}(\{- k\}, \{- \bfk \}, s) = -\frac{s}{3(k_{12}^2-s^2)(k_{34}^2-s^2)}s^4\Pi_{2,2}\,.
\label{eq:masslessflatcut}
\ee
We then want to compare it to that
of the corresponding de Sitter wavefunction, which can easily be computed from~\eqref{eq:psi3-scalar} by contracting with the $\Pi$ tensor defined in~\eqref{eq:pol-sum}:
\beq
 \psi_{\phi^4}^{(s)}(\{ k\}, \{ \bfk \}, s)+\psi^{\ast\,(s)}_{\phi^4}(\{- k\}, \{- \bfk \}, s)=\frac{s^3(k_{12}^2+2k_1k_2-s^2)(k_{34}^2+2k_3k_4-s^2)}{3(k_{12}^2-s^2)^2(k_{34}^2-s^2)^2}s^4\Pi_{2,2}\,.
 \label{eq:masslessdscut}
\ee
Notice that both~\eqref{eq:masslessflatcut} and~\eqref{eq:masslessdscut} involve the same polarization sum (as expected). 
We now want to
find a differential operator that acts on the flat space energy structure multiplying $s^4\Pi_{2,2}$ and produces the de Sitter energy structure. An operator that accomplishes this task is\footnote{The derivation of this operator is one of the slightly artisanal parts of this construction. Some general arguments motivating this form are given in~\cite{Baumann:2021fxj}. Alternatively, this operator can be obtained from inspection of the bulk time integrals that compute the wavefunction.} 
\be
\label{eq:masslessschanlift}
{\cal D}_s \equiv  -\,{\cal S}_{k_1}{\cal S}_{k_2}{\cal S}_{k_3}{\cal S}_{k_4}\left(\int{\rm d} k_{12}\, {\rm d} k_{34}\right)^2\,\left(\frac{2}{k_T}(\partial_{k_{12}}+\partial_{k_{34}})+\partial_{k_{12}}\partial_{k_{34}}\right)\,,
\ee
where all operations (including the integrals) should be understood as acting on everything to their right, and the square on the integral indicates that we should integrate twice with respect to each of $k_{12}, k_{34}$.\footnote{We are being somewhat implicit, but the integrals are definite integrals where the integration contour runs from the variable being integrated to $\infty$, as in~(5.74) of~\cite{Arkani-Hamed:2015bza}.} The
differential operator ${\cal S}_k$ is given by
\be
{\cal S}_k = 1-k\partial_k\,.
\ee
With these definitions, we can verify that
\be
{\cal D}_s \frac{s}{3(k_{12}^2-s^2)(k_{34}^2-s^2)} = -\frac{s^3(k_{12}^2+2k_1k_2-s^2)(k_{34}^2+2k_3k_4-s^2)}{3(k_{12}^2-s^2)^2(k_{34}^2-s^2)^2}\,,
\ee
as desired.
Now, the idea is that we act with the {\it same} operator on the energy structure multiplying the full flat space wavefunction. In doing so, we obtain the energy structure in de Sitter space that multiplies $s^4\Pi_{2,2}$:
\be
\begin{aligned}
f_{(2,2)}^{(s)} ={\cal D}_s \frac{1}{k_TE_LE_R} = \frac{1}{k_T}-\frac{s^2}{k_TE_LE_R}&+\frac{sk_1k_2}{k_TE_L^2 E_R}+\frac{sk_3k_4}{k_TE_LE_R^2}+\frac{2s k_1 k_2k_3k_4}{k_T^2 E_L^2E_R^2}\\
&+\frac{k_1k_2}{k_T^2E_L}+\frac{k_3k_4}{k_T^2 E_R} -\frac{s(k_1k_2+k_3k_4)}{k_T^2 E_L E_R}+\frac{2k_1k_2k_3k_4}{k_T^3 E_LE_R}\,,
\end{aligned}
\ee
which agrees with~\eqref{f22}.

\vskip4pt
We can obtain the form factors multiplying the lower-helicity polarization sums by noting  that in flat space these energy structures are just the $s\to 0$ limit of the highest-helicity form factor (with an alternating minus sign). We therefore have
\be
f_{(2,1)}^{(s)}=-f_{(2,0)}^{(s)}= \lim_{s\to 0}f_{(2,2)}^{(s)}\,,
\ee
which indeed match~\eqref{eq:f21} and~\eqref{eq:f20}.

\vskip4pt
Finally, we want to consider the uplift of $f_c^{\rm flat}$. This is slightly more subtle than the other terms. The most natural thing to do is to act with the analogue of~\eqref{eq:masslessschanlift} appropriate for contact diagrams: 
\be
\label{eq:contactlift}
{\cal D}_{k_T} \equiv  -{\cal S}_{k_1}{\cal S}_{k_2}{\cal S}_{k_3}{\cal S}_{k_4}\left(\int{\rm d} k_T\right)^2\,.
\ee
Doing this, we find the expression
\be
{\cal D}_{k_T}f_c =\frac{2k_1k_2k_3k_4(k_{12}k_{34}+s^2)}{k_T^3}+\frac{(k_1k_2k_3+k_1k_2k_4+k_1k_3k_4+k_2k_3k_4)(k_{12}k_{34}+s^2)}{k_T^2}+\cdots\,,
\label{eq:liftedcontact}
\ee
where we have only written the leading two terms in the $k_T\to 0$ limit. These match precisely those in~\eqref{eq:fcdSmassless}, but the more subleading terms we have suppressed do not. However, it is easy to understand the origin of this defect and correct it.\footnote{Another way to understand this is to note that there was ambiguity in our definition of $f_c^{\rm flat}$---we could have just as well grouped some part of it into the helicity-0 form factor, and this ambiguity affects the lifted wavefunction at ${\cal O}(k_T^{-1})$.} We expect that the full wavefunction coefficient will satisfy the MLT in~\eqref{MLT-0}, for reasons that we explain below. It is relatively straightforward to check that
\be
\partial_{k_1}\left(f^{(s)}_{(2,2)} s^4 \Pi^{(s)}_{2,2}\right)\Big\rvert_{k_1=0}=\partial_{k_1}\left(f^{(s)}_{(2,1)}s^2 \Pi^{(s)}_{2,1}\right)\Big\rvert_{k_1=0} = 0\,.
\ee
However, the helicity-$0$ piece satisfies 
\be
\partial_{k_1}\left(f^{(s)}_{(2,0)} (E_L E_R - s k_T )\Pi^{(s)}_{2,0}\right) \hspace{-1pt}\xrightarrow{k_1=0} \frac{(k_{34}^2+k_3k_4+k_2k_{34})(3(k_3-k_4)^2-s^2)(k_2(3k_2+2k_{34})+s^2)}{4s^2(k_2+k_3+k_4)^2}.
\label{eq:leftovermassless4pt}
\ee
This is certainly not zero and must be compensated by the $f_c$ term. However, the full expression~\eqref{eq:liftedcontact} satisfies the MLT by itself because of the way that we have lifted it from flat space using the ${\cal S}$ operators. We therefore need to correct it in order to cancel off~\eqref{eq:leftovermassless4pt}. The fact that~\eqref{eq:leftovermassless4pt} only contains terms with at most a $(k_2+k_3+k_4)^{-2}$ singularity indicates that we only need to correct~\eqref{eq:liftedcontact} by terms that go like $k_T^{-1}$, or are less singular in the limit that the total energy vanishes. 

\vskip4pt
In practice, one can make an ansatz including all possible terms with mass dimension three and which are no more singular than $k_{T}^{-1}$ and add them to~\eqref{eq:liftedcontact}. The answer is uniquely fixed by two requirements: the full wavefunction coefficient satisfies the MLT, and the full wavefunction coefficient vanishes in the limit that one of its external momenta is taken to be soft, as it should thanks to the shift symmetry of the scalar. The combination of these two requirements reproduces~\eqref{eq:fcdSmassless}, as we show more explicitly in Section \ref{ss:blesss4}.


\subsubsection{On the validity of the manifestly local test} \label{ss:MLT}
We now give some justification for using the MLT in this paper. The MLT is a condition that must be satisfied by all wavefunction coefficients arising from theories with manifestly local interactions involving fields whose mode functions are those of a massless scalar or graviton in de Sitter space with Bunch--Davies vacuum conditions \cite{MLT}. As written in \eqref{MLT-0}, the MLT for quartic wavefunction coefficients states that 
 \be \label{MLT}
    \frac{\partial }{\partial k_{a}} \psi_{4}\Big|_{k_{a}=0}=0\, , \quad a=1, \dots, 4 .
\ee
In this expression we consider $\psi_{4}$ as a function of $k_a$, $s$, and $t$, and hold $s$ and $t$ fixed when we differentiate with respect to $k_a$.  This condition is valid away from the physical configuration in which both $k_a$ and ${\bf k}_a$ are taken to zero, so it is distinct from cosmological soft theorems.\footnote{The MLT is also valid for light fields with $m^2<2H^2$ and generalizes to higher even spacetime dimensions \cite{Goodhew:2022ayb}.}
 
 \vskip4pt
 The MLT holds for both contact and exchange diagrams and can be derived in two complementary ways. The first comes from the fact that the bulk-to-boundary propagator for massless scalars and gravitons does not contain a term linear in the energy when expanded around $k=0$. This is the case for all conformal time, $\eta$, and so is inherited by the wavefunction coefficient as long as we keep all other variables fixed. The second method comes from demanding that $(2n-2)$-point functions that arise from a single exchange process between two $n$-point functions do not have spurious poles. The cosmological optical theorem (COT) \cite{COT} for such an exchange process constrains the energy dependence of the constituent $n$-point functions to satisfy~\eqref{MLT} in order for the $(2n-2)$-point function to be regular as the energy of the internal line goes to zero, as it should be in a manifestly local theory. The assumption of {\it manifest locality} is that the interactions involve products of fields and positive powers of their derivatives at the same spacetime point~\cite{BBBB}. This in particular assumes that the Lagrangian does not contain any inverse Laplacians. Such inverse Laplacians may arise in local theories when non-dynamical fields are integrated out~\cite{BBBB}, as we have seen above. 
 
 \vskip4pt
 Here we wish to point out and prove that the MLT is valid more generally. In particular, it applies to theories whose interaction vertices are finite in Fourier space when one of the momenta is taken to zero. To see this, consider the general interaction
 \be
 \mathcal{L}\supset \int_{\bfk_1,\ldots, \bfk_n} \hat\delta\left(\raisebox{.75pt}{$\sum$}\, \bfk_a \right) F(\bfk_1,\dots,\bfk_n)\prod_{a=1}^n \phi_a(\bfk_a)\,,
 \ee
 where $\phi_a$ are some (possibly distinct) fields of arbitrary spin (with spin indices omitted) and $F$ is some kernel that depends on the momenta. The MLT applies to external lines corresponding to the field $\phi_a$ in diagrams involving any number of such interactions as long as the interaction is soft, 
 \be \label{FiniteInteractions}
 \lim_{k_a\to 0} F(\bfk_1,\dots,\bfk_n)<\infty\,,
 \ee
 where again we assume that these fields have the usual de Sitter mode functions with Bunch--Davies vacuum conditions. The proof is the same as in Section 3.2 of \cite{MLT} and relies on the fact that the bulk-to-boundary propagator for such fields does not contain a term linear in the energy of such fields when expanded. 
 
 \vskip4pt
 This result implies the validity of the MLT for manifestly local theories, since in that case $F$ is a polynomial in the momenta $\bfk_a$, and so the above condition is satisfied. But the above argument shows that the MLT also applies in the presence of inverse Laplacians as long as they act on a product of fields, since such an interaction vertex still satisfies~\eqref{FiniteInteractions} in Fourier space. The MLT is therefore satisfied by the four-point functions of massless scalars and gravitons (as can easily be checked for the explicit expressions we write below), and we will use it as part of our bootstrap strategy in Sections~\ref{ss:blesss4} and~\ref{s:dS}.\footnote{Another argument for the validity of the MLT is that in axial gauge all of the non-localities can be shuffled into the graviton bulk-to-bulk propagator~\cite{Raju:2011mp,Albayrak:2019yve}, with all the interactions manifestly local.}

 
\subsection{The boostless bootstrap}\label{ss:blesss4}

Let us now introduce our third method for constructing four-point functions \cite{BBBB,MLT}. The strategy is again centered on the boundary---the advantage compared to the discussion in Section~\ref{ss:dSs4} is that it is more systematic and rigorous, providing a further independent check of our bulk results.
Again, we first concentrate on massless scalars exchanging a graviton, but we will use the same method to derive the pure gravity four-point function in Section~\ref{s:dS}. 

\subsubsection{Overview of approach}
The overall strategy contains three separate steps from which we can bootstrap quartic wavefunction coefficients.\footnote{We focus on the real part of the wavefunction coefficients, since for parity-even theories only the real part contributes to expectation values of field operators. See~\cite{Cabass:2021fnw} for details. Similar methods to the ones we employ here can be used to fix the imaginary parts too~\cite{MLT}.}
\begin{itemize}
    \item \textbf{Step 1: Unitarity and partial-energy poles:} ~ Late-time wavefunction coefficients that arise from unitary time evolution in the bulk spacetime must satisfy the COT,~\eqref{COT4Point}. Throughout this work we will make a distinction between the parts of wavefunction coefficients that come from time evolution and the kinematic pieces that consist of contractions between polarisation tensors and spatial momentum vectors. 
    We refer to the former as the~\textit{trimmed} contribution, which is what is constrained by the COT. 
    
    Trimmed wavefunction coefficients arising from $s$-channel exchanges can have at most three singularities: they can diverge when the total energy ($k_{T} = \sum_{a=1}^4 k_{a}$) goes to zero, or when either of the partial-energy sums ($E_{L} = k_{1}+k_{2}+s$ and $E_{R} = k_{3}+k_{4}+s$) go to zero.\footnote{This restriction follows from the choice of Bunch--Davies vacuum.}
    It was noticed in~\cite{MLT} that the COT in the form~\eqref{COT4Point} fixes the residues of $\textit{all}$ partial-energy poles. Indeed, only the first term on the LHS of~\eqref{COT4Point} has poles when $E_{L}$ or $E_R$ is taken to zero. (In the second term, the sign of the energies $k_1$ and $k_2$ have been flipped, so it is singular at different loci in energy space, where either $k_{12}-s$ or $k_{34}-s$ vanish, but regular when $E_L,E_R$ vanish.)
    Since only $\psi_4$ on the LHS of~\eqref{COT4Point} has partial-energy singularities, we can use the (known) RHS of the same equation to fix the Laurent expansion of these singularities.
    Since only the highest-helicity components of the exchanged field contribute to the cut, partial-energy poles can only arise due to the exchange of these components. From a Lagrangian point of view, this makes sense since the non-dynamical potential modes can be integrated out in favour of additional contact diagrams, as we saw in Section~\ref{ss:ssgss}, and these diagrams only give rise to total-energy poles. 
   
   The residues of partial-energy poles can be efficiently fixed using the de Sitter energy shifts derived in~\cite{MLT,Baumann:2021fxj}, which resemble the flat-space shifts employed in~\cite{Arkani-Hamed:2017fdk}, with the constituent three-point functions given as an input. We will introduce these shifts below.  
   
 \item \textbf{Step 2: The amplitude limit and leading total-energy pole:} ~ It is now well-known that cosmological wavefunction coefficients contain scattering amplitudes for the same process on their leading total-energy pole. This was noticed in~\cite{Raju:2012zr,Maldacena:2011nz}. This fact has played an important role in many approaches to bootstrapping cosmological correlators and we will employ it here to fix the residue of the leading total-energy pole of four-point functions. 
 
 For the quartic wavefunction coefficients of interest in this paper, the relationship is \cite{COT}\footnote{This differs from the result of~\cite{COT} by an overall minus sign due to different conventions for the amplitude. In the conventions of this paper, the four-point amplitude for $\mathcal{L} = \lambda \phi^4/4!$ is $\mathcal{A} =\lambda$.}
 \begin{align} \label{AmplitudeLimit}
 \lim_{k_{T} \rightarrow 0} \psi_{\phi^4}(\{ {\bf k}\}) =\frac{2}{H^2} \frac{\mathbb{e}_{4}\mathcal{A}_{4}}{k_{T}^3},
 \end{align}
 which contains both contact and exchange contributions. The theories of interest in this paper can be written to involve only interactions with net two derivatives, and therefore the order of the leading total-energy pole scales as $k_T^{-3}$.\footnote{In general, for massless fields the degree of the leading pole is given by~\cite{BBBB}
 \be \label{eq:pole-order}
 p =1 + \sum_{V} ([V ]-4),
 \ee
 where the sum is over the vertices $V$ in a given diagram and $[V]$ is the mass dimension of $V$. For partial-energy singularities, one should instead sum over the vertices associated to the subgraph whose energy is conserved.} 
 To use~\eqref{AmplitudeLimit}, we write down an ansatz for the quartic wavefunction that has an overall factor of $\mathbb{e}_{4} / k_{T}^3$ and has homogeneity degree 3 under rescaling momenta. 
 This is the correct scaling for massless scalars and gravitons required by scale invariance.
 We then demand that this ansatz satisfies~\eqref{AmplitudeLimit} for some specific amplitude $\mathcal{A}_{4}$. 
 \item \textbf{Step 3: Manifestly local test and subleading total-energy poles:} ~ The previous two steps fix all partial-energy poles along with the leading total-energy pole of a quartic wavefunction coefficient, given the three-point function and four-point amplitude as inputs. The final step is to fix the subleading total-energy poles. To do this, we demand that the full quartic wavefunction coefficient satisfies the MLT~\eqref{MLT}, which, as we discussed in Section~\ref{ss:MLT}, applies to spectator scalars and gravitons. Like the COT, the MLT constrains the trimmed part of the wavefunction only. 
 To employ the MLT, we write down an ansatz for subleading total energy poles that is consistent with scale invariance and Bose symmetry and then fix the free parameters by enforcing the MLT. 
\end{itemize}

We now apply these three steps to bootstrap the wavefunction coefficient with $s$-channel symmetries, $\psi_{\phi^4}^{(s)}$, which captures the $s$-channel exchange diagram and some contributions from contact diagrams. The other channels are given by permuting the energies and momenta. Throughout we will be working at tree-level, which implies that the quartic wavefunction coefficients we consider are rational functions \cite{Anninos:2014lwa,Goodhew:2022ayb}. 

\subsubsection{Unitarity and partial-energy poles}

Since we are considering massless scalars exchanging a graviton, the relevant three-point function is that of two massless scalars and the transverse, traceless graviton $\psi_{\phi \phi \gamma}$. This was written above in~\eqref{eq:psi3-scalar} and is given by
\be \label{TwoScalarsOneGraviton}
\psi_{\phi \phi \gamma}({\bf k}_1, {\bf k}_2,{\bf k}_3) = \frac{e_{ij}(\bfk_{3}) k_{1}^{i} k_{2}^{j}}{H^2} \tilde{\psi}_{3}(k_1, k_2, k_3), \qquad \tilde{\psi}_{3}(k_1, k_2,k_3) = \frac{1}{k_{T}^2}(k_{T}^3 -  k_{T}e_{2} - e_{3}).
\ee
We refer to $\tilde{\psi}_{3}$ as the trimmed part, which comes from the time evolution. This wavefunction coefficient can also be fixed using the MLT, where it arises as the solution with the lowest degree total-energy pole \cite{MLT}.

\vskip4pt
We can use this three-point function to fix all the partial-energy poles of $\psi_{\phi^4}^{(s)}$, as explained above. This calculation was done in~\cite{MLT}, but we repeat it here. By gluing together two copies of~\eqref{TwoScalarsOneGraviton}, the contribution to the four-point function will be a product of a tensor structure, coming from the two copies of $e_{ij}k^{i} k^{j}$, and a trimmed part which only depends on the energies $\{k \}$ and $s$. The tensor structure is
\be
\sum_{h = \pm} e_{ij}^{h}({\bf s})k_{1}^{i}k_{2}^{j} e_{lk}^{h}(-{\bf s})k_{3}^{l}k_{4}^{k},
\ee
where we sum over the two possible helicities of the exchanged graviton.  We can write this in terms of the polarisation sum $ \Pi_{2,2}^{(s)}$, given in Appendix~\ref{app:PolSums}. 
If we write the trimmed part as $f_{\text{cut}}$---which we will compute below---then we have
\be \label{WFfromCut}
\psi_{\phi^4}^{(s)}( \{ {\bf k} \} ) = \frac{f_{\text{cut}}}{12 H^2 M_{\rm Pl}^2} s^4 \Pi^{(s)}_{2,2} + \ldots,
\ee
where $\ldots$ represents terms without partial-energy poles (which will be fixed at a later stage) and we have included a factor of $H^2 / M_{\rm Pl}^2$ from the graviton power spectrum~\eqref{GravitonPS}. 

\vskip4pt
We write the trimmed part as $f_{\text{cut}}$ in~\eqref{WFfromCut} because it is fixed by cosmological cutting rules. This $s$-channel wavefunction coefficient is fundamentally  a function of the independent variables $k_{a},s,t$, but in order to write it in a way where the permutation symmetries are manifest, we write the polarisation sum in terms of $u$ as well, even though $u$ can always be eliminated by~\eqref{NL_energy_relation}.

\vskip4pt
To compute this trimmed part, we use de Sitter partial-energy shifts \cite{MLT,Baumann:2021fxj}. The trimmed wavefunction coefficient is a function of $k_{a},s$ only, and to fix it we change to a different set of independent variables.
We consider $f_{\rm cut}$ as a function of  $\{E_{L},E_{R},k_{1}k_{2},k_{3}k_{4},s\}$ so that 
$f_{\text{cut}} = f_{\text{cut}}(E_{L},E_{R},k_{1}k_{2},k_{3}k_{4},s)$,
and we define
\begin{align}
\tilde{\psi}_{3,L}(E_{L},k_{1}k_{2},s) &= \tilde{\psi}_{3}(k_1, k_2, s), \\
\tilde{\psi}_{3,R}(E_{R},k_{3}k_{4},s) &= \tilde{\psi}_{3}(k_3, k_4, s).
\end{align}
With these definitions, we can write the COT as
\be\label{NewCot}
f_{\text{cut}}(E_{L},E_{R},k_{1}k_{2},k_{3}k_{4},s) + f_{\text{cut}}^{\ast}(-E_{L}+2s,-E_{R}+2s,k_{1}k_{2},k_{3}k_{4},s) = \Xi,
\ee
where we have defined
\be
\begin{aligned}
\Xi =&\, \frac{1}{s^3}\left[\tilde{\psi}_{3,L}(E_L,k_1 k_2,s)-\tilde{\psi}_{3,L}(E_L-2s,k_1k_2,-s)\right]  \\
    &~\times \left[\tilde{\psi}_{3,R}(E_R,k_3 k_4,s)-\tilde{\psi}_{3,R}(E_R-2s,k_3k_4,-s)\right].
\end{aligned}
\ee
Here we have used the COT for contact diagrams to eliminate the complex conjugate as~\cite{COT} 
\be
\tilde{\psi}_{3, L}^{\ast}(-E_{L}+2s,k_{1}k_{2},s) = -\tilde{\psi}_{3,L}(E_{L}-2s,k_{1}k_{2},-s).
\ee
The second term on the LHS of~\eqref{NewCot} is analytic around $E_{L},E_{R}=0$ and therefore does not contribute to the residues of partial-energy poles, so we can fix the poles of the first term from knowledge of $\Xi$. To isolate the partial-energy poles, we perform a complex shift of the partial energies that keeps the total energy fixed:
\be
f_{\text{cut}}(E_{L},E_{R},k_{1}k_{2},k_{3}k_{4},s) \mapsto \tilde{f}_{\text{cut}}(z) = f_{\text{cut}}(E_{L}+z, E_{R}-z, k_{1}k_{2},k_{3}k_{4},s).
\ee
This shifted wavefunction coefficient trivially reproduces the unshifted one at $z=0$, is an analytic function in the complex $z$ plane, except for isolated poles in two locations, $z = -E_{L}$ and $z = E_{R}$, and has a Laurent expansion near these poles given by~\cite{MLT}
\begin{align}
\label{Laurentz}
    \tilde{f}_{\text{cut}}(z)&=\sum_{0<n\leq m}\dfrac{A_n(E_R,E_L,k_1k_2,k_3k_4,s)}{(z+E_L)^n}+{\cal O}(z+E_L),\\ \label{An}
    A_n &=\dfrac{1}{(m-n)!} \left[\partial^{m-n}_z (z+E_L)^m\, \Xi(E_L+z,E_R-z,k_1k_2,k_3k_4,s)\right]_{z=-E_L},
\end{align}
where $m$ is the order of the leading partial energy pole, and the expressions for the $A_{n}$ follow from~\eqref{NewCot} and are completely fixed by the three-point function. In $f_{\text{cut}}$, the expansions around $E_{L}=0$ and $E_{R}=0$ are equivalent since we are using the same three-point function for each sub-diagram. We can use the residue theorem to write
\be
  f_{\text{cut}}(E_L,E_R,k_1k_2,k_3k_4,s)=\dfrac{1}{2\pi i}\oint\limits_{{\cal C}_0}\rd z\, \dfrac{\tilde{f}_{\text{cut}}(z)}{z}\,,
\ee
where ${\cal C}_0$ is a contour that encircles the origin, oriented clockwise. We can then deform the contour and write the trimmed wavefunction coefficient as a sum over residues plus a boundary term as~\cite{MLT}
\begin{align} \label{Psi4Residue}
    f_{\text{cut}}(E_L,E_R,k_1k_2,k_3k_4,s) &=-\text{Res}\left[\dfrac{\tilde{f}_{\text{cut}}(z)}{z}\right]_{z=-E_L}-\text{Res}\left[\dfrac{\tilde{f}_{\text{cut}}(z)}{z}\right]_{z=E_R}+B \\
    & = f_{\text{Res}} + B .
\end{align}
The boundary contribution at infinity, $B$, is given by
\begin{align}\label{bound}
    B=\dfrac{1}{2\pi i}\oint\limits_{{\cal C}_\infty}\rd z\, \dfrac{\tilde{f}_{\text{cut}}(z)}{z}\,,
\end{align}
and we can use the Laurent expansion~\eqref{Laurentz} to write
\begin{align} \label{psireshere}
    f_{\text{Res}}&=\sum_{0<n\leq m}\dfrac{A_n(E_L,E_R,k_1k_2,k_3k_4,s)}{E_L^n}+\sum_{0<n\leq m}\dfrac{A_n(E_R,E_L, k_3k_4,k_1k_2,s)}{E_R^n}\,.
\end{align}
 We see that the partial-energy poles are now completely fixed by taking derivatives of the three-point function, which fixes the trimmed four-point function up to a boundary term that can only be singular when the total energy goes to zero: $E_{L}+E_{R}-2s \rightarrow 0$.\footnote{\label{ft:mlts}As explained in~\cite{MLT}, this boundary term might need to contain a simple $s^3$ term in order to satisfy the COT. This can be easily checked by taking~\eqref{Psi4Residue} and comparing the LHS and RHS of~\eqref{NewCot}. We refer the reader to \cite{MLT,Baumann:2021fxj} for further details on this procedure.}

\vskip4pt
We now turn back to the case of interest here and use~\eqref{TwoScalarsOneGraviton} to compute the $A_{n}$. The degree of the leading partial-energy poles is equivalent to the degree of the leading total-energy pole of the corresponding three-point function, which for us is two, so that $m=2$ in~\eqref{Laurentz}. We then have \cite{MLT}
\begin{align}
A_{2}(E_{L},E_{R},k_{1}k_{2},k_{3}k_{4},s) &= \frac{2  k_{1}k_{2}s}{k_{T}^2(E_{L}+E_{R})^2}\big[2k_{3}k_{4} + (E_{L}+E_{R})k_{T}\big], \\
A_{1}(E_{L},E_{R},k_{1}k_{2},k_{3}k_{4},s)&= \frac{1}{k_{T}^3(E_{L}+E_{R})^3}\sum_{n=0}^4 a_{n}s^{n}, 
\label{eq:A1exp}
\end{align}
where the coefficients appearing in the expansion~\eqref{eq:A1exp} are
\begin{align}
a_{0} &= 2k_{1}k_{2}(E_{L}+E_{R})^2[(E_{L}+E_{R})^2 + 2k_{3}k_{4}], \\
a_{1} &= -4k_{1}k_{2}(E_{L}+E_{R})[(E_{L}+E_{R})^2 - 2k_{3}k_{4}], \\
a_{2} &= -2(E_{L}+E_{R})^4 -4 (E_{L}+E_{R})^2(k_{1}k_{2}+k_{3}k_{4})-16k_{1}k_{2}k_{3}k_{4}, \\
a_{3} &= 8(E_{L}+E_{R})[(E_{L}+E_{R})^2 + k_{1}k_{2}+k_{3}k_{4}], \\
a_{4} &= -8(E_{L}+E_{R})^2,
\end{align}
and $k_{T} = E_{L}+E_{R}-2s$. If we plug these expressions into~\eqref{psireshere}, the resulting $f_{\text{Res}}$ satisfies the COT so there is no need to include any additional $s^3$ pieces (see Footnote~\ref{ft:mlts}). It follows that
\be
\begin{aligned} \label{fCut}
f_{\text{cut}} = &-\frac{2s^2}{k_{T}E_{L}E_{R}}+\frac{2s k_{1}k_{2}}{k_{T}E_{L}^2E_{R}}+\frac{2s k_{3}k_{4}}{k_{T}E_{R}^2E_{L}}+\frac{4s  k_{1}k_{2}k_{3}k_{4}}{k_{T}^2 E_{L}^2 E_{R}^2} \\ &-\frac{2s(k_{1}k_{2}+k_{3}k_{4})}{k_{T}^2 E_{L}E_{R}}
+ \frac{2k_{1}k_{2}}{k_{T}^2E_{L}}+\frac{2k_{3}k_{4}}{k_{T}^2E_{R}} + \frac{4 k_{1}k_{2}k_{3}k_{4}}{k_{T}^3E_{L}E_{R}}.
\end{aligned}
\ee
We note that in the final expression for $f_{\text{cut}}$, all spurious poles have cancelled out, and consistency dictates the need for total-energy poles even though no information about these residues was input directly. We see that these partial-energy poles match those found by computing nested time integrals in Section \ref{ss:ssgss}, given by $f_{(2,2)}^{(s)}$. 

\subsubsection{The amplitude limit and leading total-energy pole}

Having fixed all the partial-energy poles, we now ensure that the leading total-energy pole reproduces the correct scattering amplitude. 
The four-point amplitude for minimally coupled scalars is
\be \label{ScalarsAmplitude}
\mathcal{A}_4= \frac{1}{4 M_{\rm Pl}^2} \frac{(S^2+T^2+U^2)^2}{STU}.
\ee
Our strategy for constructing the full wavefunction coefficient is to work
``channel-by-channel", meaning that we initially concentrate on the part of the wavefunction coefficient that has the symmetries of an $s$-channel exchange diagram. The relevant symmetry group is $\mathbb{Z}_2 \times \mathbb{Z}_2$, which acts on the momenta as follows: ${\bf k}_{1} \leftrightarrow {\bf k}_{2}$, ${\bf k}_{3} \leftrightarrow {\bf k}_{4}$, and $({\bf k}_{1},{\bf k}_{2})  \leftrightarrow ({\bf k}_{3},{\bf k}_{4})$. We then sum over different channels at the end of the calculation. We have written~\eqref{WFfromCut} in a way that makes these symmetries manifest, using a redundant set of variables since the energies are related by~\eqref{NL_energy_relation}. We therefore need to decompose~\eqref{ScalarsAmplitude} into a sum over channels where each channel has the relevant symmetries when written in terms of $S,T,U$ (which satisfy $S+T+U=0$). When we match the $s$-channel wavefunction coefficient to the $S$-channel amplitude according to~\eqref{AmplitudeLimit}, we will respectively eliminate $u$ and $U$ such that we work with independent variables. 
We define the $S$-channel amplitude as
\be \label{MasslessScalarsAmp}
\mathcal{A}_{4}^{(S)} 
= \frac{1}{M_{\rm Pl}^2}\frac{TU}{S},
\ee
such that $\mathcal{A}_{4} = \mathcal{A}_{4}^{(S)}+\mathcal{A}_{4}^{(T)}+\mathcal{A}_{4}^{(U)}$. Such a definition is ambiguous (the invariant quantity is the residue of the $S=0$ pole), but we must make some choice and our final wavefunction coefficient will not depend on this choice.\footnote{For example, we could add to $\mathcal{A}_{4}^{(S)}$ a term proportional to $S$ without  affecting the total amplitude. This extra term would then end up being degenerate with a local term that can always be added to the wavefunction.
}

\vskip4pt
We choose to access the total-energy pole by sending $k_{4} \rightarrow -(k_{1}+k_{2}+k_{3})$ and therefore according to~\eqref{AmplitudeLimit} we must demand that the wavefunction coefficient satisfies
\be \label{AmpLimit}
\lim_{k_{T} \rightarrow 0} k_T^3 \psi_{\phi^4}^{(s)} =   \frac{2}{H^2 M_{\rm Pl}^2} k_{1}k_{2}k_{3}(k_{1}+k_{2}+k_{3})\frac{(k_{13}^2-t^2)(k_{12}^2+k_{13}^2-s^2-t^2)}{(k_{12}^2-s^2)}.
\ee
We initially take the result from step 1 of the previous section (written as a function of $k_{a},s,t$), ~\eqref{WFfromCut}, and compare its amplitude limit to~\eqref{AmpLimit}. We find that we do not recover the correct expression and that the difference between the two expressions can be written as a series in $s$ with even powers ranging from $s^{-4}$ to $s^{2}$. The $s^{-4}$, $s^{0}$ and $s^{2}$ terms have coefficients that are functions of $k_{1}$, $k_{2}$ and $k_{3}$, while the coefficient of the $s^{-2}$ term also depends linearly on $t^2$. We now need to correct~\eqref{WFfromCut} such that we recover the correct amplitude limit and we can do so by writing down an ansatz order-by-order in $s$. 

\vskip4pt
Recall that our ansatz should be a function of \textit{all} internal and external energies, such that all symmetries are manifest, then we eliminate $u$ when we go on the total-energy pole. Our ansatz needs to include terms linear in $t^2$, which means that by symmetry there is also a linear dependence on $u^2$. Our ansatz should scale as $\sim k^3$ by scale invariance and should contain an overall factor of $\mathbb{e}_{4} / k_{T}^3$ given~\eqref{AmpLimit}. We therefore add to~\eqref{WFfromCut}:
\be \label{WFfromAmp}
\Delta_{1}\psi_{\phi^4}^{(s)} = \frac{\mathbb{e}_{4}}{s^4 k_{T}^3} \text{Poly}^{(6)}(k_{1},k_{2},k_{3},k_{4},s^2,t^2,u^2),
\ee
where $\text{Poly}^{(6)}$ denotes a general polynomial of its arguments that scales as $\lambda^6$ under ${\bf k}_a \mapsto \lambda {\bf k}_a$.
The wavefunction coefficient should not be singular as $s \rightarrow 0$, as this would correspond to a level of non-locality. We need to use factors of $(k_{1}-k_{2})$ and $(k_{3}-k_{4})$ to cancel such poles. To see this we take ${\bf k}_{2} \rightarrow - {\bf k}_{1} + {\bm \epsilon}$ and ${\bf k}_{4} \rightarrow - {\bf k}_{3} - {\bm \epsilon}$ which fixes ${\bf s} \rightarrow {\bm \epsilon}$. In terms of the energies we have
\begin{align}
k_{2} \rightarrow \sqrt{k_{1}^2 - 2 {\bm \epsilon} \cdot {\bf k}_{1} + \epsilon^2}, \qquad
k_{4} \rightarrow \sqrt{k_{3}^2 + 2 {\bm \epsilon} \cdot {\bf k}_{3} + \epsilon^2}, \qquad
s \rightarrow \epsilon.
\end{align}
Due to the ${\bm \epsilon} \cdot {\bf k}_{1}$ and ${\bm \epsilon} \cdot {\bf k}_{3}$ terms, the only way to guarantee that a polynomial in the external energies cancels a pole in $s$ of order $c$, is for the polynomial to contain overall factors of the form $(k_{1}-k_{2})^{a} (k_{3}-k_{4})^{b}$ subject to $a+b=c$. With these conditions, our ansatz takes the form
\begin{align}
\text{Poly}^{(6)}(k_{1},k_{2},k_{3},k_{4},s^2,t^2,u^2)  =  \bigg[ & (k_{1}-k_{2})^4 \, \text{Poly}^{(2)}_{1}(k_{12},k_{1}k_{2},k_{34},k_{3}k_{4})  \nonumber \\
 & +  (k_{1}-k_{2})^2 (k_{3}-k_{4})^2 \,\text{Poly}^{(2)}_{2}(k_{12},k_{1}k_{2},k_{34},k_{3}k_{4}) \nonumber \\ & +  (k_{1}-k_{2})^2 s^2 \,\text{Poly}^{(2)}_{3}(k_{12},k_{1}k_{2},k_{34},k_{3}k_{4}) \nonumber  \\
 & + b_{1} (k_{1}-k_{2})(k_{3}-k_{4}) s^2 (t^2-u^2)  \nonumber \\
 & + s^4 \,  \text{Poly}^{(2)}_{4}(k_{12},k_{1}k_{2},k_{34},k_{3}k_{4})   \nonumber \\ &  + b_{2} s^6 \bigg]+ \left( {\bf k}_{1} \leftrightarrow {\bf k}_{3}, {\bf k}_{2} \leftrightarrow {\bf k}_{4} \right) .
\end{align}
The notation $\text{Poly}^{(n)}_{m}(x_1, x_2, \dots)$ denotes a general polynomial of $x_i$ with homogeneity degree $n$ under rescaling momenta, where $m$ is an index distinguishing different polynomials of the same homogeneity degree, and $b_i$ are constants. 
We have dropped any terms that depend on $(t^2+u^2)$ since they are degenerate with terms already contained in the ansatz due to the relation in~\eqref{NL_energy_relation}.

\vskip4pt
Demanding that we recover the correct amplitude in the limit $k_T\to 0$ leads to a family of solutions which only differ at $\mathcal{O}(1 / k_{T}^2)$. At this stage we can pick any of these solutions without loss of generality since we will ultimately fix the subleading total-energy poles in the next section. We choose a solution such that our wavefunction coefficient takes the form
\begin{align} \label{AfterStep2}
\psi_{\phi^4}^{(s)}  = &\frac{1}{6 H^2 M_{\rm Pl}^2} \left[ \frac{f_{\text{cut}}}{2} s^4 \Pi^{(s)}_{2,2} - \frac{2 \mathbb{e}_{4}}{k_{T}^3}(s^2 \Pi^{(s)}_{2,1} - (k_{12}k_{34}+s^2)\Pi^{(s)}_{2,0}) - \frac{2 \mathbb{e}_{4}(k_{12}k_{34}+s^2)}{k_{T}^3} \right] \\  &+ \ldots \nn,    
\end{align}
where we see the appearance of familiar polarisation sums, which are written in Appendix~\ref{app:PolSums}. Note that we have not assumed that these sums must appear, rather they are required to realise the correct amplitude limit. This expression now satisfies the COT, has the correct amplitude limit, and the $\ldots$ represents terms that both do not contribute to partial-energy poles and are subleading as $k_{T} \rightarrow 0$. These will be fixed in the next section.

\subsubsection{Manifestly local test and subleading total-energy poles} \label{sec:scalarsMLT}

Finally, we have to ensure that the wavefunction coefficient satisfies the MLT in the form~\eqref{MLT}. 
Recall that we take the derivative with respect to an external energy and then set that energy to zero while holding all other variables fixed. We can therefore work order-by-order in $s,t$ (we eliminate $u$ before taking the derivative). We find that~\eqref{AfterStep2} does not satisfy the MLT. Instead, the derivative of this object with respect to $k_{4}$ at $k_{4}=0$ can be written as a series in $s$ with even powers ranging from $s^{-4}$ to $s^{4}$. Each series coefficient depends on $k_{1},k_{2}$ and $k_{3}$, while the $s^{-2}$ and $s^{2}$ coefficients also depend linearly on $t^2$ and the $s^{0}$ coefficient also depends linearly and quadratically on $t^2$. This allows us to write down an ansatz which we add to~\eqref{AfterStep2}. We take the $t^2$ and $u^2$ terms to appear in the combination $(t^2-u^2)$ since this is how they appear in~\eqref{AfterStep2} (regardless of which solution we choose in matching to the amplitude).

\vskip4pt
Our ansatz cannot have partial-energy poles and cannot alter the leading total-energy pole since these have already been fixed. We therefore add to~\eqref{AfterStep2}:
\be \label{Step3Ansatz}
\Delta_{2} \psi^{(s)}_{\phi^4} = \frac{1}{s^4 k_{T}^2} \text{Poly}^{(9)}(k_{1},k_{2},k_{3},k_{4},s^2,t^2,u^2),
\ee
where the homogeneity degree of the polynomial is fixed by scale invariance. As before, we need to maintain the symmetries of an $s$-channel exchange diagram and ensure that there are no poles as $s \rightarrow 0$ for physical momenta. This allows us to write
\begin{align}
\text{Poly}^{(9)}(k_{1},k_{2},k_{3},k_{4},s^2,t^2,u^2) = \bigg[ & (k_{1}-k_{2})^4 \text{Poly}^{(5)}_{1}(k_{12},k_{1}k_{2},k_{34},k_{3}k_{4}) \nonumber \\
& + (k_{1}-k_{2})^2 (k_{3}-k_{4})^2 \text{Poly}^{(5)}_{2}(k_{12},k_{1}k_{2},k_{34},k_{3}k_{4})  \nonumber \\ & +  (k_{1}-k_{2})^2 s^2 \text{Poly}^{(5)}_{3}(k_{12},k_{1}k_{2},k_{34},k_{3}k_{4})  \nonumber  \\
 & +(k_{1}-k_{2})(k_{3}-k_{4}) s^2 (t^2-u^2) \text{Poly}^{(3)}_{1}(k_{12},k_{1}k_{2},k_{34},k_{3}k_{4})  \nonumber \\
 & + s^4 \text{Poly}^{(5)}_{4}(k_{12},k_{1}k_{2},k_{34},k_{3}k_{4})   \nonumber \\
 & +b_{3}(k_{1}-k_{2})(k_{3}-k_{4})s^4(t^2-u^2)k_{T}\nonumber \\
 & +b_{4}s^4(t^2-u^2)^2k_{T} \nonumber \\
 & + s^6 \text{Poly}^{(3)}_{2}(k_{12},k_{1}k_{2},k_{34},k_{3}k_{4}) \nonumber \\
 & + b_{5} s^8 k_{T} \bigg] +\left( {\bf k}_{1} \leftrightarrow {\bf k}_{3}, {\bf k}_{2} \leftrightarrow {\bf k}_{4} \right),
\end{align}
where again the superscripts on the polynomials indicate their homogeneity degree under rescaling the momenta and $b_i$ are constants. 

\vskip4pt
In the previous section we found a family of solutions that yields the correct amplitude limit (since we did not use $k_{T}$ as an independent variable), and all of these solutions differ by functions that are captured by~\eqref{Step3Ansatz}. After imposing the MLT we find three free parameters (in addition to $H$ and $M_{\rm Pl}$) and we can write the part of the wavefunction coefficient with $s$-channel symmetries as 
\be
\begin{aligned} \label{AfterStep3}
\psi_{\phi^4}^{(s)}(\{ \bfk \}) & =\frac{1}{6 H^2 M_{\rm Pl}^2} \left[ f^{(s)}_{(2,2)} s^4 \Pi^{(s)}_{2,2} + f^{(s)}_{(2,1)}s^2 \Pi^{(s)}_{2,1}+ f^{(s)}_{(2,0)} (E_L E_R - s k_T )\Pi^{(s)}_{2,0} + f_{c} \right]  \\
& + q_{1} (k_{1}^3+k_{2}^3+k_{3}^3+k_{4}^3)  \\ & + \frac{q_{2}}{s^2}\big[(k_{1}-k_{2})^2 k_{12}(k_{12}^2-k_{1}k_{2})+(k_{3}-k_{4})^2 k_{34}(k_{34}^2-k_{3}k_{4})\big]  \\
& + \frac{q_{3}}{s^4}\big[(k_{1}-k_{2})^4 k_{12}(k_{12}^2+k_{1}k_{2})+(k_{3}-k_{4})^4 k_{34}(k_{34}^2+k_{3}k_{4})\big],
\end{aligned}
\ee
where the constants $q_{i}$ are linear combinations of the constants contained in the polynomial ansatz, and the $f$s are as defined in Section~\ref{ss:ssgss}. 
We therefore see that up to these additional three terms, we match the bulk calculation. It is easy to see that these three extra terms satisfy all the conditions we have imposed: they are regular as $E_{L,R} \rightarrow 0$, they do not contribute to the leading total-energy pole, and one can straightforwardly check that they satisfy the MLT. 

\vskip4pt
The fact that these additional solutions do not have any
total-energy poles suggests that they arise from field redefinitions~\cite{BBBB}. The $q_{1}$ term is well known, and arises from taking the free theory for the massless scalar and performing the field redefinition $\phi \mapsto \phi +  \frac{H^2}{4}q_{1}\phi^3$ (this field redefinition generates a wavefunction coefficient $\psi_{4} = 3 q_{1} \sum_{a=1}^{4}k_{a}^3$, which is what we get when we add the $t$ and $u$ permutations to~\eqref{AfterStep3}). The other two require more complicated field redefinitions which must violate manifest locality given that the resulting wavefunction coefficient contains inverse powers of $s$. They are given by\footnote{We have checked that there are no other field redefinitions that could yield $1/s^2$ or $1/s^4$ terms in $\psi_{\phi^4}^{(s)}$, while also ensuring the absence of $s \rightarrow 0$ poles for physical momenta. This is consistent with what we have found using the MLT in this section.} 
\begin{align} 
\phi \mapsto  \phi & + \frac{ H^2 q_{2} }{3} \left[3 \phi \frac{1}{\partial^2}(\phi \partial^2 \phi) +  \phi \frac{1}{\partial^2} (\phi'^2) + \phi' \frac{1}{\partial^2}(\phi \phi') \right], \label{nonlocal1} \\
\phi \mapsto  \phi &+
   \frac{H^2 q_3}{3}\left[-27\partial^2\phi\frac{1}{\partial^4}(\phi\partial^2\phi)-12\phi\frac{1}{\partial^4}(\phi\partial^4\phi)+6\partial^2\phi\frac{1}{\partial^4}({\phi'}^2)+2\phi'\frac{1}{\partial^4}(\phi\partial^2\phi')\right. \nonumber \\ &\left.-4\phi\frac{1}{\partial^4}(\phi'\partial^2\phi') - 10\phi'\frac{1}{\partial^4}(\phi'\partial^2\phi)-30\partial^2\phi\frac{1}{\partial^4}(\partial_i\phi\partial_i\phi)-60\partial_i\phi\frac{1}{\partial^4}(\phi\partial^2\partial_i\phi)\right]. \label{nonlocal2}
\end{align}
Since any choice of $q_{1,2,3}$ satisfies all our requirements, we can simply set them to any value we like. However, we can also fix these additional terms by imposing a shift symmetry for $\phi$ which is present for minimally coupled massless scalars. A shift symmetry ensures that the wavefunction coefficient has a vanishing soft limit~\cite{Ghosh:2014kba,Bittermann:2022nfh}:
\be \label{SoftLimit}
\lim_{\bfk_{4} \rightarrow 0} \psi_{\phi^4} = 0.
\ee
We take $\bfk_{4} \rightarrow 0$ by sending $k_{4} \rightarrow 0$, $s \rightarrow k_{3}$, $t \rightarrow k_{2}$ (after eliminating $u$), and find that~\eqref{SoftLimit} fixes $q_{1} = q_{2} = q_{3} = 0$. The wavefunction coefficient now exactly matches the bulk expression.


\section{The graviton trispectrum} \label{s:brute}

We now turn to the case of principal interest: the graviton trispectrum. In the previous section, we employed several different methods to derive the analogous wavefunction and correlator for massless minimally coupled scalars interacting via gravity. The consistency of these results gives us confidence in them, and so we will again derive the result in these different ways as a cross-check. We again find agreement, up to terms that can be generated by field redefinitions of the graviton (which are unavoidable ambiguities).


\subsection{Lagrangian calculation} \label{ss:lagrangian}

In this section, we derive the tree-level graviton trispectrum in de Sitter spacetime for pure gravity, starting from the Einstein--Hilbert action. First, in Section~\ref{ss:constr}, after choosing a convenient gauge, we derive the constraint equations for the lapse and shift and then solve them to second order. Second, in Section~\ref{ss:L4}, we obtain the full Lagrangian to quartic order, which we then use to compute the graviton quartic wavefunction coefficient and the corresponding correlator in Section~\ref{ss:psi4}. 

 
\subsubsection{Constraint equations}\label{ss:constr}

The constraint equations~\eqref{eq:constraints} expanded out to second order, including the $\gamma_{ij}$ contributions but not the $\phi$ contributions, can be written as
\be
\begin{aligned}
	\frac{1}{4a^2}\gamma_{ij,k}\gamma_{ij,k}+\frac{1}{4}\dot{\gamma}_{ij}\dot{\gamma}_{ij}+\frac{4}{a^2}H\partial_i\partial_i\psi^{(2)}+12H^2\alpha^{(2)} &=0,\\
	2H\partial_j\alpha^{(2)}+\frac{1}{4}\left(\dot{\gamma}_{ik}\gamma_{ij,k}-\dot{\gamma}_{ik}\gamma_{ik,j}-\dot{\gamma}_{kj,i}\gamma_{ki}-\frac{2}{a^2}\partial_i\partial_i\beta_{j}^{(2)}\right)&=0.
\end{aligned}
\ee
In order to simplify some later equations, it is convenient to define $V_j$ by
\begin{equation}
V_j=\frac{1}{2}\left(-\dot{\gamma}_{ik}\gamma_{ij,k}+\dot{\gamma}_{ik}\gamma_{ik,j}+\dot{\gamma}_{kj,i}\gamma_{ki}\right),
\end{equation}
which has the divergence
\begin{equation}
    \partial_iV_i=\frac{1}{2}\partial_i\left(\dot{\gamma}_{jk}\partial_i\gamma_{jk}\right),
\end{equation}
where we have used the fact that the graviton is transverse. The solutions to the constraint equations can then be written in terms of $V_i$ as\footnote{This is the point of disagreement with~\cite{Fu:2015vja}, who have a different solution to the constraints.}
\begin{align}
	\alpha^{(2)}&=\frac{1}{4H}\partial^{-2}\partial^jV_j,\\
	\psi^{(2)}&=-\frac{a^2}{16H}\partial^{-2}\left(\frac{1}{a^2}\gamma_{ij,k}\gamma_{ij,k}+\dot{\gamma}_{ij}\dot{\gamma}_{ij}+12H\partial^{-2}\partial^jV_j\right),\\
	\beta^{(2)}_{j}&=a^2\partial^{-2}\left(\partial^{-2}\partial_j\partial_iV_i-V_j\right).
\end{align}
The presence of inverse Laplacians requires us to specify some boundary conditions for the fields, which---as all three of these terms are perturbations---are that they must vanish on the boundary of our volume. This allows us to integrate spatial derivatives by parts.

 
\subsubsection{The Lagrangian to quartic order}\label{ss:L4}

Before substituting the solutions of the constraint equations, the Lagrangian to quartic order can be written as
\be
\begin{aligned}
    \mathcal{L}&=\frac{M_{\rm Pl}^2}{2}a^3\left(-12H^2+\frac{1}{4}\left(\dot{\gamma}_{ij}\dot{\gamma}_{ij}-\frac{1}{a^2}\partial_k\gamma_{ij}\partial_k\gamma_{ij}\right)+\frac{1}{4a^2}\gamma_{ij}\partial_j\gamma_{km}\left(\partial_i\gamma_{km}-2\partial_m\gamma_{ik}\right)\right.  \\&\left.-\frac{1}{24}\dot{\gamma}_{ij}\dot{\gamma}_{km}\gamma_{ik}\gamma_{jm}+\frac{1}{24}\dot{\gamma}_{ik}\dot{\gamma}_{ij}\gamma_{jm}\gamma_{km}-\frac{1}{4}\alpha^{(2)}\left(\dot{\gamma}_{ij}\dot{\gamma}_{ij}+\frac{1}{a^2}\partial_k\gamma_{ij}\partial_k\gamma_{ij}\right)-6H^2(\alpha^{(2)})^2\right.  \\&\left.+\frac{1}{8a^2}\gamma_{ij}\left(\gamma_{km}\partial_j\gamma_{mn}\partial_k\gamma_{in}-\gamma_{ik}\partial_j\gamma_{mn}\partial_k\gamma_{mn}-\frac{2}{3}\gamma_{km}\partial_m\gamma_{jn}\partial_n\gamma_{ik}\right.\right.  \\&\left.\left.+\frac{1}{3}\gamma_{km}\partial_n\gamma_{jm}\partial_n\gamma_{ik}+\frac{4}{3}\gamma_{ik}\partial_k\gamma_{mn}\partial_n\gamma_{jm}+\frac{1}{3}\gamma_{ik}\partial_m\gamma_{kn}\partial_n\gamma_{jm}-\frac{1}{3}\gamma_{ik}\partial_n\gamma_{km}\partial_n\gamma_{jm}\right)\right. \\&\left.+\frac{1}{2a^4}\partial_j\beta^{(2)}_i\partial_j\beta^{(2)}_i-\frac{4H}{a^2}\alpha^{(2)}\partial_i\partial_i\psi^{(2)}-\frac{1}{a^2}V_i(\beta^{(2)}_i+\partial_i\psi^{(2)})\right).
\end{aligned}
\ee
We can then eliminate the lapse and shift and integrate the expression by parts in space (using the fact that $\alpha,\ \beta$ and $\psi$ vanish on the boundary) to obtain the action up to quartic order in graviton fluctuations, 
\begin{align}
    \mathcal{L}&=\frac{M_{\rm Pl}^2}{2}\left(-12a^4H^2+\frac{a^2}{4}\left(\gamma'_{ij}\gamma'_{ij}-\partial_k\gamma_{ij}\partial_k\gamma_{ij}\right)+\frac{a^2}{4}\gamma_{ij}\partial_j\gamma_{km}\left(\partial_i\gamma_{km}-2\partial_m\gamma_{ik}\right)\right. \nn\\&\left.+\frac{a^2}{4}\left(\frac{1}{6}\gamma'_{ik}\gamma'_{ij}\gamma_{jm}\gamma_{km}-\frac{1}{6}\gamma'_{ij}\gamma'_{km}\gamma_{ik}\gamma_{jm}+\frac{1}{2}\gamma_{ij}\gamma_{km}\partial_j\gamma_{mn}\partial_k\gamma_{in}-\frac{1}{2}\gamma_{ij}\gamma_{ik}\partial_j\gamma_{mn}\partial_k\gamma_{mn}\right.\right. \nn\\&\left.\left.+\frac{1}{2}\gamma_{ik}\gamma_{ij}\partial_m\gamma_{kn}\partial_n\gamma_{jm}+\gamma_{ik}\gamma_{ij}\partial_k\gamma_{mn}\partial_n\gamma_{jm}+\frac{1}{6}\gamma_{ij}\gamma_{km}\partial_n\gamma_{jm}\partial_n\gamma_{ik}-\frac{1}{6}\gamma_{ik}\gamma_{ij}\partial_n\gamma_{km}\partial_n\gamma_{jm}\right)\right. \nn \\&\left.+\frac{a^4}{2}V_j\partial^{-2}V_j-\frac{a}{32H}\partial^{-2}\partial_l\left(\gamma'_{mn}\partial_l\gamma_{mn}\right)\left(\gamma'_{ij}\gamma'_{ij}+\partial_k\gamma_{ij}\partial_k\gamma_{ij}\right) \right. \nn\\&\left. +\frac{a^2}{32}\partial^{-2}\partial_i\left(\gamma'_{jk}\partial_i\gamma_{jk}\right)\partial^{-2}\partial_l\left(\gamma'_{mn}\partial_l\gamma_{mn}\right)\right), \label{toApp}
\end{align}
where we have changed variables from cosmic time to conformal time for later convenience. As with the scalar case above, all inverse Laplacians act on a product of fields, so the MLT will be satisfied by our final wavefunction coefficient (see Section~\ref{sec:scalarsMLT}). We can then Fourier transform this expression and rewrite $\gamma_{ij}$ in terms of polarisation tensors using~\eqref{eq:gamma}, where the polarization tensors satisfy the conditions in~\eqref{eq:pol-conditions}.
The Lagrangian is then given by 
\begin{align}
    \mathcal{L}&=M_{\rm Pl}^2\int_{\bf{k}}\sum_{h} \frac{a^2}{8}\left({\gamma'}_k^h{\gamma'}_k^{h}-k^2\gamma_k^h\gamma_k^{h}\right) \nn \\ &+M_{\rm Pl}^2\int_{\bf{k}_1,\bf{k}_2,\bf{k}_3}\sum_{h_1,h_2,h_3}\frac{a^2}{8}\epsilon^{ij}_{1}k^{j}_1\epsilon^{lm}_{2}\left(k_{3}^i\epsilon^{lm}_{3}-2k_{3}^m\epsilon^{il}_{3}\right)\gamma_{k_1}^{h_1}\gamma_{k_2}^{h_2}\gamma_{k_3}^{h_3} \nn \\
    &+M_{\rm Pl}^2\int_{\textbf{k}_1, \dots, \bf{k}_4}\sum_{h_{1,2,3,4}}-\frac{a^2}{48}\gamma_{k_1}^{h_1}\gamma_{k_2}^{h_2}\gamma_{k_3}^{h_3}\gamma_{k_4}^{h_4}\left(3\epsilon^{ij}_{1}\epsilon^{lm}_{2}k_{3}^j\epsilon^{mn}_{3}k^l_{4}\epsilon^{in}_{4}-3\epsilon^{ij}_{1}\epsilon^{il}_{2}k_{3}^j\epsilon^{mn}_{3}k_{4}^l\epsilon^{mn}_{4} \right. \nn \\
    &\left.+3\epsilon^{il}_{1}\epsilon^{ij}_{2}k_{3}^m\epsilon^{ln}_{3}k_{4}^n\epsilon^{jm}_{4}+6\epsilon^{il}_{1}\epsilon^{ij}_{2}k_{3}^l\epsilon^{mn}_{3}k_{4}^n\epsilon^{jm}_{4}+k_{3}^nk_{4}^n\epsilon^{ij}_{1}\epsilon^{lm}_{2}\epsilon^{jm}_{3}\epsilon^{il}_{4}-k_{3}^nk_{4}^n\epsilon^{il}_{1}\epsilon^{ij}_{2}\epsilon^{lm}_{3}\epsilon^{jm}_{4}\right)\nn\\
   &+\frac{a^2}{48}{\gamma_{k_1}^{h_1}}'\gamma_{k_2}^{h_2}{\gamma_{k_3}^{h_3}}'\gamma_{k_4}^{h_4}\left[\epsilon^{ij}_{1}\epsilon^{jm}_{2}\epsilon^{ik}_{3}\epsilon^{km}_4-\epsilon^{ij}_1\epsilon^{ik}_2\epsilon^{km}_3\epsilon^{jm}_4+\frac{3(k_{1}^ik_{2}^i+k_2^2)(k_{3}^jk_{4}^j+k_4^2)}{4s^4}\epsilon^{pq}_1\epsilon^{pq}_2\epsilon^{mn}_3\epsilon^{mn}_4\right. \nn \\
    &\left.-\frac{3}{s^2}\left(\epsilon^{in}_1k_{2}^n\epsilon^{ij}_2-\epsilon^{in}_1k_{2}^j\epsilon^{in}_2-k_{1}^i\epsilon^{kj}_1\epsilon^{ki}_2\right)\left(\epsilon^{lm}_3k_{4}^m\epsilon^{lj}_4-\epsilon^{lm}_3k_{4}^j\epsilon^{lm}_4-k_{3}^l\epsilon^{mj}_3\epsilon^{lm}_4\right)\right] \nn \\
    &-\frac{a}{64H}{\gamma'}_{k_1}^{h_1}\gamma_{k_2}^{h_2}\left({\gamma'}_{k_3}^{h_3}{\gamma'}_{k_4}^{h_4}-k_{3}^pk_{4}^p\gamma_{k_3}^{h_3}\gamma_{k_4}^{h_4}\right)\frac{(k_{1}^lk_{2}^l+k_2^2)}{s^2}\epsilon^{mn}_1\epsilon^{mn}_2\epsilon^{ij}_3\epsilon^{ij}_4,
\end{align}
where, for brevity, we have introduced the notation
\begin{equation}
    \epsilon^{ij}_a=\epsilon^{h_a}_{ij}(\textbf{k}_a).
\end{equation}

Given the Lagrangian, we can now calculate the cubic and quartic wavefunction coefficients. It turns out to be most convenient to just directly compute the on-shell action, evaluated on the classical solution with the correct boundary conditions. To get the action to fourth order, we need the following perturbative solution to the equations of motion
\be
    \gamma_k^h=K_k^s\bar{\gamma}^h_k+\sum_{h'}\int  \rd\eta' \left.\frac{\delta \mathcal{L}_{3}}{\delta \gamma_k^h}\right\rvert_{\gamma_k^h=K_k^{h'}\bar{\gamma}^{h'}_k}G_k^{hh'}(\eta,\eta')={\gamma_k^h}^{(0)}+\delta \gamma_k^h,
    \label{eq:1stordersoln}
\ee
where the bulk-to-boundary and bulk-to-bulk propagators are respectively given by the expressions~\eqref{eq:bbdyprop} and~\eqref{eq:bbulkprop}.
After some redefinition of dummy variables, the variation of the cubic action $\mathcal{L}_{3}$ is given by
\be
\begin{aligned}
    \frac{\delta \mathcal{L}_{3}}{\delta \gamma_k^h}=M_{\rm Pl}^2\int_{\textbf{k}^3}\sum_{h_4,h_5,h_6}\delta^{hh_4}\delta(\textbf{k}_4-\textbf{k})\frac{a^2}{8}\gamma_{k_5}^{h_5}\gamma_{k_6}^{h_6}\epsilon^{ij}_4&\left(2\epsilon^{il}_5\epsilon^{jm}_6 k_{5}^mk_{6}^l+4\epsilon^{lm}_5\epsilon^{il}_6k_{6}^jk_{6}^m \right.\\&\left.-2\epsilon^{lm}_5\epsilon^{ij}_6k_{6}^lk_{6}^m-\epsilon^{lm}_5\epsilon^{lm}_6k_{6}^i k_{6}^j\right).
\end{aligned}
\ee
Given this expression, we can substitute~\eqref{eq:1stordersoln} into the action and read off the relevant wavefunction coefficients.


\subsubsection{The bispectrum}

We can extract the cubic wavefunction coefficient, which is given by
\cite{Maldacena:2002vr,Maldacena:2011nz}
\be \label{GravitonCubicWF}
{\rm Re} \, \psi_{\gamma^3}({\bf k}_{i}; \epsilon_{i}) =-  \frac{M_{\rm Pl}^2  }{4 H^2}  \left(k_T-\frac{e_2}{k_T} -\frac{e_3}{k_T^2}\right) \big[  \epsilon_1 \ccdot {\bf k}_2  \,\epsilon_2 \ccdot \epsilon_3 + \epsilon_2 \ccdot {\bf k}_3  \,\epsilon_1 \ccdot \epsilon_3 + \epsilon_3 \ccdot {\bf k}_1  \,\epsilon_1 \ccdot \epsilon_2 \big]^2,
\ee
and the bispectrum is related to this by
\be
  \big\langle \gamma^{s_1}_{{\bf k}_1} \gamma^{s_2}_{ {\bf k}_2}  \gamma^{s_3}_{{\bf k}_3}  \big\rangle' =-2 {\rm Re} \, \psi_{\gamma^3}({\bf k}_{i}; \epsilon_{i}) \prod_{a=1}^3  P_{\gamma}(k_a).
\ee
We can also compute this correlator from the Hamiltonian using the in-in formalism. These correlators are completely fixed as well by de Sitter symmetries~\cite{Maldacena:2011nz}. Compared to the three-point function for two massless scalars and a graviton, cf.,~\eqref{eq:psi3-scalar}, the only difference lies in the tensor structure. The energy dependence is the same since the bulk-to-boundary propagators for scalars and gravitons are identical.

\subsubsection{The quartic wavefunction coefficient}\label{ss:psi4}

We now turn to compute the quartic wavefunction coefficient for gravitons from the above Lagrangian. There are essentially two complicated aspects of the calculation: the first is simply managing the contractions of polarizations and momenta, while the second is evaluating the time integrals that capture the time evolution of the system.

\vskip4pt
There are five distinct time integrals that we must evaluate to find the quartic wavefunction coefficient. Four are relevant for contact contributions and are given by
\begin{align}
    \int\rd\eta\, a^2 K_{k_1}(\eta)K_{k_2}(\eta)K_{k_3}(\eta)K_{k_4}(\eta) &=-i\frac{-2\mathbb{e}_4- \mathbb{e}_3 k_T-\mathbb{e}_2 k_T^2+k_T^4}{H^2k_T^3}-\frac{1}{H^2\eta_0},\\
    \int\rd\eta\, a^2 K_{k_1}'(\eta)K_{k_2}(\eta)K_{k_3}'(\eta)K_{k_4}(\eta) &=-i\frac{k_1^2k_3^2\left(k_T^2+k_T(k_2+k_4)+2k_2k_4\right)}{H^2k_T^3},\\
    \int\rd\eta\, a K_{k_1}'(\eta)K_{k_2}(\eta)K_{k_3}'(\eta)K_{k_4}'(\eta) &=-2i\frac{k_1^2k_3^2k_4^2k_T+3k_1^2k_2k_3^2k_4^2}{Hk_T^4},\\
     \int\rd\eta\, a K_{k_1}'(\eta)K_{k_2}(\eta)K_{k_3}(\eta)K_{k_4}(\eta) &=i\frac{6k_1 \mathbb{e}_4+2k_1^2k_T^3-3k_1^3k_T^2+2k_1^2k_T \mathbb{e}_2+2k_1^4k_T}{Hk_T^4},
\end{align}
where we have only provided expressions for one ordering of momenta since the others can be obtained by permutation.  
The final integral is relevant for exchange contributions and is identical to that for massless scalars~\eqref{f22},
\be
     -iH^2\int\rd\eta\,\rd\eta' a(\eta)^2 a(\eta')^2 K_{k_1}(\eta)K_{k_2}(\eta)K_{k_3}(\eta')K_{k_4}(\eta') G_{s}(\eta,\eta') =
f^{(s)}_{(2,2)}.
\ee

\vskip4pt
We now have all the ingredients we need to compute the wavefunction coefficient: we read off the tensor structures from the Lagrangian and use the above results for the time evolution. We add together all diagrams (as always, the exchange diagram gets an additional factor of one-half from the 4th order correction to the quadratic action~\cite{COT,Goon:2018fyu}). The final quartic wavefunction coefficient can then be written as
\be
\label{eq:4gWFdS}
\psi_{\gamma^4}(\{ {\bf k} \} )=\frac{M_{\rm Pl}^2}{16 H^2} \Big[ \psi_{\gamma^4}^{(s)}(\{ {\bf k} \} )+\psi_{\gamma^4}^{(t)}(\{ {\bf k} \} )+\psi_{\gamma^4}^{(u)}(\{ {\bf k} \} )\Big],
\ee
where we have again split the answer into contributions from different channels for convenience. The $s$-channel contribution is
\be
\begin{aligned}\label{final}
& \psi_{\gamma^4}^{(s)}(\{ {\bf k} \} )  = M_{\rm Pl}^2 H^2 \psi_{\phi^4}^{(s)} (\epsilon_1 \ccdot \epsilon_2)^2(\epsilon_3 \ccdot \epsilon_4)^2+2f_{(2,2)}^{(s)} \left[ s^2 \Pi^{(s)}_{1,1} (\epsilon_1 \ccdot \epsilon_2)(\epsilon_3 \ccdot \epsilon_4) W_s   +2 W_s^2 \right]  \\
&\quad -2 f_{2,0}^{(s)} \Pi^{(s)}_{1,0}(\epsilon_1 \ccdot \epsilon_2)(\epsilon_3 \ccdot \epsilon_4) W_s + g_{\rm cont.}^{(s)}(\epsilon_1 \ccdot \epsilon_3)(\epsilon_1 \ccdot \epsilon_4)(\epsilon_2 \ccdot \epsilon_3)(\epsilon_2 \ccdot \epsilon_4)-2 f_{\rm cont.} W_s W_s^{(c)},
\end{aligned}
\ee
where  $f_{(2,2)}^{(s)}$ and $f^{(s)}_{(2,0)}$ are defined as in~\eqref{f22} and~\eqref{eq:f20}, and we have defined the following polarization structures:
\begin{align}
    W_s & = (\epsilon_1 \ccdot \epsilon_2)\big[ (\epsilon_3 \ccdot k_1)(\epsilon_4 \ccdot k_2)-(\epsilon_3 \ccdot k_2)(\epsilon_4 \ccdot k_1) \big]+(\epsilon_3 \ccdot \epsilon_4)\big[ (\epsilon_1 \ccdot k_3)(\epsilon_2 \ccdot k_4)-(\epsilon_1 \ccdot k_4)(\epsilon_2 \ccdot k_3) \big] \nn \\
    \label{eq:Wsdef}
    & \hspace{.5cm}+ \big[(\epsilon_1 \ccdot k_2) \epsilon_2-(\epsilon_2 \ccdot k_1) \epsilon_1\big]\cdot \left[(\epsilon_3 \ccdot k_4) \epsilon_4-(\epsilon_4 \ccdot k_3) \epsilon_3\right], \\
     W_s^{(c)} & =(\epsilon_1 \ccdot \epsilon_4)(\epsilon_2 \ccdot \epsilon_3)-(\epsilon_1 \ccdot \epsilon_3)(\epsilon_2 \ccdot \epsilon_4),
     \label{eq:Wcdef}
\end{align}
along with the contact interaction form factors
\begin{align}
    f_{\rm cont.} & = \prod_{i=1}^4 (1-k_i \partial_{k_i})k_T (\log k_T -1) =\frac{2 \mathbb{e}_{4}}{k_T^3}+\frac{\mathbb{e}_{3}}{k_T^2} +\frac{\mathbb{e}_{2}}{k_T}-k_T,  \\
    g_{\rm cont}^{(s)} & = -(k_{12} k_{34}+s^2)\left[\frac{4 \mathbb{e}_{4}}{k_T^3}-\frac{(k_{12}-k_{34})(k_{12}^2+2k_1 k_2-k_{34}^2-2 k_3 k_4 )}{2 k_T^2}+3 \frac{k_1 k_2 +k_3 k_4}{k_T}-\frac{3 k_T}{2} \right] \nn \\
    & \hspace{13pt}-\frac{4 \mathbb{e}_{4}}{k_T} +\frac{8 k_{12}^2 k_{34}^2 -4 k_{12}^4-4k_{34}^4-2 k_1 k_2 k_{34}^2-2 k_3 k_4 k_{12}^2+8 k_1 k_2 k_{12}^2+8 k_3 k_4 k_{34}^2 }{3 k_T} \nn \\
    &\hspace{13pt} + \frac{14}{3} k_T (k_1 k_2 + k_3 k_4) - \frac{10 k_T^3}{9}+\frac{16}{9} \sum_{a=1}^4k_a^3.
\end{align}
Notice that the 
scalar quartic wavefunction coefficient, $\psi_{\phi^4}^{(s)}$---which we computed in Section~\ref{ss:ssgss}---appears,
and we remind the reader that the various polarisation sums are given in Appendix~\ref{app:PolSums}. Equation~\eqref{eq:4gWFdS} is the main result of this paper. A decompressed version of this expression is included in an ancillary text file.

\vskip4pt
Note that we have only written the real part of the wavefunction coefficient, as this is what contributes to the trispectrum via
\begin{align}
  \big\langle \gamma^{h_1}_{{\bf k}_1} \gamma^{h_2}_{{\bf k}_2}  \gamma^{h_3}_{{\bf k}_3} \gamma^{h_4}_{{\bf k}_4} \big\rangle' =\,& \Bigg[4 \sum_{3 \, \, {\rm perms}} \sum_h P_{\gamma}(s) \, {\rm Re} \, \psi_{\gamma^{h_1}\gamma^{h_2}\gamma^{h} }({\bf k}_{1},{\bf k}_{2},-{\bf s}) \,   {\rm Re} \, \psi_{\gamma^{h_3}\gamma^{h_4}\gamma^{h} }({\bf k}_{3},{\bf k}_{4},{\bf s})\nonumber \\
& -2 \, {\rm Re} \, \psi_{\gamma^4} (\{ {\bf k} \} ) \Bigg] \prod_{a=1}^4  P_{\gamma}(k_a)   .
\end{align}
We have also calculated this trispectrum directly from the Hamiltonian using the in-in formalism and obtained the same result.\footnote{Since no time derivatives appear at cubic order, the quartic-order Hamiltonian can be written as minus the quartic-order Lagrangian.} We note that the IR-divergent contributions that appear in the above expressions for contact diagram time integrals drop out when we take the real part of the wavefunction.\footnote{Indeed in those expressions the divergent parts have real coefficients which are then multiplied by a factor of $-i$ coming from the Feynman rules.}

\vskip4pt
We can perform several consistency checks on the result~\eqref{final}.
 On general grounds, we expect that tree-level wavefunction coefficients of massless scalars and gravitons in de Sitter space are rational functions of the external energies $ \{k_{1},k_{2},k_{3},k_{4}\}$ and internal energies $ \{s,t,u\}$ \cite{Goodhew:2022ayb} (see also \cite{Anninos:2014lwa} and the connection to holographic renormalization \cite{Skenderis:2002wp}) with poles at vanishing partial and total energy configurations. Therefore, $  \psi_{4} $ should only have poles in the variables $k_{T}$ and $ E_{L,R}^{(s,t,u)}$, which is indeed the case for our result. Moreover, adapting~\eqref{eq:pole-order} to this case, we expect that the wavefunction should scale as
\be
\lim_{k_{T}\to0}\psi_{4}= \mathcal{O}(k_T^{-3}), \qquad\quad \lim_{E_{L,R}\to 0}\psi_{4}= \mathcal{O}(E_{L,R}^{-2}) ,
\ee
in the vicinity of its singularities,
which indeed agrees with~\eqref{final}. It also has the correct amplitude limit on the leading $k_T$ pole.

\vskip4pt
Given the result~\eqref{final}, it is natural to try to cast it in a simpler form. For example,
by an appropriate use of partial fractions, one should always be able to write the result in such a way that every term has at most two poles, e.g., $  \{k_{T},E_{L}\} $ or $  \{k_{T},E_{R}\} $ or $ \{E_{L},E_{R}\}$. This re-writing comes at the cost of introducing spurious poles in the denominator and so, without further input, the resulting expressions are not particularly illuminating. It is also natural to eliminate polarisation vectors in favour of cosmological spinor helicity variables~\cite{Maldacena:2011nz}, but this also does not provide a drastic simplification (in contrast to the analogous story in flat space), essentially because helicity scaling does not completely fix the structure of spinor brackets.


\subsection{Lifting from flat space to de Sitter space} 
\label{ss:dSgggglift}

We would now like to reproduce the de Sitter four graviton wavefunction~\eqref{eq:4gWFdS} using the same heuristic method employed in Section~\ref{ss:dSs4} for the massless scalar four-point function. We take as our starting point the four-graviton wavefunction in flat space and then transform it to its de Sitter counterpart. 
The four-point graviton wavefunction coefficient in flat space is 
\be
\label{eq:flatggggcorr2}
\begin{aligned}
\psi_{\gamma^4}^{{\rm flat}} = &~\psi^{(s)\,\rm flat}_{\phi^4}(\epsilon_1\cdot \epsilon_2)^2(\epsilon_3\cdot\epsilon_4)^2+\frac{1}{k_TE_LE_R}\bigg[2{\cal P}^{(s)}_1(\epsilon_1\cdot \epsilon_2)(\epsilon_3\cdot\epsilon_4) W_s+4W_s^2\bigg]\\
&+\psi^{(t)\,\rm flat}_{\phi^4}(\epsilon_1\cdot \epsilon_3)^2(\epsilon_2\cdot\epsilon_4)^2+\frac{1}{k_TE^{(t)}_LE^{(t)}_R}\bigg[2{\cal P}^{(t)}_1(\epsilon_1\cdot \epsilon_3)(\epsilon_2\cdot\epsilon_4) W_t+4W_t^2\bigg]
\\
&+\psi^{(u)\,\rm flat}_{\phi^4}(\epsilon_1\cdot \epsilon_4)^2(\epsilon_3\cdot\epsilon_2)^2+\frac{1}{k_TE^{(u)}_LE^{(u)}_R}\bigg[2{\cal P}^{(u)}_1(\epsilon_1\cdot \epsilon_4)(\epsilon_3\cdot\epsilon_2) W_u+4W_u^2\bigg]
\\
&+\frac{1}{k_T}\bigg[\Big(-2(k_{12}k_{34}+s^2)+\lambda k_T^2\Big)(\epsilon_1\cdot \epsilon_3)(\epsilon_1\cdot \epsilon_4)(\epsilon_2\cdot \epsilon_3)(\epsilon_2\cdot \epsilon_4)\\
&~~~~~~~~
+\Big(-2(k_{13}k_{24}+t^2)+\lambda k_T^2\Big)(\epsilon_1\cdot \epsilon_2)(\epsilon_1\cdot \epsilon_4)(\epsilon_3\cdot \epsilon_2)(\epsilon_3\cdot \epsilon_4)\\
&~~~~~~~~
+\Big(-2(k_{14}k_{23}+u^2)+\lambda k_T^2\Big)(\epsilon_1\cdot \epsilon_3)(\epsilon_1\cdot \epsilon_2)(\epsilon_4\cdot \epsilon_3)(\epsilon_4\cdot \epsilon_2)\\
&~~~~~~~~
-2 W_s W_s^{(c)}-2 W_t W_t^{(c)}-2 W_u W_u^{(c)}\bigg]
\,,
\end{aligned}
\ee
where $W_s$ and $W_s^{(c)}$ are defined as in~\eqref{eq:Wsdef} and~\eqref{eq:Wcdef}, and we have defined
\be
{\cal P}_1^{(s)} = s^2 \Pi_{1,1}^{(s)} - E_{L}E_{R} \Pi_{1,0}^{(s)}\,,
\ee
where the $t$ and $u$-channel versions can be obtained by permutation.
Additionally, we note that the flat space scalar wavefunction of massless scalars exchanging a graviton $\psi^{(s)\,\rm flat}_{\phi^4}$~\eqref{eq:flatspacespin2exc} enters into this formula, along with its $t$ and $u$-channel permutations. 

\vskip4pt
As in the scalar case, the wavefunction~\eqref{eq:flatggggcorr2} can itself be bootstrapped by requiring that it has the right singularities and residues, along with requiring that, when we suitably average over polarizations, it reduces to the massless scalar result (which is why $\psi^{{\rm flat}}_{\phi^4}$ appears multiplying some polarization structures). 
Note that there is an inherent ambiguity in that these requirements do not uniquely specify~\eqref{eq:flatggggcorr2}. Indeed, there is a free parameter, $\lambda$.\footnote{We could also introduce another free parameter $\tilde{\lambda}$ by  multiplying $(\epsilon_1\cdot \epsilon_2)^2(\epsilon_3\cdot\epsilon_4)^2$ (and the other two permutations) by $\tilde{\lambda} k_{T}$. This would cause the scalar correlator multiplying this tensor structure to not vanish in the soft limit, so we do not include this term here, but we will encounter this possibility in Section~\ref{s:dS}.} In position space, the contribution proportional to this parameter is a contact term (localized at coincident points), and the precise contact terms required to appear in the wavefunction are fixed by our choice of the Ward identity that we require~\eqref{eq:flatggggcorr2} to satisfy~\cite{Baumann:2020dch}.\footnote{From the bulk point of view, different choices of contact terms appearing in the Ward identity correspond to different choices of field variables. For example, using the canonically normalized variable $g_{\mu\nu} = \eta_{\mu\nu}+2\gamma_{\mu\nu}/M_{\rm Pl}$, one finds $\lambda =10/3$.}

\vskip4pt
We now want to contemplate lifting~\eqref{eq:flatggggcorr2} to de Sitter space. To start, we compute the cut of the flat space graviton four-point function, 
\begin{align}
\nonumber
\psi_{\gamma^4}^{{\rm flat}}(\{ k\})+\psi_{\gamma^4}^{\ast\,{\rm flat}}(\{- k\}) = &-\frac{2s}{(k_{12}^2-s)(k_{34}^2-s^2)} \bigg[\frac{1}{6}s^4\Pi_{2,2}^{(s)}(\epsilon_1\cdot \epsilon_2)^2(\epsilon_3\cdot\epsilon_4)^2\\
&+2s^2\Pi_{1,1}^{(s)}(\epsilon_1\cdot \epsilon_2)(\epsilon_3\cdot\epsilon_4) W_s+4W_s^2\bigg]+t+u~{\rm channels}\,.
\label{eq:g4flatcut}
\end{align}
We can then compare it with the corresponding de Sitter formula, which we can compute using the graviton three-point function in de Sitter space~\eqref{GravitonCubicWF}.
With this, we compute the cut of the graviton four-point function to be
\begin{align}
\nonumber
\psi_{\gamma^4}(\{ k\})+\psi^*_{\gamma^4}(\{- k\})=\,&\frac{2s^3(k_{12}^2+2k_1k_2-s^2)(k_{34}^2+2k_3k_4-s^2)}{(k_{12}^2-s^2)^2(k_{34}^2-s^2)^2}\bigg[\frac{1}{6}s^4\Pi_{2,2}^{(s)}(\epsilon_1\cdot \epsilon_2)^2(\epsilon_3\cdot\epsilon_4)^2\\
&+2s^2\Pi_{1,1}^{(s)}(\epsilon_1\cdot \epsilon_2)(\epsilon_3\cdot\epsilon_4) W_s+4W_s^2\bigg]+t+u~{\rm channels}\,.
\label{eq:g4dscut}
\end{align}

In both~\eqref{eq:g4flatcut} and~\eqref{eq:g4dscut}, we have suppressed the contributions from the $t$ and $u$ exchange channels, but it will be straightforward to reintroduce them via permutation once we have constructed the $s$-channel contribution. It is simplest to start with the function multiplying the $(\epsilon_1\cdot \epsilon_2)^2(\epsilon_3\cdot\epsilon_4)^2$ tensor structure. In flat space, this factor is the four-point function of massless scalars exchanging a graviton, $\psi_{\phi^4}^{(s)\, \rm flat}$. The natural thing to do is to promote this to its de Sitter space version, which we constructed in Section~\ref{ss:dSs4}. Next, we consider the second line of~\eqref{eq:g4flatcut}; we want to transform the cut of the form factor into its de Sitter counterpart. These are the same energy factors that appeared in the scalar four-point function, so if we act with the operator ${\cal D}_s$, we generate
\be
{\cal D}_s\frac{1}{k_T E_L E_R} = f_{(2,2)}^{(s)}\,.
\ee
Finally, in order for the total energy singularity to be Lorentz invariant, we have to include the helicity-0 polarization sum multiplying the tensor structure $(\epsilon_1\cdot \epsilon_2)(\epsilon_3\cdot\epsilon_4) W_s$. In flat space, this is part of ${\cal P}_1^{(s)}$ and the form factor for this piece is (minus) the $s\to0$ limit of the helicity-$1$ form factor, so we replace this form factor in de Sitter by $ -f_{(2,2)}^{(s)}\Big\rvert_{s=0} = -f_{(2,0)}^{(s)}$. Putting this all together, we obtain the $s$-channel contribution to the graviton four-point function:
\be
\begin{aligned}
\psi_{\gamma^4}^{(s)} = 
~\psi_{\phi^4}^{(s)}(\epsilon_1\cdot \epsilon_2)^2(\epsilon_3\cdot\epsilon_4)^2&+f^{(s)}_{(2,2)}\bigg[2s^2\Pi_{1,1}(\epsilon_1\cdot \epsilon_2)(\epsilon_3\cdot\epsilon_4) W_s+4W_s^2\bigg]\\
&-f^{(s)}_{(2,0)}\bigg[2\Pi_{1,0}(\epsilon_1\cdot \epsilon_2)(\epsilon_3\cdot\epsilon_4) W_s\bigg]\,.
\end{aligned}
\label{eq:schangggglift}
\ee
We can then check that this reproduces the exchange part of~\eqref{final}. We can obtain the $t$ and $u$-channel exchange contributions by permuting momenta.

\vskip4pt
Finally, we turn to the contribution from contact interactions. We notice that there are two different form factors appearing in~\eqref{eq:flatggggcorr2}:
\begin{align}
    g_c^{(s)\,{\rm flat}} &= -\frac{2(k_{12}k_{34}+s^2)}{k_T}+\lambda k_T\,,\\
    f_c^{\rm flat} &= \frac{1}{k_T}\,.
\end{align}
We can transmute these to their de Sitter counterparts by acting with~\eqref{eq:contactlift}:
\begin{align}
    -{\cal D}_{k_T}g_c^{(s)\,{\rm flat}} &= g_{\rm cont}^{(s)}+\frac{2+\lambda}{3}(k_1^3+k_2^3+k_3^3+k_4^3)\,,\\
    {\cal D}_{k_T}f_c^{\rm flat} &= f_{\rm cont}\,.
\end{align}
Note that in this case we do not have to correct the result of the lifting procedure to satisfy the MLT, because~\eqref{eq:schangggglift} already satisfies it by itself (this is a consequence of the fact that we have directly used the de Sitter massless scalar four-point function as a building block, which itself satisfies the MLT). We can then construct the $s$-channel contribution to the graviton four-point function
\be
\label{eq:liftedgggg}
\begin{aligned}
\psi_{\gamma^4}^{(s)} = 
~&\psi_{\phi^4}^{(s)}(\epsilon_1\cdot \epsilon_2)^2(\epsilon_3\cdot\epsilon_4)^2+f^{(s)}_{(2,2)}\bigg[2s^2\Pi_{1,1}(\epsilon_1\cdot \epsilon_2)(\epsilon_3\cdot\epsilon_4) W_s+4W_s^2\bigg]\\
&-f^{(s)}_{(2,0)}\bigg[2\Pi_{1,0}(\epsilon_1\cdot \epsilon_2)(\epsilon_3\cdot\epsilon_4) W_s\bigg]+g_{\rm cont}^{(s)}(\epsilon_1\cdot \epsilon_4)(\epsilon_2\cdot \epsilon_3)(\epsilon_2\cdot \epsilon_4)
-2 f_{\rm cont}W_s W_s^{(c)}\\
&+\frac{2+\lambda}{3}(k_1^3+k_2^3+k_3^3+k_4^3)(\epsilon_1\cdot \epsilon_4)(\epsilon_2\cdot \epsilon_3)(\epsilon_2\cdot \epsilon_4)\,.
\end{aligned}
\ee
We see that this agrees precisely with~\eqref{final} if we fix $\lambda=-2$. It is not surprising that there is a free parameter, because it corresponds to the ambiguity in our choice of bulk field variables. By summing this along with the $t$ and $u$ channels, we produce the graviton four-point wavefunction.

\vskip4pt
This derivation has the virtue of relative simplicity. Nevertheless, one could question whether the final result is unique. In the following section, we systematically construct the most general wavefunction that has the correct singularities and which satisfies the MLT, reproducing both~\eqref{eq:liftedgggg} and the result of the bulk calculation.


\subsection{The boostless bootstrap} \label{s:dS}

We now return to our third method and follow the three-step procedure introduced in Section~\ref{ss:blesss4} to bootstrap the graviton four-point function. Recall that the strategy is as follows: in step 1 we use the Cosmological Optical Theorem (COT)~\cite{COT} and energy shifts~\cite{MLT,Baumann:2021fxj} to fix all partial-energy poles, in step 2 we fix the leading total-energy pole by demanding that we recover the correct scattering amplitude, and in step 3 we use the MLT~\cite{MLT} to fix the subleading total-energy poles. Throughout we use scale invariance, by demanding that the wavefunction coefficient scales as $k^3$, and Bose symmetry. As in Section~\ref{ss:blesss4}, we directly construct the part of the wavefunction coefficient with the symmetries of an $s$-channel diagram, from which we can get the full wavefunction coefficient by adding the other two permutations of momenta. 

\subsubsection{Unitarity and partial-energy poles}

We start by fixing the partial-energy poles, for which we need the graviton three-point function~\eqref{GravitonCubicWF}. The trimmed part of this wavefunction coefficient coincides with that of two massless scalars and a graviton. As a consequence, the structure of the partial-energy poles for the graviton four-point function is the same as for massless scalars exchanging a graviton---we already computed this using partial-energy shifts in Section~\ref{ss:blesss4}, with the result given by~\eqref{fCut}. To complete step 1 we need to compute the tensor structure that arises from an $s$-channel exchange diagram. We can find this from the cubic wavefunction coefficient by using the COT, as in \eqref{eq:g4dscut}.

\vskip4pt
After step 1 our wavefunction coefficient is therefore
\begin{align}\label{4GravitonsStep1}
\psi_{\gamma^4}^{(s)} = \frac{M_{\rm Pl}^2}{16 H^{2}} \left[\frac{s^4}{12} \Pi^{(s)}_{2,2} (\epsilon_1 \ccdot \epsilon_2)^2(\epsilon_3 \ccdot \epsilon_4)^2 +  s^2 \Pi^{(s)}_{1,1} (\epsilon_1 \ccdot \epsilon_2)(\epsilon_3 \ccdot \epsilon_4) W_{s} + 2 W_{s}^2 \right] f_{\text{cut}} + \ldots,
\end{align}
where here $\ldots$ denotes terms that must be regular in the limits $E_{L,R} \rightarrow 0$. The partial-energy poles match what we found in Section \ref{ss:lagrangian}.
 
\subsubsection{The amplitude limit and leading total-energy pole} 

We now fix the leading total-energy pole by demanding that we recover the correct scattering amplitude. The amplitude is a sum of the $S,T$, and $U$ channel contributions along with a contact contribution:\footnote{This amplitude has an extra overall factor of $(M_{\rm Pl}/2)^4$ compared to the amplitude for canonically normalised graviton perturbations. This is to match the normalization used for $\gamma_{ij}$.}
\be
\mathcal{A}_{4} = \frac{M_{\rm Pl}^2}{16}\left( \mathcal{A}_{4}^{(S)}+\mathcal{A}_{4}^{(T)}+\mathcal{A}_{4}^{(T)}+\mathcal{A}_{4}^{(c)}\right).
\ee
The $S$-channel contribution is given by\footnote{As before, the split into different channels is somewhat arbitrary, but the choice of split does not affect the final answer.}
\be \label{SchannAmp}
\mathcal{A}_{4}^{(S)} = \frac{T U}{S} (\varepsilon_{1} \cdot \varepsilon_{2})^2(\varepsilon_{3} \cdot \varepsilon_{4})^2 - \frac{2}{S}(U-T) (\varepsilon_{1} \cdot \varepsilon_{2})(\epsilon_{3} \cdot \varepsilon_{4})\mathcal{W}_{S} - \frac{4}{S}\mathcal{W}_{S}^2 ,
\ee
with the $T$ and $U$ channels obtained from this expression, while the contact contribution is
\be \label{ContactAmp}
\mathcal{A}_{4}^{(c)} = \left[2S(\varepsilon_{1} \cdot \varepsilon_{3})(\varepsilon_{1} \cdot \varepsilon_{4})(\varepsilon_{2} \cdot \varepsilon_{3})(\varepsilon_{2} \cdot \varepsilon_{4})-2\mathcal{W}_{S}\mathcal{W}_{S}^{(c)}\right] + 2 ~ \text{perms}.
\ee
We have defined the following combinations of polarization vectors and momenta
\begin{align}
    \mathcal{W}_S & = (\varepsilon_1 \ccdot \varepsilon_2)\big[ (\varepsilon_3 \ccdot k_1)(\varepsilon_4 \ccdot k_2)-(\varepsilon_3 \ccdot k_2)(\varepsilon_4 \ccdot k_1) \big]+(\varepsilon_3 \ccdot \varepsilon_4)\big[ (\varepsilon_1 \ccdot k_3)(\varepsilon_2 \ccdot k_4)-(\varepsilon_1 \ccdot k_4)(\varepsilon_2 \ccdot k_3) \big] \nn \\
    & \hspace{13pt}+ \big[(\varepsilon_1 \ccdot k_2)\, \varepsilon_2-(\varepsilon_2 \ccdot k_1)\, \varepsilon_1\big]\cdot \big[(\varepsilon_3 \ccdot k_4)\, \varepsilon_4-(\varepsilon_4 \ccdot k_3)\, \varepsilon_3\big], \\
     \mathcal{W}_S^{(c)} & =(\varepsilon_1 \ccdot \varepsilon_4)(\varepsilon_2 \ccdot \varepsilon_3)-(\varepsilon_1 \ccdot \varepsilon_3)(\varepsilon_2 \ccdot \varepsilon_4).
\end{align}
Note that these are conceptually slightly different from $W_{s}$ and $W_{s}^{(c)}$, since we write the amplitude in a manifestly Lorentz invariant form using the four-vector $\varepsilon_{\mu}$. We could write each contribution to this amplitude in axial gauge (with $h_{0 \mu} = 0$) using the three-dimensional polarisation vector $\epsilon_{i}$ rather than $\varepsilon_{\mu}$, and indeed when we match to the amplitude we do so channel-by-channel, so in practice we are matching a de Sitter Feynman diagram for the wavefunction to a flat-space one for the amplitude with both computed in axial gauge. Of course the final answer in both cases is gauge invariant, once we include all diagrams, but thinking in terms of axial gauge gives us a transparent way to track tensor structures (since they are the same in both spacetimes).

\vskip4pt
As in Section~\ref{ss:blesss4}, we have written the $S$-channel amplitude using $S,\ T$, and $U$ such that it has the desired symmetries. When we match our wavefunction coefficient to the amplitude on the leading total-energy pole, we will eliminate $U$ from the amplitude and $u$ from the wavefunction coefficient so that the variables we are using are independent.
We now demand that our wavefunction coefficient satisfies~\eqref{AmplitudeLimit} and we treat each different tensor structure independently.

\vskip4pt
The three tensor structures of~\eqref{SchannAmp} are contained in~\eqref{4GravitonsStep1} so we first make sure each energy-dependent coefficient yields the correct amplitude. For the $(\epsilon_{1} \cdot \epsilon_{2})^2(\epsilon_{3} \cdot \epsilon_{4})^2$ structure we notice that, up to an overall factor, this part of the amplitude matches that of minimally coupled massless scalars, cf.,~\eqref{MasslessScalarsAmp}, while the corresponding part of~\eqref{4GravitonsStep1} also matches---up to the same factor---what we found after step 1 for massless scalars. We can therefore use our result from Section~\ref{ss:blesss4}. Up to subleading total-energy poles, which we will fix in the next section, we therefore take 
\be
\frac{f_{\text{cut}}}{12} s^4 \Pi^{(s)}_{2,2} \rightarrow \frac{1}{6} \left[ \frac{f_{\text{cut}}}{2} s^4 \Pi^{(s)}_{2,2} - \frac{2 \mathbb{e}_{4}}{k_{T}^3}(s^2 \Pi^{(s)}_{2,1} - (k_{12}k_{34}+s^2)\Pi^{(s)}_{2,0}) - \frac{2 \mathbb{e}_{4}(k_{12}k_{34}+s^2)}{k_{T}^3} \right]
\ee
in~\eqref{4GravitonsStep1}. 

\vskip4pt
Next we consider the $(\epsilon_{1} \cdot \epsilon_{2})(\epsilon_{3} \cdot \epsilon_{4}) W_{s}$ structure. If we compare the amplitude limit from~\eqref{4GravitonsStep1} to the correct amplitude we find that they differ by a single term which has an overall factor of $s^{-2}$, and which is independent of $t^2$. We therefore need to correct this part of~\eqref{4GravitonsStep1} and our ansatz is completely fixed up to an overall factor: first, we need to include an overall factor of $\mathbb{e}_{4}/(s^2 k_{T}^3)$. The remaining dependence must scale as $\sim k^2$, by scale invariance, but it also needs to be anti-symmetric in ${\bf k}_{1} \leftrightarrow {\bf k}_{2}$ and ${\bf k}_{3} \leftrightarrow {\bf k}_{4}$, since the tensor structure in this case is anti-symmetric. We also need to cancel the $s \rightarrow 0$ pole for physical momenta. All of these considerations together imply that we need to include an overall factor of $(k_{1}-k_{2})(k_{3}-k_{4})$. If we fix the overall coefficient by matching to the amplitude limit we need to take 
\be
f_{\text{cut}} s^2 \Pi_{1,1}^{(s)} \rightarrow f_{\text{cut}} s^2 \Pi_{1,1}^{(s)} - \frac{4 \mathbb{e}_{4}}{  k_{T}^3} \Pi_{1,0}^{(s)}
\ee
in~\eqref{4GravitonsStep1}. Lastly, for the $S$-channel exchange we have the $W_{s}^2$ tensor structure and in this case we see that the amplitude limit from~\eqref{4GravitonsStep1} is already the correct one.

\vskip4pt
We now move to the contact interaction terms and we again concentrate on the terms with $S$-channel symmetries. These are the two written in~\eqref{ContactAmp}, and we see that these two tensor structures do not appear in~\eqref{4GravitonsStep1}. First consider the structure $(\epsilon_{1} \cdot \epsilon_{3})(\epsilon_{1} \cdot \epsilon_{4})(\epsilon_{2} \cdot \epsilon_{3})(\epsilon_{2} \cdot \epsilon_{4})$. Our energy-dependent ansatz that needs to multiply this structure should contain an overall factor of $\mathbb{e}_{4}/k_{T}^3$ along with another factor, scaling as $\sim k^2$, which on the total energy pole reduces to $S$. This means that we need to include both $s^{0}$ and $s^{2}$ terms in the ansatz. By fixing the ansatz up to subleading total-energy poles, we make the following addition to~\eqref{4GravitonsStep1}:
\be
\Delta_{1} \psi_{\gamma^4}^{(s)} = -\frac{M_{\rm Pl}^2}{16 H^2}\frac{4 \mathbb{e}_{4}}{k_{T}^3}(k_{12}k_{34}+s^2) (\epsilon_{1} \cdot \epsilon_{3})(\epsilon_{1} \cdot \epsilon_{4})(\epsilon_{2} \cdot \epsilon_{3})(\epsilon_{2} \cdot \epsilon_{4}).
\ee
Finally, we need to add the $W_{s} W_{s}^{(c)}$ tensor structure to the wavefunction to match the second tensor structure in~\eqref{ContactAmp}. This structure already scales as $\sim k^2$ and once we include the overall factor of $\mathbb{e}_{4}/k_{T}^3$ the only freedom left is in the overall coefficient. Fixing this by matching to the amplitude means that we add to~\eqref{4GravitonsStep1}:
\be
\Delta_{2} \psi_{\gamma^4}^{(s)} = - \frac{M_{\rm Pl}^2}{16 H^2} \frac{4 \mathbb{e}_{4}}{k_{T}^3} W_s W_s^{(c)}.
\ee
The full expression after fixing the total-energy poles and matching to the amplitude is therefore
\begin{align}\label{4GravitonsStep2}
\frac{16 H^2}{M_{\rm Pl}^2}\psi_{\gamma^4}^{(s)} &= \left[ \frac{f_{\text{cut}}}{12} s^4 \Pi^{(s)}_{2,2} - \frac{ \mathbb{e}_{4}}{3 k_{T}^3}(s^2 \Pi^{(s)}_{2,1} - (k_{12}k_{34}+s^2)\Pi^{(s)}_{2,0}) - \frac{ \mathbb{e}_{4}(k_{12}k_{34}+s^2)}{3 k_{T}^3} \right] (\epsilon_1 \ccdot \epsilon_2)^2(\epsilon_3 \ccdot \epsilon_4)^2 \nonumber \\ &+  \left[ f_{\text{cut}} s^2 \Pi_{1,1}^{(s)} - \frac{4 \mathbb{e}_{4}}{ k_{T}^3} \Pi_{1,0}^{(s)} \right] (\epsilon_1 \ccdot \epsilon_2)(\epsilon_3 \ccdot \epsilon_4) W_{s}+ 2  f_{\text{cut}} W_{s}^2 \nonumber \\
& - \frac{4 \mathbb{e}_{4}}{k_{T}^3}(k_{12}k_{34}+s^2) (\epsilon_{1} \cdot \epsilon_{3})(\epsilon_{1} \cdot \epsilon_{4})(\epsilon_{2} \cdot \epsilon_{3})(\epsilon_{2} \cdot \epsilon_{4})  -  \frac{4 \mathbb{e}_{4}}{k_{T}^3} W_s W_s^{(c)} + \ldots.
\end{align}
Here $\ldots$ indicates additional terms that are regular as $E_{L,R} \rightarrow 0$ and subleading in the total-energy limit. We note that the amplitude limit has done some heavy lifting for us by tying together the contact and exchange contributions, as required by diffeomorphism invariance.  

\subsubsection{Manifestly local test and subleading total-energy poles}
We now turn to the third and final step, where we use the MLT to fix the subleading total-energy poles. Recall that the MLT requires the wavefunction coefficient to satisfy~\eqref{MLT}, where we take the derivative with respect to one external energy and set that energy to zero while holding all other variables fixed. The MLT follows from the form of the bulk-to-boundary propagator for the massless graviton, so it should hold for each tensor structure separately. We therefore work with each tensor structure in~\eqref{4GravitonsStep2} separately. 

\vskip4pt
We start with the $(\epsilon_1 \cdot \epsilon_2)^2(\epsilon_3 \cdot \epsilon_4)^2$ structure. The coefficient after the first two steps is equivalent to what we found for massless scalars after the first two steps, so the corrections we need to add to satisfy the MLT are the same. In this case we therefore simply promote the energy-dependent coefficient to~\eqref{AfterStep3}, which we recall has three free parameters. 
Now, consider the $(\epsilon_1 \cdot \epsilon_2)(\epsilon_3 \cdot \epsilon_4) W_{s}$ structure. We find that the coefficient does not satisfy the MLT. Taking the derivative with respect to $k_{4}$ followed by setting $k_{4} = 0$ yields a function with $s^{-2}$, $s^{0}$ and $s^2$ terms. The coefficient of $s^{-2}$ is a function of $k_{1},k_{2}$ and $k_{3}$ only, the $s$-independent part depends on $k_{1},k_{2},k_{3}$ and $t^{2}$, while the $s^2$ term again depends only on $k_{1},k_{2}$ and $k_{3}$. We therefore need to write down an ansatz which we add to~\eqref{4GravitonsStep2}. Recall that we need to ensure the absence of poles as $s \rightarrow 0$, so whenever we have a $t^{2}$ dependence we also need to include $u^2$, and we take them to appear in the combination $(t^{2}-u^{2})$ since this is how they appear in~\eqref{4GravitonsStep2}. Finally, this tensor structure's  energy-dependent coefficient needs to be anti-symmetric under the permutations ${\bf k}_{1} \leftrightarrow {\bf k}_{2}$ and ${\bf k}_{3} \leftrightarrow {\bf k}_{4}$. We therefore add to~\eqref{4GravitonsStep2}:
\begin{align}
\frac{16 H^2}{M_{\rm Pl}^2}\Delta_{3} \psi_{\gamma^4}^{(s)} = \frac{1}{s^2 k_{T}^2}\big[(k_{1}-k_{2})(k_{3}-k_{4}) &\text{Poly}^{(3)}_{3}(k_{12},k_{1}k_{2},k_{34},k_{3}k_{4})  + \left( {\bf k}_{1} \leftrightarrow {\bf k}_{3}, {\bf k}_{2} \leftrightarrow {\bf k}_{4} \right) \nn  \\ &+ b_{6} s^{2}(t^2-u^2)k_{T}\big] (\epsilon_1 \ccdot \epsilon_2)(\epsilon_3 \ccdot \epsilon_4) W_{s}.
\end{align}
Note that the symmetries of our ansatz do not allow for an independent $s^2$ term. 

\vskip4pt
The tensor structure $W_{s}^2 $ is relatively simple. We find that its coefficient does not satisfy the MLT, but the correction that we need to add only needs to depend on the four external energies. Given that the tensor structure scales as $\sim k^4$, the energy-dependence of our ansatz needs to scale as $\sim k^{-1}$ by scale invariance. There is then a unique choice, up to an overall coefficient, that adheres to the pole structure of the full wavefunction coefficient. We therefore add to~\eqref{4GravitonsStep2}:
\be
\frac{16 H^2}{M_{\rm Pl}^2}\Delta_{4} \psi_{\gamma^4}^{(s)} =  \frac{b_{7}}{k_{T}}W_{s}^2.
\ee
The story for the final two structures proceeds in the same way and we need to add the following two corrections:
\begin{align}
\frac{16 H^2}{M_{\rm Pl}^2}\Delta_{5} \psi_{\gamma^4}^{(s)} = & \frac{1}{k_{T}^2}\big[(\text{Poly}^{(5)}_{5}(k_{12},k_{1}k_{2},k_{34},k_{3}k_{4}) + s^2 \text{Poly}^{(3)}_{4}(k_{12},k_{1}k_{2},k_{34},k_{3}k_{4}))  \nonumber \\  & + \left( {\bf k}_{1} \leftrightarrow {\bf k}_{3}, {\bf k}_{2} \leftrightarrow {\bf k}_{4} \right) \big] (\epsilon_{1} \cdot \epsilon_{3})(\epsilon_{1} \cdot \epsilon_{4})(\epsilon_{2} \cdot \epsilon_{3})(\epsilon_{2} \cdot \epsilon_{4}), \\
\frac{16 H^2}{M_{\rm Pl}^2}\Delta_{6} \psi_{\gamma^4}^{(s)} & = \frac{1}{k_{T}^2}[\text{Poly}^{(3)}_{5}(k_{12},k_{1}k_{2},k_{34},k_{3}k_{4}) + \left( {\bf k}_{1} \leftrightarrow {\bf k}_{3}, {\bf k}_{2} \leftrightarrow {\bf k}_{4} \right)] W_s W_s^{(c)}.
\end{align}
Imposing the MLT for each tensor structure, and adding the other two permutations, then yields
\be
\psi_{\gamma^4}=\frac{M_{\rm Pl}^2}{16 H^2} \left[ \psi_{\gamma^4}^{(s)}+\psi_{\gamma^4}^{(t)}+\psi_{\gamma^4}^{(u)}\right],
\ee
where
\be
\begin{aligned}\label{finalBoot}
\psi_{\gamma^4}^{(s)} & = M_{\rm Pl}^2 H^2 \psi_{\phi^4}^{(s)}(\epsilon_1 \ccdot \epsilon_2)^2(\epsilon_3 \ccdot \epsilon_4)^2+ 2[f_{(2,2)}^{(s)}s^2 \Pi^{(s)}_{1,1} - f_{2,0}^{(s)} \Pi^{(s)}_{1,0}] (\epsilon_1 \ccdot \epsilon_2)(\epsilon_3 \ccdot \epsilon_4) W_s  \\  & 
\hspace{12pt}+4 f_{(2,2)}^{(s)} W_s^2  
 + \left(g_{\rm cont.}^{(s)}+r_{1} \sum_{a=1}^{4} k_{a}^3 \right)(\epsilon_1 \ccdot \epsilon_3)(\epsilon_1 \ccdot \epsilon_4)(\epsilon_2 \ccdot \epsilon_3)(\epsilon_2 \ccdot \epsilon_4)-2 f_{\rm cont.} W_s W_s^{(c)}.
\end{aligned}
\ee

In this expression $\psi_{\phi^4}^{(s)}$ is given by~\eqref{AfterStep3}. We see that this result matches the one derived in Section~\ref{ss:lagrangian} and given in~\eqref{final} up to four free parameters which are not tied to $M_{\rm Pl}$, and which do not have any poles. These are the constants $q_{1},q_{2},q_{3}$ in~\eqref{AfterStep3} and the new constant $r_{1}$. The $q_{1}$ term can be generated by the field redefinition $\gamma_{ij} \mapsto \gamma_{ij} + \frac{H^2 q_{1}}{4}\gamma_{kl}^2 \gamma_{ij}$, while the $r_{1}$ term can be generated by $\gamma_{ij} \mapsto \gamma_{ij} + \frac{H^2 r_{1}}{4} \gamma_{il} \gamma_{lk} \gamma_{kj}$. The other two terms can be shifted by using the same field redefinitions we used in~\eqref{nonlocal1} and~\eqref{nonlocal2}, with the addition of indices for the graviton:  
\begin{align} 
\gamma_{ij} \mapsto  \gamma_{ij} &+ \frac{ H^2 q_{2} }{3} \left[3 \gamma_{ij} \frac{1}{\partial^2}(\gamma_{kl} \partial^2 \gamma_{kl}) +  \gamma_{ij} \frac{1}{\partial^2} (\gamma_{kl}'^2) + \gamma'_{ij} \frac{1}{\partial^2}(\gamma_{kl} \gamma'_{kl}) \right],  \\
\gamma_{ij} \mapsto  \gamma_{ij} &+
   \frac{H^2 q_3}{3}\left[-27\partial^2\gamma_{ij}\frac{1}{\partial^4}(\gamma_{kl}\partial^2\gamma_{kl})-12\gamma_{ij}\frac{1}{\partial^4}(\gamma_{kl}\partial^4\gamma_{kl}) \right. \nonumber \\&\hspace{44pt}\left.+6\partial^2\gamma_{ij}\frac{1}{\partial^4}({\gamma'_{kl}}^2) +2\gamma'_{ij}\frac{1}{\partial^4}(\gamma_{kl}\partial^2\gamma'_{kl})-4\gamma_{ij}\frac{1}{\partial^4}(\gamma'_{kl}\partial^2\gamma'_{kl}) \right.  \\\nonumber
   &\left. 
   \hspace{44pt}- 10\gamma'_{ij}\frac{1}{\partial^4}(\gamma'_{kl}\partial^2\gamma_{kl}) -30\partial^2\gamma_{ij}\frac{1}{\partial^4}(\partial_m \gamma_{kl}\partial_m\gamma_{kl})-60\partial_m\gamma_{ij}\frac{1}{\partial^4}(\gamma_{kl}\partial^2\partial_m \gamma_{kl})\right]. 
\end{align}
We could also construct similar non-local field redefinitions that generate energy-dependent coefficients of the $(\epsilon_1 \cdot \epsilon_3)(\epsilon_1 \cdot \epsilon_4)(\epsilon_2 \cdot \epsilon_3)(\epsilon_2 \cdot \epsilon_4)$ tensor structure involving inverse powers of $s$. This would simply require different contractions between the fields in the above field redefinitions. We did not include terms of this form in our ansatz when we imposed the MLT condition since for this tensor structure no inverse powers of $s$ arose after steps 1 and 2. They can however be dealt with similarly and the resulting free parameters can be removed by field redefinitions. If we were to make a particular choice of field variables, e.g., by enforcing a particular representation of the Ward identity or soft theorem, then we could use this to fix some of these free parameters, as in the quartic scalar calculation.


\section{Conclusions}
\label{sec:conclusions}

In this work, we have computed the graviton four-point correlator and quartic wavefunction coefficient at tree level in de Sitter space for pure Einstein gravity. We derived our result in several different ways and found agreement among them. A few natural directions to pursue present themselves:
\begin{itemize}
    
    \item A consistency check of our result is to ensure that it respects the appropriate soft-graviton theorem, e.g., from \cite{Hinterbichler:2013dpa}. We have found it challenging to prove this, for a few reasons. First, the relevant soft theorem in \cite{Hinterbichler:2013dpa} for a correlator with soft and hard tensors is written in components. To compare to our result one needs to make a specific choice of polarizations, such as the helicity basis, and take derivatives of polarization tensors with respect to their spatial momentum argument, which is subtle because of the arbitrariness in defining this dependence and the necessary discontinuities. This problem appears first for the trispectrum and it is not present in the soft theorem for the bispectrum, where all contractions of polarization tensors can be written in terms of the external energies (see, e.g., \cite{Cabass:2021fnw}). Finally, a naive matching of the coefficients of the polarization tensor structures does not seem to satisfy the trispectrum soft theorem and we could not exclude that different structures cancel each other due to dimension-dependent identities. We postpone further investigations to the future.

    \item The graviton four-point function should display appropriate invariance under the isometries of de Sitter spacetime. For correlators at the future conformal boundary, this invariance should reduce to the set of conformal (and current conservation) Ward identities. It would be nice to check explicitly that our result indeed satisfies these constraints.

    \item In general we expect that in simple and highly constrained theories---such as pure gravity---higher-order observables should be fixed in terms of the lowest-order interaction. For example, in flat space all $n$-particle graviton amplitudes can in principle be derived from the three-graviton amplitude via BCFW recursion relations~\cite{BCFW,Benincasa:2007qj}. Because of this, one might hope to find variables in which the graviton four-point function we compute is not much more complicated than the graviton three-point function. Our final result~\eqref{final} does not yet realize this hope. It would be nice to find a more natural set of variables, analogous to spinor-helicity variables for maximally helicity-violating amplitudes~\cite{Parke:1986gb}. Cosmological spinor-helicity variables~\cite{Maldacena:2011nz}, as used in \cite{Li:2018wkt}, do not seem to achieve this simplification. One possibility is that simplifications can be achieved by using a Mellin space representation \cite{Sleight:2019mgd, Sleight:2019hfp, Sleight:2021iix}; this would also facilitate a comparison to the anti-de Sitter space result \cite{Raju:2012zs}, as reviewed in \cite{ Sleight:2021plv}. It would be interesting to further pursue these directions.
\end{itemize}

We close with a small historical parable. In a truly heroic computation, Parke and Taylor computed the six-gluon scattering amplitude in 1985, and presented it in~\cite{Parke:1985ax}, ending their discussion with the phrase ``...we hope to obtain a simple analytic form for the answer, making our result not only an experimentalist's, but also a theorist's delight." Less than a year later, they found a spectacular such formula~\cite{Parke:1986gb}, which catalyzed the modern amplitudes revolution. We share a similar hope to find a drastic simplification of the formulas presented in this paper. We should be so lucky.


\paragraph{Acknowledgements} We thank Soner Albayrak, Daniel Baumann, Giovanni Cabass, Zongzhe Du, Carlos Duaso Pueyo, Aaron Hillman, Kurt Hinterbichler, Sadra Jazayeri, Savan Kharel, Hayden Lee, Guilherme Pimentel, and Akhil Premkumar for useful discussions. HG is supported jointly by the Science and Technology Facilities Council through a postgraduate studentship and the Cambridge Trust Vice Chancellor's Award. The work of AJ is supported in part by DOE (HEP) Award DE-SC0009924. EP has been supported in part by the research program VIDI with Project No. 680-47-535, which is (partly) financed by the Netherlands Organisation for Scientific Research (NWO). DS is supported by a UKRI Stephen Hawking Fellowship [grant number EP/W005441/1] and a Nottingham Research Fellowship from the University of Nottingham. For the purpose of open access, the authors have applied a CC BY public copyright licence to any Author Accepted Manuscript version arising.


\appendix 

\section{Polarisation sums}\label{app:PolSums}

In this appendix, we introduce some of the formulas and formalism required to deal with exchanges of spinning particles in de Sitter space. See~\cite{Goon:2018fyu,Baumann:2019oyu,Baumann:2020dch,Baumann:2021fxj} for further details.

\vskip4pt
The transverse-traceless spin-2 projector is
\be \label{eq:pol-sum}
\Pi_{ijkl}^{{\bf k}} = \frac{1}{2} \sum_s e^s_{ij}({\bf k}) e^s_{kl}(-{\bf k})= \frac{1}{2} \pi_{i k}^{{\bf k}} \pi_{j l}^{{\bf k}} +\frac{1}{2} \pi_{i l}^{{\bf k}}  \pi_{j k}^{{\bf k}} -\frac{1}{2} \pi_{i j}^{{\bf k}}  \pi_{k l}^{{\bf k}} ,
\ee
where $\pi_{ij}^{{\bf k}}  = \delta_{ij} - \frac{1}{k^2} k_i k_j $. It is useful to define the following contraction of this:
\begin{align}
 \Pi_{2,2}^{(s)} \equiv\,& \frac{24}{s^4}\Pi_{ijkl}^{\bf s} k^1_i k^2_j k^3_k k^4_l \\
 =\, &-\frac{3}{4}+3\frac{\sum_{a=1}^4 k_a^2}{2s^2}-3\frac{4(k_1^2+k_2^2)(k_3^2+k_4^2)-2k_1^2k_2^2-2k_3^2k_4^2+\sum_{a=1}^4 k_a^4-2(t^2-u^2)^2}{4s^4} \nn \\
& +3 \frac{ 2 (t^2-u^2)(k^2_1-k^2_2)(k^2_3-k^2_4)+(k_3^2+k_4^2)(k_1^2-k_2^2)^2+(k_1^2+k_2^2)(k_3^2-k_4^2)^2}{2s^6} \nn \\
&+\frac{3(k_1-k_2)^2 (k_3-k_4)^2 k_{12}^2 k_{34}^2}{4 s^8}.
\end{align}
We also define 
\begin{align}
\Pi_{2,1}^{(s)} & \equiv \frac{3(k_1-k_2)(k_3-k_4)}{s^4}\pi^{\bf s}_{ij} (k_1^i-k_2^i)(k_3^j-k_4^j) \\
& = \frac{3(k_1-k_2)(k_3-k_4)}{s^6} \left[s^2(t^2-u^2) + k_{12} k_{34}(k_1-k_2)(k_3-k_4) \right],\\
\Pi_{2,0}^{(s)} & \equiv \frac{1}{4} \left(1- \frac{3(k_1-k_2)^2}{s^2} \right)\left(1- \frac{3(k_3-k_4)^2}{s^2} \right),
\end{align}
which can be written in terms of contractions of the rank-2 projectors onto lower helicities. Lastly, we also define
\begin{align}
    \Pi_{1,1}^{(s)} & \equiv \frac{k_{12} k_{34}(k_1-k_2)(k_3-k_4)+s^2(t^2-u^2)}{s^4}, \\
    \Pi_{1,0}^{(s)} & \equiv \frac{(k_1-k_2)(k_3-k_4)}{s^2}.
\end{align}
When checking that we recover the correct amplitude limit on the leading total-energy pole, it is useful to have expressions for these sums in that limit. Individually the limits are not particularly illuminating but the following combinations have neat limits:
\begin{align}
s^4 \Pi_{2,2}^{(s)} - E_{L}E_{R}s^2 \Pi_{2,1}^{(s)}+E_{L}^2 E_{R}^2 \Pi^{(s)}_{2,0} & \xrightarrow{k_{T} \rightarrow 0} S^2P_{2}\left(1+\frac{2 T}{S} \right), \\
s^2 \Pi_{1,1}^{(s)} - E_{L}E_{R} \Pi_{1,0}^{(s)} & \xrightarrow{k_{T} \rightarrow 0} -SP_{1}\left(1+\frac{2 T}{S} \right).
\end{align}
Here we recognise the Legendre polynomials associated with the exchange of spinning particles. 
\bibliographystyle{JHEP}
\bibliography{refs}

\providecommand{\href}[2]{#2}\begingroup\raggedright\begin{thebibliography}{10}

\bibitem{Weinberg:1965nx}
S.~Weinberg, \emph{{Infrared photons and gravitons}},
  \href{https://doi.org/10.1103/PhysRev.140.B516}{\emph{Phys. Rev.} {\bfseries
  140} (1965) B516}.

\bibitem{Bern:2019prr}
Z.~Bern, J.J.~Carrasco, M.~Chiodaroli, H.~Johansson and R.~Roiban, \emph{{The
  Duality Between Color and Kinematics and its Applications}},
  \href{https://arxiv.org/abs/1909.01358}{{\ttfamily 1909.01358}}.

\bibitem{Benincasa:2007xk}
P.~Benincasa and F.~Cachazo, \emph{{Consistency Conditions on the S-Matrix of
  Massless Particles}},  \href{https://arxiv.org/abs/0705.4305}{{\ttfamily
  0705.4305}}.

\bibitem{Schuster:2008nh}
P.C.~Schuster and N.~Toro, \emph{{Constructing the Tree-Level Yang-Mills
  S-Matrix Using Complex Factorization}},
  \href{https://doi.org/10.1088/1126-6708/2009/06/079}{\emph{JHEP} {\bfseries
  06} (2009) 079} [\href{https://arxiv.org/abs/0811.3207}{{\ttfamily
  0811.3207}}].

\bibitem{McGady:2013sga}
D.A.~McGady and L.~Rodina, \emph{{Higher-spin massless $S$-matrices in
  four-dimensions}},
  \href{https://doi.org/10.1103/PhysRevD.90.084048}{\emph{Phys. Rev. D}
  {\bfseries 90} (2014) 084048}
  [\href{https://arxiv.org/abs/1311.2938}{{\ttfamily 1311.2938}}].

\bibitem{Pajer:2020wnj}
E.~Pajer, D.~Stefanyszyn and J.~Supe\l{}, \emph{{The Boostless Bootstrap:
  Amplitudes without Lorentz boosts}},
  \href{https://doi.org/10.1007/JHEP12(2020)198}{\emph{JHEP} {\bfseries 12}
  (2020) 198} [\href{https://arxiv.org/abs/2007.00027}{{\ttfamily
  2007.00027}}].

\bibitem{Hertzberg:2020yzl}
M.P.~Hertzberg, J.A.~Litterer and M.~Sandora, \emph{{Symmetries from locality.
  II. Gravitation and Lorentz boosts}},
  \href{https://doi.org/10.1103/PhysRevD.102.025023}{\emph{Phys. Rev. D}
  {\bfseries 102} (2020) 025023}
  [\href{https://arxiv.org/abs/2005.01744}{{\ttfamily 2005.01744}}].

\bibitem{Creminelli:2014wna}
P.~Creminelli, J.~Gleyzes, J.~Nore\~na and F.~Vernizzi, \emph{{Resilience of
  the standard predictions for primordial tensor modes}},
  \href{https://doi.org/10.1103/PhysRevLett.113.231301}{\emph{Phys. Rev. Lett.}
  {\bfseries 113} (2014) 231301}
  [\href{https://arxiv.org/abs/1407.8439}{{\ttfamily 1407.8439}}].

\bibitem{Bordin:2017hal}
L.~Bordin, G.~Cabass, P.~Creminelli and F.~Vernizzi, \emph{{Simplifying the EFT
  of Inflation: generalized disformal transformations and redundant
  couplings}}, \href{https://doi.org/10.1088/1475-7516/2017/09/043}{\emph{JCAP}
  {\bfseries 09} (2017) 043}
  [\href{https://arxiv.org/abs/1706.03758}{{\ttfamily 1706.03758}}].

\bibitem{Bartolo:2017szm}
N.~Bartolo and G.~Orlando, \emph{{Parity breaking signatures from a
  Chern-Simons coupling during inflation: the case of non-Gaussian
  gravitational waves}},
  \href{https://doi.org/10.1088/1475-7516/2017/07/034}{\emph{JCAP} {\bfseries
  07} (2017) 034} [\href{https://arxiv.org/abs/1706.04627}{{\ttfamily
  1706.04627}}].

\bibitem{Bordin:2020eui}
L.~Bordin and G.~Cabass, \emph{{Graviton non-Gaussianities and Parity Violation
  in the EFT of Inflation}},
  \href{https://doi.org/10.1088/1475-7516/2020/07/014}{\emph{JCAP} {\bfseries
  07} (2020) 014} [\href{https://arxiv.org/abs/2004.00619}{{\ttfamily
  2004.00619}}].

\bibitem{Bartolo:2020gsh}
N.~Bartolo, L.~Caloni, G.~Orlando and A.~Ricciardone, \emph{{Tensor
  non-Gaussianity in chiral scalar-tensor theories of gravity}},
  \href{https://doi.org/10.1088/1475-7516/2021/03/073}{\emph{JCAP} {\bfseries
  03} (2021) 073} [\href{https://arxiv.org/abs/2008.01715}{{\ttfamily
  2008.01715}}].

\bibitem{Cabass:2021fnw}
G.~Cabass, E.~Pajer, D.~Stefanyszyn and J.~Supe\l{}, \emph{{Bootstrapping Large
  Graviton non-Gaussianities}},
  \href{https://arxiv.org/abs/2109.10189}{{\ttfamily 2109.10189}}.

\bibitem{Maldacena:2011nz}
J.M.~Maldacena and G.L.~Pimentel, \emph{{On graviton non-Gaussianities during
  inflation}}, \href{https://doi.org/10.1007/JHEP09(2011)045}{\emph{JHEP}
  {\bfseries 09} (2011) 045} [\href{https://arxiv.org/abs/1104.2846}{{\ttfamily
  1104.2846}}].

\bibitem{Fu:2015vja}
T.-F.~Fu and Q.-G.~Huang, \emph{{The four-point correlation function of
  graviton during inflation}},
  \href{https://doi.org/10.1007/JHEP07(2015)132}{\emph{JHEP} {\bfseries 07}
  (2015) 132} [\href{https://arxiv.org/abs/1502.02329}{{\ttfamily
  1502.02329}}].

\bibitem{Li:2018wkt}
S.Y.~Li, Y.~Wang and S.~Zhou, \emph{{KLT-Like Behaviour of Inflationary
  Graviton Correlators}},
  \href{https://doi.org/10.1088/1475-7516/2018/12/023}{\emph{JCAP} {\bfseries
  12} (2018) 023} [\href{https://arxiv.org/abs/1806.06242}{{\ttfamily
  1806.06242}}].

\bibitem{Raju:2012zs}
S.~Raju, \emph{{Four Point Functions of the Stress Tensor and Conserved
  Currents in AdS$_4$/CFT$_3$}},
  \href{https://doi.org/10.1103/PhysRevD.85.126008}{\emph{Phys. Rev. D}
  {\bfseries 85} (2012) 126008}
  [\href{https://arxiv.org/abs/1201.6452}{{\ttfamily 1201.6452}}].

\bibitem{Albayrak:2019yve}
S.~Albayrak and S.~Kharel, \emph{{Towards the higher point holographic momentum
  space amplitudes. Part II. Gravitons}},
  \href{https://doi.org/10.1007/JHEP12(2019)135}{\emph{JHEP} {\bfseries 12}
  (2019) 135} [\href{https://arxiv.org/abs/1908.01835}{{\ttfamily
  1908.01835}}].

\bibitem{Raju:2012zr}
S.~Raju, \emph{{New Recursion Relations and a Flat Space Limit for AdS/CFT
  Correlators}}, \href{https://doi.org/10.1103/PhysRevD.85.126009}{\emph{Phys.
  Rev. D} {\bfseries 85} (2012) 126009}
  [\href{https://arxiv.org/abs/1201.6449}{{\ttfamily 1201.6449}}].

\bibitem{Raju:2011mp}
S.~Raju, \emph{{Recursion Relations for AdS/CFT Correlators}},
  \href{https://doi.org/10.1103/PhysRevD.83.126002}{\emph{Phys. Rev. D}
  {\bfseries 83} (2011) 126002}
  [\href{https://arxiv.org/abs/1102.4724}{{\ttfamily 1102.4724}}].

\bibitem{Mata:2012bx}
I.~Mata, S.~Raju and S.~Trivedi, \emph{{CMB from CFT}},
  \href{https://doi.org/10.1007/JHEP07(2013)015}{\emph{JHEP} {\bfseries 07}
  (2013) 015} [\href{https://arxiv.org/abs/1211.5482}{{\ttfamily 1211.5482}}].

\bibitem{Bzowski:2011ab}
A.~Bzowski, P.~McFadden and K.~Skenderis, \emph{{Holographic predictions for
  cosmological 3-point functions}},
  \href{https://doi.org/10.1007/JHEP03(2012)091}{\emph{JHEP} {\bfseries 03}
  (2012) 091} [\href{https://arxiv.org/abs/1112.1967}{{\ttfamily 1112.1967}}].

\bibitem{Bzowski:2012ih}
A.~Bzowski, P.~McFadden and K.~Skenderis, \emph{{Holography for inflation using
  conformal perturbation theory}},
  \href{https://doi.org/10.1007/JHEP04(2013)047}{\emph{JHEP} {\bfseries 04}
  (2013) 047} [\href{https://arxiv.org/abs/1211.4550}{{\ttfamily 1211.4550}}].

\bibitem{Bzowski:2013sza}
A.~Bzowski, P.~McFadden and K.~Skenderis, \emph{{Implications of conformal
  invariance in momentum space}},
  \href{https://doi.org/10.1007/JHEP03(2014)111}{\emph{JHEP} {\bfseries 03}
  (2014) 111} [\href{https://arxiv.org/abs/1304.7760}{{\ttfamily 1304.7760}}].

\bibitem{Bzowski:2019kwd}
A.~Bzowski, P.~McFadden and K.~Skenderis, \emph{{Conformal $n$-point functions
  in momentum space}},
  \href{https://doi.org/10.1103/PhysRevLett.124.131602}{\emph{Phys. Rev. Lett.}
  {\bfseries 124} (2020) 131602}
  [\href{https://arxiv.org/abs/1910.10162}{{\ttfamily 1910.10162}}].

\bibitem{Kundu:2014gxa}
N.~Kundu, A.~Shukla and S.P.~Trivedi, \emph{{Constraints from Conformal
  Symmetry on the Three Point Scalar Correlator in Inflation}},
  \href{https://doi.org/10.1007/JHEP04(2015)061}{\emph{JHEP} {\bfseries 04}
  (2015) 061} [\href{https://arxiv.org/abs/1410.2606}{{\ttfamily 1410.2606}}].

\bibitem{Kundu:2015xta}
N.~Kundu, A.~Shukla and S.P.~Trivedi, \emph{{Ward Identities for Scale and
  Special Conformal Transformations in Inflation}},
  \href{https://doi.org/10.1007/JHEP01(2016)046}{\emph{JHEP} {\bfseries 01}
  (2016) 046} [\href{https://arxiv.org/abs/1507.06017}{{\ttfamily
  1507.06017}}].

\bibitem{Arkani-Hamed:2015bza}
N.~Arkani-Hamed and J.~Maldacena, \emph{{Cosmological Collider Physics}},
  \href{https://arxiv.org/abs/1503.08043}{{\ttfamily 1503.08043}}.

\bibitem{Shukla:2016bnu}
A.~Shukla, S.P.~Trivedi and V.~Vishal, \emph{{Symmetry constraints in
  inflation, $\alpha$-vacua, and the three point function}},
  \href{https://doi.org/10.1007/JHEP12(2016)102}{\emph{JHEP} {\bfseries 12}
  (2016) 102} [\href{https://arxiv.org/abs/1607.08636}{{\ttfamily
  1607.08636}}].

\bibitem{Arkani-Hamed:2017fdk}
N.~Arkani-Hamed, P.~Benincasa and A.~Postnikov, \emph{{Cosmological Polytopes
  and the Wavefunction of the Universe}},
  \href{https://arxiv.org/abs/1709.02813}{{\ttfamily 1709.02813}}.

\bibitem{Arkani-Hamed:2018kmz}
N.~Arkani-Hamed, D.~Baumann, H.~Lee and G.L.~Pimentel, \emph{{The Cosmological
  Bootstrap: Inflationary Correlators from Symmetries and Singularities}},
  \href{https://doi.org/10.1007/JHEP04(2020)105}{\emph{JHEP} {\bfseries 04}
  (2020) 105} [\href{https://arxiv.org/abs/1811.00024}{{\ttfamily
  1811.00024}}].

\bibitem{Baumann:2019oyu}
D.~Baumann, C.~Duaso~Pueyo, A.~Joyce, H.~Lee and G.L.~Pimentel, \emph{{The
  cosmological bootstrap: weight-shifting operators and scalar seeds}},
  \href{https://doi.org/10.1007/JHEP12(2020)204}{\emph{JHEP} {\bfseries 12}
  (2020) 204} [\href{https://arxiv.org/abs/1910.14051}{{\ttfamily
  1910.14051}}].

\bibitem{Albayrak:2018tam}
S.~Albayrak and S.~Kharel, \emph{{Towards the higher point holographic momentum
  space amplitudes}},
  \href{https://doi.org/10.1007/JHEP02(2019)040}{\emph{JHEP} {\bfseries 02}
  (2019) 040} [\href{https://arxiv.org/abs/1810.12459}{{\ttfamily
  1810.12459}}].

\bibitem{Sleight:2019mgd}
C.~Sleight, \emph{{A Mellin Space Approach to Cosmological Correlators}},
  \href{https://doi.org/10.1007/JHEP01(2020)090}{\emph{JHEP} {\bfseries 01}
  (2020) 090} [\href{https://arxiv.org/abs/1906.12302}{{\ttfamily
  1906.12302}}].

\bibitem{Sleight:2019hfp}
C.~Sleight and M.~Taronna, \emph{{Bootstrapping Inflationary Correlators in
  Mellin Space}}, \href{https://doi.org/10.1007/JHEP02(2020)098}{\emph{JHEP}
  {\bfseries 02} (2020) 098}
  [\href{https://arxiv.org/abs/1907.01143}{{\ttfamily 1907.01143}}].

\bibitem{Baumgart:2019clc}
M.~Baumgart and R.~Sundrum, \emph{{De Sitter Diagrammar and the Resummation of
  Time}}, \href{https://doi.org/10.1007/JHEP07(2020)119}{\emph{JHEP} {\bfseries
  07} (2020) 119} [\href{https://arxiv.org/abs/1912.09502}{{\ttfamily
  1912.09502}}].

\bibitem{Gorbenko:2019rza}
V.~Gorbenko and L.~Senatore, \emph{{$\lambda \phi^4$ in dS}},
  \href{https://arxiv.org/abs/1911.00022}{{\ttfamily 1911.00022}}.

\bibitem{Mirbabayi:2020vyt}
M.~Mirbabayi, \emph{{Markovian dynamics in de Sitter}},
  \href{https://doi.org/10.1088/1475-7516/2021/09/038}{\emph{JCAP} {\bfseries
  09} (2021) 038} [\href{https://arxiv.org/abs/2010.06604}{{\ttfamily
  2010.06604}}].

\bibitem{Cohen:2020php}
T.~Cohen and D.~Green, \emph{{Soft de Sitter Effective Theory}},
  \href{https://doi.org/10.1007/JHEP12(2020)041}{\emph{JHEP} {\bfseries 12}
  (2020) 041} [\href{https://arxiv.org/abs/2007.03693}{{\ttfamily
  2007.03693}}].

\bibitem{Green:2020whw}
D.~Green and R.A.~Porto, \emph{{Signals of a Quantum Universe}},
  \href{https://doi.org/10.1103/PhysRevLett.124.251302}{\emph{Phys. Rev. Lett.}
  {\bfseries 124} (2020) 251302}
  [\href{https://arxiv.org/abs/2001.09149}{{\ttfamily 2001.09149}}].

\bibitem{Green:2020ebl}
D.~Green and E.~Pajer, \emph{{On the Symmetries of Cosmological
  Perturbations}},
  \href{https://doi.org/10.1088/1475-7516/2020/09/032}{\emph{JCAP} {\bfseries
  09} (2020) 032} [\href{https://arxiv.org/abs/2004.09587}{{\ttfamily
  2004.09587}}].

\bibitem{BBBB}
E.~Pajer, \emph{{Building a Boostless Bootstrap for the Bispectrum}},
  \href{https://doi.org/10.1088/1475-7516/2021/01/023}{\emph{JCAP} {\bfseries
  01} (2021) 023} [\href{https://arxiv.org/abs/2010.12818}{{\ttfamily
  2010.12818}}].

\bibitem{Baumann:2020dch}
D.~Baumann, C.~Duaso~Pueyo, A.~Joyce, H.~Lee and G.L.~Pimentel, \emph{{The
  Cosmological Bootstrap: Spinning Correlators from Symmetries and
  Factorization}},
  \href{https://doi.org/10.21468/SciPostPhys.11.3.071}{\emph{SciPost Phys.}
  {\bfseries 11} (2021) 071}
  [\href{https://arxiv.org/abs/2005.04234}{{\ttfamily 2005.04234}}].

\bibitem{Sleight:2020obc}
C.~Sleight and M.~Taronna, \emph{{From AdS to dS exchanges: Spectral
  representation, Mellin amplitudes, and crossing}},
  \href{https://doi.org/10.1103/PhysRevD.104.L081902}{\emph{Phys. Rev. D}
  {\bfseries 104} (2021) L081902}
  [\href{https://arxiv.org/abs/2007.09993}{{\ttfamily 2007.09993}}].

\bibitem{Sleight:2021iix}
C.~Sleight and M.~Taronna, \emph{{On the consistency of (partially-)massless
  matter couplings in de Sitter space}},
  \href{https://doi.org/10.1007/JHEP10(2021)156}{\emph{JHEP} {\bfseries 10}
  (2021) 156} [\href{https://arxiv.org/abs/2106.00366}{{\ttfamily
  2106.00366}}].

\bibitem{Baumann:2021fxj}
D.~Baumann, W.-M.~Chen, C.~Duaso~Pueyo, A.~Joyce, H.~Lee and G.L.~Pimentel,
  \emph{{Linking the singularities of cosmological correlators}},
  \href{https://doi.org/10.1007/JHEP09(2022)010}{\emph{JHEP} {\bfseries 09}
  (2022) 010} [\href{https://arxiv.org/abs/2106.05294}{{\ttfamily
  2106.05294}}].

\bibitem{Albayrak:2020fyp}
S.~Albayrak, S.~Kharel and D.~Meltzer, \emph{{On duality of color and
  kinematics in (A)dS momentum space}},
  \href{https://doi.org/10.1007/JHEP03(2021)249}{\emph{JHEP} {\bfseries 03}
  (2021) 249} [\href{https://arxiv.org/abs/2012.10460}{{\ttfamily
  2012.10460}}].

\bibitem{Albayrak:2020isk}
S.~Albayrak, C.~Chowdhury and S.~Kharel, \emph{{Study of momentum space scalar
  amplitudes in AdS spacetime}},
  \href{https://doi.org/10.1103/PhysRevD.101.124043}{\emph{Phys. Rev. D}
  {\bfseries 101} (2020) 124043}
  [\href{https://arxiv.org/abs/2001.06777}{{\ttfamily 2001.06777}}].

\bibitem{Armstrong:2020woi}
C.~Armstrong, A.E.~Lipstein and J.~Mei, \emph{{Color/kinematics duality in
  AdS$_{4}$}}, \href{https://doi.org/10.1007/JHEP02(2021)194}{\emph{JHEP}
  {\bfseries 02} (2021) 194}
  [\href{https://arxiv.org/abs/2012.02059}{{\ttfamily 2012.02059}}].

\bibitem{Meltzer:2021zin}
D.~Meltzer, \emph{{The inflationary wavefunction from analyticity and
  factorization}},
  \href{https://doi.org/10.1088/1475-7516/2021/12/018}{\emph{JCAP} {\bfseries
  12} (2021) 018} [\href{https://arxiv.org/abs/2107.10266}{{\ttfamily
  2107.10266}}].

\bibitem{Bonifacio:2021azc}
J.~Bonifacio, E.~Pajer and D.-G.~Wang, \emph{{From amplitudes to contact
  cosmological correlators}},
  \href{https://doi.org/10.1007/JHEP10(2021)001}{\emph{JHEP} {\bfseries 10}
  (2021) 001} [\href{https://arxiv.org/abs/2106.15468}{{\ttfamily
  2106.15468}}].

\bibitem{Hogervorst:2021uvp}
M.~Hogervorst, J.a.~Penedones and K.S.~Vaziri, \emph{{Towards the
  non-perturbative cosmological bootstrap}},
  \href{https://arxiv.org/abs/2107.13871}{{\ttfamily 2107.13871}}.

\bibitem{DiPietro:2021sjt}
L.~Di~Pietro, V.~Gorbenko and S.~Komatsu, \emph{{Analyticity and unitarity for
  cosmological correlators}},
  \href{https://doi.org/10.1007/JHEP03(2022)023}{\emph{JHEP} {\bfseries 03}
  (2022) 023} [\href{https://arxiv.org/abs/2108.01695}{{\ttfamily
  2108.01695}}].

\bibitem{Sleight:2021plv}
C.~Sleight and M.~Taronna, \emph{{From dS to AdS and back}},
  \href{https://doi.org/10.1007/JHEP12(2021)074}{\emph{JHEP} {\bfseries 12}
  (2021) 074} [\href{https://arxiv.org/abs/2109.02725}{{\ttfamily
  2109.02725}}].

\bibitem{Goodhew:2021oqg}
H.~Goodhew, S.~Jazayeri, M.H.~Gordon~Lee and E.~Pajer, \emph{{Cutting
  cosmological correlators}},
  \href{https://doi.org/10.1088/1475-7516/2021/08/003}{\emph{JCAP} {\bfseries
  08} (2021) 003} [\href{https://arxiv.org/abs/2104.06587}{{\ttfamily
  2104.06587}}].

\bibitem{Premkumar:2021mlz}
A.~Premkumar, \emph{{Regulating Loops in dS}},
  \href{https://arxiv.org/abs/2110.12504}{{\ttfamily 2110.12504}}.

\bibitem{Bzowski:2022rlz}
A.~Bzowski, P.~McFadden and K.~Skenderis, \emph{{A handbook of holographic
  4-point functions}},  \href{https://arxiv.org/abs/2207.02872}{{\ttfamily
  2207.02872}}.

\bibitem{Pethybridge:2021rwf}
B.~Pethybridge and V.~Schaub, \emph{{Tensors and spinors in de Sitter space}},
  \href{https://doi.org/10.1007/JHEP06(2022)123}{\emph{JHEP} {\bfseries 06}
  (2022) 123} [\href{https://arxiv.org/abs/2111.14899}{{\ttfamily
  2111.14899}}].

\bibitem{Green:2022slj}
D.~Green, Y.~Huang and C.-H.~Shen, \emph{{Inflationary Adler Conditions}},
  \href{https://arxiv.org/abs/2208.14544}{{\ttfamily 2208.14544}}.

\bibitem{Pimentel:2022fsc}
G.L.~Pimentel and D.-G.~Wang, \emph{{Boostless cosmological collider
  bootstrap}}, \href{https://doi.org/10.1007/JHEP10(2022)177}{\emph{JHEP}
  {\bfseries 10} (2022) 177}
  [\href{https://arxiv.org/abs/2205.00013}{{\ttfamily 2205.00013}}].

\bibitem{Jazayeri:2022kjy}
S.~Jazayeri and S.~Renaux-Petel, \emph{{Cosmological Bootstrap in Slow
  Motion}},  \href{https://arxiv.org/abs/2205.10340}{{\ttfamily 2205.10340}}.

\bibitem{Armstrong:2022csc}
C.~Armstrong, H.~Gomez, R.~Lipinski~Jusinskas, A.~Lipstein and J.~Mei,
  \emph{{Effective field theories and cosmological scattering equations}},
  \href{https://doi.org/10.1007/JHEP08(2022)054}{\emph{JHEP} {\bfseries 08}
  (2022) 054} [\href{https://arxiv.org/abs/2204.08931}{{\ttfamily
  2204.08931}}].

\bibitem{Cabass:2022rhr}
G.~Cabass, S.~Jazayeri, E.~Pajer and D.~Stefanyszyn, \emph{{Parity violation in
  the scalar trispectrum: no-go theorems and yes-go examples}},
  \href{https://arxiv.org/abs/2210.02907}{{\ttfamily 2210.02907}}.

\bibitem{Armstrong:2022jsa}
C.~Armstrong, H.~Gomez, R.~Lipinski~Jusinskas, A.~Lipstein and J.~Mei,
  \emph{{New recursions for tree-level correlators in (Anti) de Sitter space}},
   \href{https://arxiv.org/abs/2209.02709}{{\ttfamily 2209.02709}}.

\bibitem{Armstrong:2022vgl}
C.~Armstrong, A.~Lipstein and J.~Mei, \emph{{Enhanced Soft Limits in de Sitter
  Space}},  \href{https://arxiv.org/abs/2210.02285}{{\ttfamily 2210.02285}}.

\bibitem{Mirbabayi:2022gnl}
M.~Mirbabayi and F.~Riccardi, \emph{{Probing de Sitter from the horizon}},
  \href{https://arxiv.org/abs/2211.11672}{{\ttfamily 2211.11672}}.

\bibitem{Maldacena:2002vr}
J.M.~Maldacena, \emph{{Non-Gaussian features of primordial fluctuations in
  single field inflationary models}},
  \href{https://doi.org/10.1088/1126-6708/2003/05/013}{\emph{JHEP} {\bfseries
  05} (2003) 013} [\href{https://arxiv.org/abs/astro-ph/0210603}{{\ttfamily
  astro-ph/0210603}}].

\bibitem{Benincasa:2019vqr}
P.~Benincasa, \emph{{Cosmological Polytopes and the Wavefuncton of the Universe
  for Light States}},  \href{https://arxiv.org/abs/1909.02517}{{\ttfamily
  1909.02517}}.

\bibitem{Hillman:2021bnk}
A.~Hillman and E.~Pajer, \emph{{A differential representation of cosmological
  wavefunctions}}, \href{https://doi.org/10.1007/JHEP04(2022)012}{\emph{JHEP}
  {\bfseries 04} (2022) 012}
  [\href{https://arxiv.org/abs/2112.01619}{{\ttfamily 2112.01619}}].

\bibitem{MLT}
S.~Jazayeri, E.~Pajer and D.~Stefanyszyn, \emph{{From locality and unitarity to
  cosmological correlators}},
  \href{https://doi.org/10.1007/JHEP10(2021)065}{\emph{JHEP} {\bfseries 10}
  (2021) 065} [\href{https://arxiv.org/abs/2103.08649}{{\ttfamily
  2103.08649}}].

\bibitem{Baumann:2022jpr}
D.~Baumann, D.~Green, A.~Joyce, E.~Pajer, G.L.~Pimentel, C.~Sleight et~al.,
  \emph{{Snowmass White Paper: The Cosmological Bootstrap}},  in \emph{{2022
  Snowmass Summer Study}}, 3, 2022
  [\href{https://arxiv.org/abs/2203.08121}{{\ttfamily 2203.08121}}].

\bibitem{Benincasa:2022gtd}
P.~Benincasa, \emph{{Amplitudes meet Cosmology: A (Scalar) Primer}},
  \href{https://arxiv.org/abs/2203.15330}{{\ttfamily 2203.15330}}.

\bibitem{COT}
H.~Goodhew, S.~Jazayeri and E.~Pajer, \emph{{The Cosmological Optical
  Theorem}}, \href{https://doi.org/10.1088/1475-7516/2021/04/021}{\emph{JCAP}
  {\bfseries 04} (2021) 021}
  [\href{https://arxiv.org/abs/2009.02898}{{\ttfamily 2009.02898}}].

\bibitem{Seery:2006vu}
D.~Seery, J.E.~Lidsey and M.S.~Sloth, \emph{{The inflationary trispectrum}},
  \href{https://doi.org/10.1088/1475-7516/2007/01/027}{\emph{JCAP} {\bfseries
  01} (2007) 027} [\href{https://arxiv.org/abs/astro-ph/0610210}{{\ttfamily
  astro-ph/0610210}}].

\bibitem{Seery:2008ax}
D.~Seery, M.S.~Sloth and F.~Vernizzi, \emph{{Inflationary trispectrum from
  graviton exchange}},
  \href{https://doi.org/10.1088/1475-7516/2009/03/018}{\emph{JCAP} {\bfseries
  03} (2009) 018} [\href{https://arxiv.org/abs/0811.3934}{{\ttfamily
  0811.3934}}].

\bibitem{Chen:2006nt}
X.~Chen, M.-x.~Huang, S.~Kachru and G.~Shiu, \emph{{Observational signatures
  and non-Gaussianities of general single field inflation}},
  \href{https://doi.org/10.1088/1475-7516/2007/01/002}{\emph{JCAP} {\bfseries
  01} (2007) 002} [\href{https://arxiv.org/abs/hep-th/0605045}{{\ttfamily
  hep-th/0605045}}].

\bibitem{Pajer:2016ieg}
E.~Pajer, G.L.~Pimentel and J.V.S.~Van~Wijck, \emph{{The Conformal Limit of
  Inflation in the Era of CMB Polarimetry}},
  \href{https://doi.org/10.1088/1475-7516/2017/06/009}{\emph{JCAP} {\bfseries
  06} (2017) 009} [\href{https://arxiv.org/abs/1609.06993}{{\ttfamily
  1609.06993}}].

\bibitem{Anninos:2014lwa}
D.~Anninos, T.~Anous, D.Z.~Freedman and G.~Konstantinidis, \emph{{Late-time
  Structure of the Bunch-Davies De Sitter Wavefunction}},
  \href{https://doi.org/10.1088/1475-7516/2015/11/048}{\emph{JCAP} {\bfseries
  11} (2015) 048} [\href{https://arxiv.org/abs/1406.5490}{{\ttfamily
  1406.5490}}].

\bibitem{Goon:2018fyu}
G.~Goon, K.~Hinterbichler, A.~Joyce and M.~Trodden, \emph{{Shapes of gravity:
  Tensor non-Gaussianity and massive spin-2 fields}},
  \href{https://doi.org/10.1007/JHEP10(2019)182}{\emph{JHEP} {\bfseries 10}
  (2019) 182} [\href{https://arxiv.org/abs/1812.07571}{{\ttfamily
  1812.07571}}].

\bibitem{Weinberg:2005vy}
S.~Weinberg, \emph{{Quantum contributions to cosmological correlations}},
  \href{https://doi.org/10.1103/PhysRevD.72.043514}{\emph{Phys. Rev. D}
  {\bfseries 72} (2005) 043514}
  [\href{https://arxiv.org/abs/hep-th/0506236}{{\ttfamily hep-th/0506236}}].

\bibitem{Karateev:2017jgd}
D.~Karateev, P.~Kravchuk and D.~Simmons-Duffin, \emph{{Weight Shifting
  Operators and Conformal Blocks}},
  \href{https://doi.org/10.1007/JHEP02(2018)081}{\emph{JHEP} {\bfseries 02}
  (2018) 081} [\href{https://arxiv.org/abs/1706.07813}{{\ttfamily
  1706.07813}}].

\bibitem{Cespedes:2020xqq}
S.~C\'espedes, A.-C.~Davis and S.~Melville, \emph{{On the time evolution of
  cosmological correlators}},
  \href{https://doi.org/10.1007/JHEP02(2021)012}{\emph{JHEP} {\bfseries 02}
  (2021) 012} [\href{https://arxiv.org/abs/2009.07874}{{\ttfamily
  2009.07874}}].

\bibitem{Melville:2021lst}
S.~Melville and E.~Pajer, \emph{{Cosmological Cutting Rules}},
  \href{https://doi.org/10.1007/JHEP05(2021)249}{\emph{JHEP} {\bfseries 05}
  (2021) 249} [\href{https://arxiv.org/abs/2103.09832}{{\ttfamily
  2103.09832}}].

\bibitem{Goodhew:2022ayb}
H.~Goodhew, \emph{{Rational Wavefunctions in de Sitter Spacetime}},
  \href{https://arxiv.org/abs/2210.09977}{{\ttfamily 2210.09977}}.

\bibitem{Ghosh:2014kba}
A.~Ghosh, N.~Kundu, S.~Raju and S.P.~Trivedi, \emph{{Conformal Invariance and
  the Four Point Scalar Correlator in Slow-Roll Inflation}},
  \href{https://doi.org/10.1007/JHEP07(2014)011}{\emph{JHEP} {\bfseries 07}
  (2014) 011} [\href{https://arxiv.org/abs/1401.1426}{{\ttfamily 1401.1426}}].

\bibitem{Bittermann:2022nfh}
N.~Bittermann and A.~Joyce, \emph{{Soft limits of the wavefunction in
  exceptional scalar theories}},
  \href{https://arxiv.org/abs/2203.05576}{{\ttfamily 2203.05576}}.

\bibitem{Skenderis:2002wp}
K.~Skenderis, \emph{{Lecture notes on holographic renormalization}},
  \href{https://doi.org/10.1088/0264-9381/19/22/306}{\emph{Class. Quant. Grav.}
  {\bfseries 19} (2002) 5849}
  [\href{https://arxiv.org/abs/hep-th/0209067}{{\ttfamily hep-th/0209067}}].

\bibitem{Hinterbichler:2013dpa}
K.~Hinterbichler, L.~Hui and J.~Khoury, \emph{{An Infinite Set of Ward
  Identities for Adiabatic Modes in Cosmology}},
  \href{https://doi.org/10.1088/1475-7516/2014/01/039}{\emph{JCAP} {\bfseries
  01} (2014) 039} [\href{https://arxiv.org/abs/1304.5527}{{\ttfamily
  1304.5527}}].

\bibitem{BCFW}
R.~Britto, F.~Cachazo, B.~Feng and E.~Witten, \emph{{Direct proof of tree-level
  recursion relation in Yang-Mills theory}},
  \href{https://doi.org/10.1103/PhysRevLett.94.181602}{\emph{Phys. Rev. Lett.}
  {\bfseries 94} (2005) 181602}
  [\href{https://arxiv.org/abs/hep-th/0501052}{{\ttfamily hep-th/0501052}}].

\bibitem{Benincasa:2007qj}
P.~Benincasa, C.~Boucher-Veronneau and F.~Cachazo, \emph{{Taming Tree
  Amplitudes In General Relativity}},
  \href{https://doi.org/10.1088/1126-6708/2007/11/057}{\emph{JHEP} {\bfseries
  11} (2007) 057} [\href{https://arxiv.org/abs/hep-th/0702032}{{\ttfamily
  hep-th/0702032}}].

\bibitem{Parke:1986gb}
S.J.~Parke and T.R.~Taylor, \emph{{An Amplitude for $n$ Gluon Scattering}},
  \href{https://doi.org/10.1103/PhysRevLett.56.2459}{\emph{Phys. Rev. Lett.}
  {\bfseries 56} (1986) 2459}.

\bibitem{Parke:1985ax}
S.J.~Parke and T.R.~Taylor, \emph{{Gluonic Two Goes to Four}},
  \href{https://doi.org/10.1016/0550-3213(86)90230-0}{\emph{Nucl. Phys. B}
  {\bfseries 269} (1986) 410}.

\end{thebibliography}\endgroup

\end{document}